\documentclass[sigconf]{acmart}

\usepackage{ifpdf}
\ifpdf
\else
\fi

\usepackage{graphicx}
\usepackage{algorithm}
\usepackage{algorithmic}
\usepackage{hyperref}
\usepackage{multirow}
\usepackage{booktabs}
\usepackage{flushend}
\usepackage{soul}

  \renewenvironment{thebibliography}[1]{%
    \begin{oldthebibliography}{#1}%
      \setlength{\parskip}{0ex}%
      \setlength{\itemsep}{0ex}%
  }%
  {%
    \end{oldthebibliography}%
  }

\def\BibTeX{{\rm B\kern-.05em{\sc i\kern-.025em b}\kern-.08em
    T\kern-.1667em\lower.7ex\hbox{E}\kern-.125emX}}

\renewcommand{\em}{\it}

\newcommand{\ignore}[1]{}
\newcommand{\inst}[1]{\texttt{#1}}

\def\cfigure[#1,#2,#3]{
\begin{figure}[t]
\begin{center}

\includegraphics[width=3.4in]{#1} 
 
\vspace*{-2mm}
\caption[]{#2
} \label{#3}
 
\end{center}
\vspace*{-0.22in}
\end{figure}}

\def\cfiguredouble[#1,#2,#3,#4]{
\begin{figure}
\begin{center}
\includegraphics[width=3.4in]{#1} 
\\(a)\\
\includegraphics[width=3.4in]{#2} 
\\(b)\\
\caption[]{#3} \label{#4}
\end{center}
\vspace*{-0.2in}
\end{figure}}

\def\cfiguretemp[#1,#2,#3]{
\begin{figure}
\vspace*{0mm}
\begin{center}

\includegraphics[width=3.5in]{#1} 
 
\vspace*{-3mm}\caption[]{#2
} \label{#3}
 
\vspace*{-5mm}
\end{center}
\vspace*{-2mm}
\end{figure}}

\def\wfigure[#1,#2,#3]{
\begin{figure*}
\vspace*{0mm}
\begin{center}
 
\includegraphics[width=6in]{#1} 
 
\vspace*{-3mm}\caption[]{#2
} \label{#3}
 
\vspace*{-5mm}
\end{center}
\vspace*{-2mm}
\end{figure*}}

\def\threefigure[#1,#2,#3,#4,#5]{
\begin{figure*}
\vspace*{0mm}
\begin{center}

\begin{tabular}{ccc}
\includegraphics[width=2in]{#1} & \includegraphics[width=2in]{#2} &  \includegraphics[width=2in]{#3} \\
(a) & (b) & (c) \\
\end{tabular}

\vspace*{-3mm}\caption[]{#4
} \label{#5}

\vspace*{-5mm}
\end{center}
\vspace*{-2mm}
\end{figure*}}

\def\dcfigure[#1,#2,#3,#4,#5,#6]{
{
\begin{figure*}
\vspace*{0.2in}\
\begin{center}
\begin{minipage}[c]{3in}{
\includegraphics[width=3in]{#1} 
\vspace*{-3mm}\caption[]{#2} \label{#3} \
}\end{minipage}\hspace*{0.5in}\
\begin{minipage}[c]{3in}{
\includegraphics[width=3in]{#4} 
\vspace*{-3mm}\caption[]{#5}\label{#6} \
}\end{minipage}
\end{center}
\vspace*{-0.4in}\
\end{figure*}
}
}

\def\qcfigure[#1,#2,#3,#4,#5,#6]{
{
\begin{figure*}
\vspace*{0.2in}\
\begin{center}
\begin{minipage}[c]{3in}{
\includegraphics[width=3in]{#1} 
\vspace*{-3mm}
}
\end{minipage}\hspace*{0.5in}\
\begin{minipage}[c]{3in}{
\includegraphics[width=3in]{#2} 
\vspace*{-3mm}
}\end{minipage}

\begin{minipage}[c]{3in}{
\includegraphics[width=3in]{#3} 
\vspace*{-3mm}
}
\end{minipage}\hspace*{0.5in}\
\begin{minipage}[c]{3in}{
\includegraphics[width=3in]{#4} 
\vspace*{-3mm}
}\end{minipage}
\end{center}
\caption[]{#5}\label{#6}
\end{figure*}
}
}

\def\twfigureabc[#1,#2,#3,#4,#5]{
{
\tiny
\begin{figure}
\begin{center}
\begin{minipage}[c]{3.4in}{
\includegraphics[width=3.4in]{#1} 
\vspace*{-4mm}
}
\end{minipage}
\\(a)\\

\begin{minipage}[c]{3.4in}{
\includegraphics[width=3.4in]{#2} 
\vspace*{-4mm}
}\end{minipage}
\\(b)\\

\begin{minipage}[c]{3.4in}{
\includegraphics[width=3.4in]{#3} 
\vspace*{-4mm}
}
\end{minipage}
\\(c)\\
\end{center}
\vspace*{-6mm}
\caption[]{#4}
\label{#5}
\end{figure}
}
}

\def\twfigure[#1,#2,#3,#4,#5]{
{
\begin{figure}
\vspace*{0.2in}\
\begin{center}
\begin{minipage}[c]{6.5in}{
\includegraphics[width=6.5in]{#1} 
\vspace*{-3mm}
}
\end{minipage}

\begin{minipage}[c]{6.5in}{
\includegraphics[width=6.5in]{#2} 
\vspace*{-3mm}
}\end{minipage}

\begin{minipage}[c]{6.5in}{
\includegraphics[width=6.5in]{#3} 
\vspace*{-3mm}
}
\end{minipage}
\end{center}
\caption[]{#4}\label{#5}
\end{figure}
}
}

\def\dwfigure[#1,#2,#3,#4]{
{
\begin{figure*}
\vspace*{0.2in}\
\begin{center}
\begin{minipage}[c]{3.5in}{
\includegraphics[width=3.5in]{#1} 
\vspace*{-3mm}
}
\end{minipage}

\begin{minipage}[c]{3.5in}{
\includegraphics[width=3.5in]{#2} 
\vspace*{-3mm}
}\end{minipage}

\end{center}
\caption[]{#3}\label{#4}
\end{figure*}
}
}

\def\dssfigure[#1,#2,#3,#4,#5,#6]{
{
\begin{figure*}
\vspace*{0.2in}\
\begin{center}
\begin{minipage}[c]{4in}{
\includegraphics[width=4in]{#1}
\vspace*{-3mm}\caption[]{#2} \label{#3} \
}\end{minipage}\hspace*{0.5in}\
\begin{minipage}[c]{2in}{
\includegraphics[width=2in]{#4}
\vspace*{-3mm}\caption[]{#5}\label{#6} \
}\end{minipage}
\end{center}
\vspace*{-0.4in}\
\end{figure*}
}
}

\def\dsfigure[#1,#2,#3,#4,#5,#6]{
{
\begin{figure*}
\vspace*{0.2in}\
\begin{center}
\begin{minipage}[c]{3in}{
\includegraphics[width=3in]{#1}
\vspace*{-3mm}\caption[]{#2} \label{#3} \
}\end{minipage}\hspace*{0.5in}\
\begin{minipage}[c]{3in}{
\hspace*{0.5in}\
\includegraphics[height=3in]{#4}
\vspace*{-3mm}\caption[]{#5}\label{#6} \
}\end{minipage}
\end{center}
\vspace*{-0.4in}\
\end{figure*}
}
}

\def\dsyfigure[#1,#2,#3,#4,#5,#6]{
{
\begin{figure*}
\vspace*{0.2in}\
\begin{center}
\begin{minipage}[c]{2.5in}{
\includegraphics[height=2.5in]{#1}
\vspace*{-3mm}\caption[]{#2} \label{#3} \
}\end{minipage}\hspace*{0.5in}\
\begin{minipage}[c]{2.5in}{
\includegraphics[height=2.5in]{#4}
\vspace*{-3mm}\caption[]{#5}\label{#6} \
}\end{minipage}
\end{center}
\vspace*{-0.4in}\
\end{figure*}
}
}

\def\dyfigure[#1,#2,#3,#4,#5,#6]{
{
\begin{figure*}
\vspace*{0.2in}\
\begin{center}
\begin{minipage}[c]{3in}{
\includegraphics[height=3in]{#1} 
\vspace*{-3mm}\caption[]{#2} \label{#3} \
}\end{minipage}\hspace*{0.5in}\
\begin{minipage}[c]{3in}{
\includegraphics[height=3in]{#4} 
\vspace*{-3mm}\caption[]{#5}\label{#6} \
}\end{minipage}
\end{center}
\vspace*{-0.4in}\
\end{figure*}
}
}

\def\dyoldfigure[#1,#2,#3,#4,#5,#6]{
{
\begin{figure*}
\vspace*{0.2in}\
\begin{center}
\begin{minipage}[c]{3in}{
\epsfysize=2.0in\
\hspace{0.5in}\
\epsfbox{#1}
\vspace*{-3mm}\caption[]{#2} \label{#3} \
}\end{minipage}\hspace*{0.25in}\
\begin{minipage}[c]{3in}{
\epsfysize=2.0in\
\hspace{0.5in}\
\epsfbox{#4}
\vspace*{-3mm}\caption[]{#5}\label{#6} \
}\end{minipage}
\end{center}
\vspace*{-0.4in}\
\end{figure*}
}
}

\def\wpfigure[#1,#2,#3,#4]{
\begin{figure*}
\vspace*{4mm}
\begin{center}

\includegraphics[width=#4]{#1} 

\vspace*{-3mm}\caption[]{#2
} \label{#3}

\vspace*{-5mm}
\end{center}
\end{figure*}}

\def\wprfigure[#1,#2,#3,#4,#5]{
\begin{figure*}
\vspace*{4mm}
\begin{center}

\includegraphics[width=#4, angle=#5]{#1} 

\vspace*{-3mm}\caption[]{#2
} \label{#3}

\vspace*{-5mm}
\end{center}
\end{figure*}}

\def\DoubleFigureWSlide[#1,#2,#3,#4,#5,#6,#7,#8,#9]{
\begin{figure*}
\vspace*{#9}
\begin{center}
\begin{minipage}{#4}
\includegraphics[width=#4]{#1}
\vspace*{-3mm}\caption{#2
}\label{#3}
\end{minipage}
\hspace{2em}
\begin{minipage}{#8}
\includegraphics[width=#8]{#5}
\vspace*{-3mm}\caption{#6
}\label{#7}
\end{minipage}
\vspace*{-5mm}
\end{center}
\end{figure*}
}

\def\DoubleFigureW[#1,#2,#3,#4,#5,#6,#7,#8]{
\begin{figure*}
\vspace*{0in}
\begin{center}
\begin{minipage}{#4}
\includegraphics[width=#4]{#1}
\vspace*{-3mm}\caption{#2
}\label{#3}
\end{minipage}
\hspace{2em}
\begin{minipage}{#8}
\includegraphics[width=#8]{#5}
\vspace*{-3mm}\caption{#6
}\label{#7}
\end{minipage}
\vspace*{-5mm}
\end{center}
\end{figure*}
}

\def\DoubleFigureWHack[#1,#2,#3,#4,#5,#6,#7,#8]{
\begin{figure*}
\vspace*{0in}
\begin{center}
\begin{minipage}{3in}
\includegraphics[width=#4]{#1}
\vspace*{-3mm}\caption{#2
}\label{#3}
\end{minipage}
\hspace{2em}
\begin{minipage}{3in}
\includegraphics[width=#8]{#5}
\vspace*{-3mm}\caption{#6
}\label{#7}
\end{minipage}
\vspace*{-5mm}
\end{center}
\end{figure*}
}

\def\ddcfigure[#1,#2,#3,#4]{
\begin{figure*}[t]
\vspace*{0.2in}\
\begin{center}
\begin{minipage}[c]{3in}{
\includegraphics[width=3in]{#1} 
}\end{minipage}\hspace{0.5in}\
\begin{minipage}[c]{3in}{
\includegraphics[width=3in]{#2} 
}\end{minipage}\vspace*{-0.10in} \caption[]{#3}\label{#4}
\end{center}
\vspace*{-0.4in}\
\end{figure*}
}

\def\dddcfigure[#1,#2,#3,#4]{
\begin{figure*}
\vspace*{0.2in}\
\begin{center}
\begin{tabular}{cc}
\includegraphics[width=3in]{#1} &
\includegraphics[width=3in]{#2} \\
(a) & (b) \\
\end{tabular}\vspace*{-0.10in}\caption[]{#3}\label{#4}
\end{center}
\vspace*{-0.4in}\
\end{figure*}
}

\def\qqcfigure[#1,#2,#3,#4,#5,#6]{
\begin{figure*}[t]
\begin{center}
\begin{tabular}{cc}
\includegraphics[width=3.3in]{#1} &
\includegraphics[width=3.3in]{#2} \\
(a) & (b) \\
\includegraphics[width=3.3in]{#3} &
\includegraphics[width=3.3in]{#4} \\
(c) & (d) \\
\end{tabular}\vspace*{-0.10in}\caption[]{#5}\label{#6}
\end{center}
\vspace*{-0.4in}\
\end{figure*}
}
\def\qqcfiguresinglecol[#1,#2,#3,#4,#5,#6]{
\begin{figure}[t]
\begin{center}
\begin{tabular}{cc}
\hspace*{-0.2in}\includegraphics[width=2in]{#1} &
\hspace*{-0.4in}\includegraphics[width=2in]{#2} \\
\hspace*{-0.2in}(a) & 
\hspace*{-0.4in}(b) \\
\hspace*{-0.2in}\includegraphics[width=2in]{#3} &
\hspace*{-0.2in}\includegraphics[width=2in]{#4} \\
\hspace*{-0.2in}(c) & 
\hspace*{-0.4in}(d) \\
\end{tabular}\vspace*{-0.10in}\caption[]{#5}\label{#6}
\end{center}
\vspace*{-0.4in}\
\end{figure}
}

\def\qqcfigureInAColumn[#1,#2,#3,#4,#5,#6]{
\begin{figure}[t]
\begin{center}
\includegraphics[width=3.3in]{#1} \\
(a)\\
\includegraphics[width=3.3in]{#2} \\
(b) \\
\includegraphics[width=3.3in]{#3} \\
(c) \\
\includegraphics[width=3.3in]{#4} \\
(d) \\
\vspace*{-0.10in}\caption[]{#5}\label{#6}
\end{center}
\vspace*{-0.5in}\
\end{figure}
}

\def\sixfigure[#1,#2,#3,#4,#5,#6,#7,#8]{
\begin{figure*}[t]
\vspace*{0.2in}\
\begin{center}
\begin{tabular}{cc}
\includegraphics[width=3in]{#1} &
\includegraphics[width=3in]{#2} \\
(a) & (b) \\
\includegraphics[width=3in]{#3} &
\includegraphics[width=3in]{#4} \\
(c) & (d) \\
\includegraphics[width=3in]{#5} &
\includegraphics[width=3in]{#6} \\
(e) & (f) \\
\end{tabular}\vspace*{-0.10in}\caption[]{#7}\label{#8}
\end{center}
\vspace*{-0.4in}\
\end{figure*}
}

\def\ddcfigureSlide[#1,#2,#3,#4,#5]{
\begin{figure*}
\vspace*{#5}\
\begin{center}
\begin{minipage}[c]{3in}{
\includegraphics[height=3in]{#1} 
}\end{minipage}\hspace{0.5in}\
\begin{minipage}[c]{3in}{
\includegraphics[height=3in]{#2} 
}\end{minipage}\vspace*{-0.10in} \caption[]{#3}\label{#4}
\end{center}
\vspace*{-0.4in}\
\end{figure*}
}

\def\dcfigureSingleCol[#1,#2,#3,#4]{
\begin{figure}
\begin{center}
\begin{tabular}{cc}
\hspace*{-0.2in}\includegraphics[width=2.4in]{#2} &
\hspace*{-0.2in}\includegraphics[width=1.2in]{#1} \\
\hspace*{-0.2in}(a) & 
\hspace*{-0.2in}(b) \\
\end{tabular}\vspace*{-0.10in}\caption[]{#3}\label{#4}
\end{center}
\vspace*{-0.4in}\
\end{figure}
}

\def\cxfigure[#1,#2,#3]{
\begin{figure}
\vspace*{4mm}
\begin{center}
 
\epsfxsize=2.5in\
\epsfbox{#1}\
 
\vspace*{-0.10in}\caption[]{#2
} \label{#3}
 
\vspace*{-5mm}
\end{center}
\vspace*{-2mm}
\end{figure}}

\newcommand{\x}{$\times$}

\newif\ifremark
\long\def\remark#1{
\ifremark%
        \begingroup%
        \dimen0=\columnwidth
        \advance\dimen0 by -1in%
        \setbox0=\hbox{\parbox[b]{\dimen0}{\protect\em #1}}
        \dimen1=\ht0\advance\dimen1 by 2pt%
        \dimen2=\dp0\advance\dimen2 by 2pt%
        \vskip 0.25pt%
        \hbox to \columnwidth{%
                \vrule height\dimen1 width 3pt depth\dimen2%
                \hss\copy0\hss%
                \vrule height\dimen1 width 3pt depth\dimen2%
        }%
        \endgroup%
\fi}

\remarktrue

\setcopyright{none}
\renewcommand\footnotetextcopyrightpermission[1]{This paper is a pre-print of a paper in the 2021 SC, the International Conference for High Performance Computing, Networking, Storage and Analysis. 
Please refer to the conference proceedings for the most complete version.}
\begin{document}



\newcommand{\mm}{mm$^2$}
\newcommand{\figtitle}[1]{\textbf{#1}}
\newcommand{\us}{$\mu$s}
\newcommand{\fixme}[1]{#1}
\newcommand{\adrian}[1]{{\color{green}\textbf{#1}}}
\newcommand{\laura}[1]{{\color{pink}\textbf{#1}}}
\newcommand{\joel}[1]{{\color{red}\textbf{#1}}}
\newcommand{\ameen}[1]{{\color{blue}\textbf{#1}}}
\newcommand{\arup}[1]{{\color{yellow}\textbf{#1}}}
\newcommand{\hungwei}[1]{{\color{blue}\textbf{#1}}}
\newcommand{\jay}[1]{{\color{red}\textbf{#1}}}
\newcommand{\Bella}[1]{{\color{blue}\textbf{\textit{#1}}}}

\newcommand{\note}[2]{{\color{red}\fixme{$\ll$ #1 $\gg$ #2}}}
\newcommand{\myitem}[1]{\hspace*{-\parindent}\textbf{#1}\hspace*{\parindent}}
\newcommand{\KaleidoSSD}[1]{Morpheus-SSD}
\newcommand{\GPTPU}[1]{GPETPU}
\newcommand{\Tensorizer}[1]{Tensorizer}
\newcommand{\CTPU}[1]{OpenCtpu}
\newcommand{\ctpu}[1]{openctpu}
\newcommand{\etpu}[1]{Edge TPU}
\newcommand{\KaleidoStorageCaps}[1]{MORPHEUS}
\newcommand{\KaleidoSSDCaps}[1]{MORPHEUS-SSD}
\newcommand{\KaleidoApp}[1]{StorageApp}
\newcommand{\KInit}[1]{MInit}
\newcommand{\KDeinit}[1]{MDeinit}
\newcommand{\KRead}[1]{MRead}
\newcommand{\KWrite}[1]{MWrite}
\newcommand{\SSDD}[1]{NVMe-P2P}
\newcommand{\Hippogriff}[1]{Gullfoss}
\newcommand{\Donard}[1]{NVMe-P2P}
\newcommand{\smallcapsHippogriff}[1]{gullfoss}
\newcommand{\allcapsHippogriff}[1]{GULLFOSS}

\newcommand{\CMF}[1]{{\color{orange}{#1}}} 
\newcommand{\CMFdel}[1]{} 
\newcommand{\CMFcom}[1]{{\color{orange}\textbf{\textit{[CMF: #1]}}}} 
\newcommand{\speedup}[1]{2.46\x{}}
\newcommand{\energysaving}[1]{40\%}
\newcommand{\rv}[1]{{{#1}}}
\newcommand{\ignoreexact}[1]{}


\title{GPTPU: Accelerating Applications using Edge Tensor Processing Units
}
\date{}
\author{Kuan-Chieh Hsu and Hung-Wei Tseng
	\\ University of California, Riverside
	\\ \{khsu037, htseng\}@ucr.edu}
\thispagestyle{firstpage}
\begin{abstract}
Neural network (NN) accelerators have been integrated
into a wide-spectrum of computer systems to accommodate the 
rapidly growing demands for artificial intelligence
(AI) and machine learning (ML) applications.
NN accelerators share the idea of providing native hardware support for operations 
on multidimensional tensor data. 
Therefore, NN accelerators are theoretically tensor processors that can
improve system performance for any problem that uses tensors as inputs/outputs.
Unfortunately, commercially available NN accelerators only expose computation
capabilities through AI/ML-specific interfaces. Furthermore, NN accelerators
reveal very few hardware design details, so applications cannot easily leverage the tensor
operations NN accelerators provide.

This paper introduces General-Purpose Computing on Edge Tensor Processing Units (\GPTPU{}),
an open-source, open-architecture framework that allows the developer and research communities
to discover opportunities that NN accelerators enable
for applications. \GPTPU{} includes a powerful programming interface
with efficient runtime system-level support---similar to
that of CUDA/OpenCL in GPGPU computing---to bridge
the gap between application demands and mismatched hardware/software interfaces.

We built \GPTPU{} machine uses Edge Tensor Processing Units (\etpu{}s), which
are widely available and representative of many commercial NN accelerators.
\ignore{
Our prototype \GPTPU{} machine uses Edge Tensor Processing Units (\etpu{}s), which
are widely available and reprepresentative of many commercial NN accelerators.
To allow \GPTPU{} to efficiently utilize the underlying \etpu{}s, we
identify missing details in \etpu{} datasheets.}
We identified several novel use cases
and revisited the algorithms.
By leveraging the underlying \etpu{}s to
perform tensor-algorithm-based compute kernels, our results reveal 
that \GPTPU{} can achieve a \speedup{} speedup over high-end CPUs and reduce energy consumption
by \energysaving{}.

\end{abstract}

\maketitle
\section{Introduction}
\label{sec:introduction}
\ignore{
The neural network (NN) renaissance has driven research and
development in hardware accelerators for artificial intelligence
(AI) and machine learning (ML) workloads. Perhaps most indicative of this
accelerated expansion are Apple's Neural Engine and Google's Tensor Processing Units (TPUs)~\cite{TPU}
that are commercially available in various platforms, ranging from
mobile/embedded devices to cloud servers. 
These NN accelerators' power/energy efficiency is orders-of-magnitude 
better than that of conventional vector processors (e.g., Graphics Processing
Units [GPUs]) for the same workloads.

The demand for artificial intelligence (AI) and machine learning (ML) applications has exploded in recent years, and the increase 
in AI/ML workloads has led to significant
research advances in, and the commercialization of, neural network (NN)
accelerators, including the widely used Tensor Processing Unit (TPU)~\cite{TPU}. 
}
The demand for artificial intelligence (AI) and machine learning (ML) applications has exploded in recent years, and the increase 
in AI/ML workloads has led to significant
research advances in neural network (NN)
accelerators, including Google's Edge Tensor Processing Units (\etpu{}s)~\cite{EdgeTPUM2}
and Apple's Neural Engines~\cite{AppleM1} that are already
presented as auxiliary hardware components in commodity systems.
These NN accelerators' power/energy efficiency is orders-of-magnitude 
better than that of conventional vector processors (e.g., Graphics Processing
Units [GPUs]) for the same workloads.
Despite the differences among microarchitectures, most NN accelerators are
essentially matrix processors that take tensors/matrices as inputs, generate tensors/matrices as outputs, and provide operators that facilitate NN computations. 

Two decades ago, graphics processing units (GPUs) were just domain-specific
accelerators used for shading and rendering. But intensive research into high-performance 
algorithms, architectures, systems, and
compilers~\cite{ryoo2008optimization,volkov2008benchmarking,garland2008parallel,lee2009openmp,narasiman2011improving,yang2010gpgpu,ryoo2008program,baskaran2008compiler,zhang2011fly,jablin2011automatic} and the availability 
of frameworks like CUDA~\cite{CUDA} and OpenCL~\cite{OpenCL}, have revolutionized GPUs and transformed them into high-performance, general-purpose vector processors.
We expect a similar revolution to take place with NN accelerators---a revolution that will create
general-purpose matrix processors for a broader spectrum of applications.
However, democratizing these NN accelerators for non-AI/ML workloads
will require the system framework and the programmer to tackle the following
issues:

\ignore{
At present, however, NN accelerators only expose hardware/software interfaces
aligned with AI/ML workloads, and programmers are restricted to using NN accelerators with a limited set of applications through 
domain-specific language frameworks. 
}
(1) The microarchitectures and instructions of NN accelerators are optimized for
NN workloads, instead of general matrix/tensor algebra. These auxiliary
NN accelerators focus on latency per inference, but not yet on delivering computation
throughput comparable to GPUs. Naively mapping
conventional matrix/tensor algorithms to AI/ML operations will lead to sub-optimal performance.
\ignore{
Naively treating
these accelerators as matrix/vector multipliers and applying conventional
matrix/tensor algorithms will lead to sub-optimal performance. }
(2) Because many AI/ML applications are error tolerant, NN accelerators typically trade accuracy for
area/energy-efficiency; 
when such a trade-off produces undesirable results, additional mechanisms are needed to make
adjustments.
(3) The programming interfaces of existing NN accelerators are specialized
for developling AI/ML applications. 
Existing frameworks 
expose very few details about the hardware/software interfaces of NN accelerators,
so programmers are 
unable to customize computation and the application can suffer from significant performance
overhead due to adjusting the parameters/data bound to the supported ML
models.
(4) Tensor algorithms are traditionally 
time-consuming, so programmers have tailored compute kernels
in favor of scalar/vector processing.
Such tailoring
makes applications
unable to take advantage of tensor operators without revisiting 
algorithms.

This paper bridges the gap between general-purpose programmability and domain-specific
NN accelerators by presenting a full-stack system architecture that enables General-Purpose Computing on 
\etpu{}s (\GPTPU{}). \GPTPU{} tackles all the aforementioned challenges through
providing a programming interface, a runtime system, compiler and libraries. 
With the system this paper proposes, programmers will be able to explore the enormous potential
of the matrix processing model inherent in \etpu{}, a commercially available 
accelerator that can be part of a system-on-module (SOM) or be easily attached to 
various forms of computer systems. A commercialized \etpu{} can inference ML models 
at 4~TOPS (tera operations per second) with only 2~W of power consumption.
The design that \GPTPU{} demonstrates can also work for NN accelerators sharing similar architectures. 
\ignore{
Nonetheless, the challenges of designing a \GPTPU{} platform are non-trivial because of the need to tackle the following issues: 
}

\GPTPU{} provides a programming framework, including an \etpu{}-specific C/C++ extension, \CTPU{}
and a runtime system. \GPTPU{} offloads programmers from directly interacting
with the accelerator's hardware to focus on the design of tensor-based algorithms
for applications. 
\CTPU{} achieves more programming
flexibility than existing domain-specific 
interfaces by exposing high-level tensor/matrix algebra operators (e.g.,
matrix multiply) and low-level accelerator operators (e.g., convolution and multiplication-accumulation)
to the programmer, so programmers can design arbitrary tensor
algorithms or customize operators that cannot be easily achieved using domain-specific
languages. 
\ignoreexact{\CTPU{} also allows the programmer to specify the desired
accuracy modes (i.e., approximate or exact computing) for operators.}
\ignore{
In contrast to \CTPU{}, domain-specific language frameworks only expose their underlying hardware
accelerators as abstractions of ML models. }
\ignore{
An essential feature of the \GPTPU{} runtime system is that it dynamically translates and reshapes input data to
fulfill the requirements and make the most efficient use of the underlying NN accelerator. 
The \GPTPU{}
runtime also schedules computation tasks and distributes prepared data to available
NN accelerators. The \GPTPU{} backend runtime library
may be implemented to quantize and calibrate input datasets and computation
results, thereby minimizing the impact of limited precision on NN accelerators.
}

The core of the \GPTPU{} runtime system is our proposed {\em \Tensorizer{}}, a
module that dynamically evaluates input data and transforms data into ML models 
that the underlying \etpu{}s or other NN accelerators can efficiently perform inference operations on. 
\Tensorizer{} handles quantization and calibration of input datasets and computation
results, thereby minimizing the impact of limited precision on NN
accelerators. The \GPTPU{}
runtime also schedules computation tasks and distributes prepared data to available
NN accelerators in a manner that minimizes the data movement overhead. 
\ignoreexact{ when the programmer chooses to approximate results. 
In the case of exact computing, \Tensorizer{} carefully layouts the bits
from input data to minimize the amount of computation and data movements. }

Despite the \etpu{}'s promising energy efficiency and
recently open-sourced C++ API, documentation is vague regarding the \etpu{}'s hardware/software
interface and architecture. This lack of detail complicates the
design of systems that fully exploit the \etpu{}'s capabilities.
To develop \GPTPU{}, we measured the performance
of available \etpu{} operators, reverse-engineered the \etpu{} hardware/software 
interface for data exchanges, and analyzed the \etpu{} architecture.
We applied our understanding of the
\etpu{} to optimize the backend runtime system for efficient task creation and data
transformation. We then built a prototype \GPTPU{} system with 8
\etpu{}s to allow concurrent \GPTPU{} task execution.

We demonstrate the potential of the \GPTPU{} system by modifying
several key applications for financial computing, linear algebra, physics
simulations and graph analytics. 
By revisiting
the algorithms at the heart of these applications and using \CTPU{}, we show that \GPTPU{} can simplify compute kernels;
\GPTPU{} preserves the nature of the application's tensor/matrix inputs and
avoids explicit decompositions of datasets and algorithms into
vector or scalar data. When used with the \GPTPU{}-integrated applications, our
prototype \GPTPU{} system exhibits a \speedup{} speedup and \energysaving{} reduction in energy consumption
relative to modern CPUs.

By introducing the \GPTPU{} system architecture, this paper makes the
following contributions:
(1) The paper introduces a novel full-stack
system architecture
to efficiently support general-purpose programming on Edge NN accelerators.
(2) The paper characterizes the capabilities and previously unidentified architectural details
of an inferencing hardware so that future research may exploit and optimize the
\GPTPU{} concept.
(3) The paper proposes and implements \Tensorizer{} to
demonstrate the mechanism of dynamically and transparently mapping operators to
NN models and \etpu{} instructions that lead to efficient use of underlying NN
accelerators. 
(4) The paper demonstrates algorithm design for non-NN applications on NN
accelerators by revisiting application algorithms to wisely use available
accelerator instructions.
(5) The paper provides an open-source framework working on commercially available
components that will allow the community to reproduce the proposed system
architecture and explore additional applications on the
\GPTPU{} platform (to be released during the conference).
(6) The paper shows the performance and energy-consumption
benefits of the prototype \GPTPU{} system.

\section{Background}
\label{sec:background}
This section briefly highlights TPU architectures and introduces
alternative NN accelerators where \GPTPU{} can potentially work.

\subsection{TPU Architecture}
\label{sec:tpu}
\ignore{
At the core of all popular NN models is the computation of values in a network of artificial
neurons. Each neuron represents the outcome of a series of sums over a set
of products (of weights or parameters) with inputs passing through
nonlinear functions. Variants of NN models differ in terms of
nonlinear functions, network sizes, numbers of layers, and neuron weights.
As most ML applications 
take matrix/tensor inputs, NN tasks amount to matrix/tensor operations
that use NN parameters/weights and previous iteration outcomes.

TPUs accelerate these NN tasks for modern ML applications by creating a systolic array
that performs operations on the units of matrices/tensors. }
As most NN applications 
take matrix/tensor inputs and iteratively update parameters/weights from previous
outcomes, the TPU microarchitecture accelerates NN tasks for modern 
ML applications by creating a systolic array
that performs operations on the units of matrices/tensors.
For inferencing tasks, the TPU treats one of the input matrices as the
trained model and the other as the samples to predict or classify.
Taking matrices/tensors as
the default inputs and outputs makes
the TPU architecture and its corresponding execution model fundamentally different from
conventional CPU/GPU architectures that compute on scalar/vector
data pairs. TPUs also incorporate large on-chip memory to hold the intermediate results that later
iterations reuse. Unlike conventional processors, TPUs do not contain
on-chip instruction caches but simply use a CISC-style
instruction-set architecture and rely on the host program to issue
instructions through the system interconnect.
And whereas conventional processors aim for precise
computation results, TPU matrix units only support operations on a limited level
of precision that is sufficient to satisfy the demands of modern ML
applications while significantly reducing both TPU costs and energy requirements.
\ignore{
Both types of TPUs are programmable through domain-specific language
framework TensorFlow Lite (TFLite), but full-fledged TensorFlow~\cite{tensorflow2015-whitepaper,I2TF}
is only available on Cloud TPUs; however, the
hardware-accelerated matrix operators and compiler-accelerated linear
algebra~\cite{XLA} are hidden behind the programming
interface. For example, TFLite only allows the programmer to
specify input/output tensors and parameters for ML
functions (e.g., \texttt{train}, \texttt{inference}, and \texttt{loss}).
}

\subsection{\etpu{}}
\label{sec:etpu}
This paper uses \etpu{}s, the trimmed-down versions of the Google Cloud TPU
to build our system. Compared with Cloud
versions, \etpu{}s contain smaller data memory (i.e., 8~MB). 
\ignore{(i.e., only 8~MB v.s. 24~MB
in the earliest version of Cloud TPUs and up to 8~GB per core in later
versions). 
}
The documented peak TOPS of \etpu{} is 4~TOPS under a 2~W TDP,
while Cloud TPUs can achieve up to 90~TOPS under a 250~W TDP.

Although Google Cloud TPUs offer higher performance, we chose the \etpu{}s for the
following reasons: (1) The \etpu{} hardware is publicly available,
whereas Cloud TPU hardware is available exclusively through
Google services; our use of \etpu{}s will therefore allow the
developer and research communities to easily replicate, utilize,
and optimize the system that this paper describes.
(2) The current version of the \etpu{} software framework has a
partially open-sourced C++ backend that enables language
front-end and runtime-system customization, whereas the Google Cloud TPU only provides
a TensorFlow front-end without the
backend source code. (3) Each \etpu{} offers better performance
per watt than Cloud TPUs (i.e., 2~TOPS/W v.s. 0.36~TOPS/W) with just 2~W of power consumption
and costs as little as USD 29,
making a platform like \GPTPU{} applicable to a broader range of computer systems than would \CMFdel{not }be possible with the Google Cloud TPU alone.

\subsection{Alternatives to TPUs}
\label{sec:other_accelerators}


In addition to TPUs, several other hardware-accelerator architectures can improve tasks in AI/ML workloads. HyGCN~\cite{HyGCN}, Caffeine~\cite{Caffeine}, Chain-NN~\cite{Chain-NN}, Tensaurus~\cite{Tensaurus}, Eyeriss~\cite{Eyeriss}, Tangram~\cite{Tangram}, SNNAP~\cite{SNNAP}, AccPar~\cite{AccPar}, Wei et al.~\cite{auto}, and Kung et al.~\cite{PSCNN} all adopt a systolic-array-based design, just as TPUs do.

DianNao~\cite{DianNao, DaDianNao}, MAERI~\cite{MAERI}, Cambricon~\cite{CambriconISA}, FlexFlow~\cite{FlexFlow}, ScaleDeep~\cite{ScaleDeep}, MnnFast~\cite{MnnFast}, TIE~\cite{TIE}, UCNN~\cite{UCNN}, CirCNN~\cite{CirCNN}, HyPar~\cite{HyPar}, Park et al.~\cite{Scale-Out}, Sharma et al.~\cite{DNNFPGA}, Alwani et al.~\cite{FLCNN}, Song et al.~\cite{song2018towards}, Shortcut Mining~\cite{ShortcutMining}, VIP~\cite{VIP}, and Simba~\cite{Simba} focus on memory 
bandwidth optimizations, chip layout, data reuses, workload balancing, and reducing inter-chiplet 
communications in their NN-accelerator architectures. 

\ignore{
DianNao~\cite{DianNao, DaDianNaoMAERI}, ~\cite{MAERI}, FlexFlow~\cite{FlexFlow}, TIE~\cite{TIE}, and Song et al.~\cite{song2018towards} 
explore the parallelism of ML-model dataflows and PE-layout re-design  in the NN accelerator.
Shortcut Mining~\cite{ShortcutMining} reuses shortcut data in NN models by using a buffering scheme to reduce off-chip data movement.
Simba~\cite{Simba} balances workloads and reduces inter-chiplet communication in MCMs to achieve scalability.
VIP~\cite{VIP} leverages a 3D-stacked memory system to alleviate high memory demand.
}


Recent advances
in near-data/in-memory processing now allow data-intensive NN computation to occur without explicitly moving data through bandwidth-limited system
interconnects. 
Neural Cache~\cite{NeuralCache}, TensorDIMM~\cite{TensorDIMM}, Manna~\cite{Manna}, DRISA~\cite{DRISA}, TETRIS~\cite{TETRIS}, Neural Cache~\cite{NeuralCache}, NAND-Net~\cite{NAND-Net}, SCOPE~\cite{SCOPE}, Wang  et al.~\cite{BPICA}, Liu et al.~\cite{PIM}, and Imani et al.~\cite{DLANYT} 
place logic in memory controllers for volatile SRAM/DRAM memory modules.
FPSA~\cite{FPSA}, and LerGAN~\cite{LerGAN}, Sparse ReRAM engine~\cite{SReRAME}, PRIME~\cite{PRIME}, PipeLayer~\cite{PipeLayer}, PUMA~\cite{PUMA}, Bojnordi et al.~\cite{MBM}, Zhang et al.~\cite{EHECNP},
and FloatPIM~\cite{FloatPIM} use the physical 
characteristics of resistive random-access memory (ReRAM) technologies to create NN
accelerators.
\ignore{; since the total current of each bitline
represents a multiplication-accumulation result, the resistance and conductance may be programmed within cells and inputs may be mapped into the supply voltage of each
wordline.} 

Regrettably, the aforementioned academic NN accelerators are not currently in
production. And commercially available ML accelerators such as Khadas VIM3~\cite{Khadas}, Rockchip
RK1808~\cite{RockchipRK1808}, Sophon BM1880~\cite{BM1880}, HiKey970~\cite{hikey970}, and 
Jetson Nano~\cite{jetsonnano} lack the performance and energy-efficiency of \etpu{}s.

Though NN accelerators have different
microarchitectures, they all use the tensor/matrix as the basic
unit of processing, and they all have limited precision, 
just like \etpu{}s. The architecture, programming interface, methods, and policies 
embodied in \GPTPU{} can easily be adapted to different NN accelerators as long as the accelerators expose their instructions appropriately.

~\ignore{NGPUs~\cite{YazdanbakhshNGPU},
NPUs~\cite{EsmaeilzadehNPU}, GANAX~\cite{GANAX}, DianNao~\cite{DianNao},
Minerva~\cite{Minerva}, FA3C~\cite{FA3C}, Laconic~\cite{Laconic}, SCNN~\cite{SCNN}, Simba~\cite{Simba}, SparTen~\cite{SparTen}, eCNN~\cite{eCNN}, ExTensor~\cite{ExTensor}, Manna~\cite{Manna}, Diffy~\cite{Diffy}, HyGCN~\cite{HyGCN}, SIGMA~\cite{SIGMA}, PIXEL~\cite{PIXEL}, E-RNN~\cite{E-RNN}, VIP~\cite{VIP}}
~

\section{Characterizing \etpu{}s}
\label{sec:arch}
To directly measure the characteristics of \etpu{}s and determine their
performance numbers, we built a customized machine with \etpu{}s attached. This
section describes the architecture of our \GPTPU{} testbed and reports
the key performance characteristics of \etpu{}s that serve as the basis for our \GPTPU{} system and
application designs.

\subsection{The prototype \etpu{} accelerated machine}
The TPU architecture relies heavily on the CPU to distribute instructions, so our custom-built \GPTPU{} hardware prototype
aims at minimizing communication latency while efficiently using the
limited system-interconnect bandwidth. To achieve this goal, the \GPTPU{} prototype machine uses \etpu{}s in
PCIe M.2 form factors; the \etpu{}s are attached directly to the PCIe system
interconnect to allow lower latency and better bandwidth compared
to other \etpu{} interconnect options, such as USB 3.0.

\begin{table*}[t]
\centering
\scriptsize
\begin{tabular}{|l|r|r|l|}
\hline
          & OPS	  & RPS & \\
Operator  & (ops per second)& (results per second) & Description\\
\hline
\texttt{conv2D}	&10268.80	&168240326.89 &  2D Convolution on a matrix  \\
\texttt{FullyConnected}	&51924.96	&6646394.57& Input vector multiplies a weight
matrix\\
\texttt{sub}	&6273.28	&82871343.60&  Pair-wise subtraction on two matrices\\
\texttt{add}	&6203.52	&98293633.48&  Pair-wise addition on two matrices\\
\texttt{mul}	&14515.84	&216469999.54& Pair-wise multiplication on two matrices\\
\texttt{crop}	&4867.96 &1562904391.76& Remove all unwanted elements outside of a sub-matrix from a given 2D matrix and return the sub-matrix\\
\texttt{ext}	&1604.78 &3637240203.38& Pad a matrix to the target dimensionality and return the padded matrix\\
\texttt{mean}	&408.54	&408.54& Count the average value of all elements in the
matrix\\
\texttt{max}	&477.08	&477.08& Find the maximum value within a matrix\\
\texttt{tanh}	&3232.31	&2148232470.28& Perform tanh function on a matrix pair-wisely\\
\texttt{ReLu}	&11194.26	&4043196115.38& Leave only non-zero values on a matrix pair-wisely\\
\hline
\end{tabular}
\caption{The maximum OPS and RPS
for each \etpu{} operator/instruction}
\label{table:goodput}
\vspace*{-0.2in}
\end{table*}
Each M.2 Edge TPU is designed to occupy only a single PCIe 2.0 lane, whereas most expansion slots that 
physically connect to the processor use multiple lanes. 
To efficiently use the precious PCIe lanes from the processor and the
limited expansion slots,
we built quad-EdgeTPU PCIe expansion cards 
using QNAP QM2-4P-384A~\cite{QNAP}. Each quad-EdgeTPU PCIe
card contains 4\x{} M.2 \etpu{}s with M.2 slots connected to a PCIe switch. The 
PCIe switch evenly divides the PCIe lanes (attached to each expansion slot) to 
four \etpu{}s and makes all \etpu{}s available to the host processor.

The current \GPTPU{} hardware prototype contains an AMD Ryzen 3700X CPU
with a Matisse
microarchitecture that can reach a max-boost clock speed of 4.4 GHz with
32~MB LLC and 24\x{} PCIe lanes available to all
peripherals. Excluding the expansion slots used for essential peripheral
devices, our hardware prototype can host 8\x{} M.2 \etpu{}s, and
each \etpu{} connects to the processor with just one hop (i.e., the PCIe
switch) in the middle. 
The machine also contains 64~GB DDR4 main memory and
an NVMe SSD.
\rv{In addition to the hardware specifications, the prototype machine
runs Ubuntu Linux 16.04 with kernel version 4.15. 
}

\subsection{Characterizing \etpu{} instructions}
\label{sec:characteristics}

Due to the long
interconnect latency and the absence of instruction caches on \etpu{}s,
coped with the variable number of cycles and different types of input/output data
resulting from the available CISC instructions, the \GPTPU{} library, runtime system, and applications
must use \etpu{} instructions wisely to achieve the best performance. 
The released \etpu{} performance numbers are available only in TOPS
(tera operations per second) 
and IPS (inferences per second). However, neither TOPS nor IPS provides sufficient insight for general-purpose
software design because (1) TOPS or IPS is highly task specific, and (2) IPS is only
representative for inferencing but not for other workloads~\cite{9177369}.

We therefore use the \emph{RPS (results per second)} as an additional metric to assess the
benefits of each available \etpu{} operator/instruction.
We define RPS as the amount of final result values an \etpu{} can generate within a second.
We measure the OPS, RPS, and data-exchange rate of each tensor arithmetic instruction
as follows: First, we begin timing the program and send the input datasets with size $s$ to the 
target \etpu{}. Second, we issue and execute the same operator 10,000 times and measure the end-to-end 
latency ($t_{1}$) as well as the total number of result values ($r_{1}$) generated since the timer started. 
Third, we repeat the first and  second step, but this time we execute the operator
20,000 times with the same input to get the end-to-end latency ($t_{2}$)
and the total number of generated result values ($r_{2}$). Finally, we calculate the
OPS of instructions/operators using Equation~\ref{eq1}, their RPSs using Equation~\ref{eq2}, and
the data-exchange rate using Equation~\ref{eq3}.

\begin{equation}
\label{eq1}
\footnotesize
OPS_{operation} = \frac{10,000}{t_2 - t_1}
\end{equation}
\begin{equation}
\label{eq2}
\footnotesize
RPS_{operation} = \frac{r_2 - r_1}{t_2-t_1}
\end{equation}
\begin{equation}
\label{eq3}
\footnotesize
Data\mbox{-}Exchange\ Rate = \frac{s}{t_1 - (t_2 - t_1)}
\vspace{-0.2in}
\end{equation}
\\

Table~\ref{table:goodput} lists the RPS and the OPS of each
\etpu{} instruction. The results lead to three observations on \etpu{}s. 
(1) \inst{Conv2D} (convolution) achieves a very high RPS given the high
amount of operations required in equivalent CPU/GPU implementations, a hint
that \etpu{} optimizes its microarchitecture for this instruction, 
(2) the OPS and RPS vary significantly for different types of instructions, and
(3) the OPS and RPS are not strongly correlated because output varies; for example, instructions 
like \texttt{sub} generate outputs with the same dimensions as their inputs, but instructions like
\texttt{FullyConnected} only produce vectors.

Our measurements also show that
data-exchange performance does not vary among different types of instructions,
but simply correlates with data size; transmitting 1~MB of data to an
\etpu{} takes around 6~ms, while transmitting 8~MB of data that
completely fill the on-chip memory takes 48~ms. The latency of copying data
between the host main memory and \etpu{}'s on-chip memory is significantly
longer than any \etpu{} instruction. 

\subsection{Characterizing \etpu{} data and model formats}
\label{sec:decipher}
\etpu{} instructions ordinarily take two types of data inputs: (1) a tensor
used for input datasets to be inferenced and (2) a model that the TFLite
framework must generate and compile. Both types of inputs must be quantized before
the host program sends them to the \etpu{} for computation. As \GPTPU{} needs to use both 
types of inputs to achieve general-purpose computing, the \GPTPU{} runtime library must translate one
of the instruction inputs as a model for the \etpu{}.

The current \etpu{} toolchain only allows the user to generate models
by invoking the \etpu{} compiler in the Python-based TFLite. 
With TFLite, translating a 2K \x{} 2K matrix into a model takes 2.7
seconds on our testbed. This latency does not create issues for inferencing
tasks in ML applications, as inferencing tasks tend to reuse the same model for
continuously changing inputs, and the overhead of creating models is amortized
as input datasets scale. However, such amortization does not stand for many
applications outside the ML realm.
Unfortunately, neither the \etpu{} compiler code nor the \etpu{} model encoding has been
released, so we have been unable to optimize the \etpu{} model-creation
overhead.

To compensate for this lack of information, we reverse-engineered the \etpu{} model formats by creating 
models with different inputs, dimensions, and value ranges.
We examined the models generated with the different inputs, and
we identified the following key characteristics that allowed us to optimize the \GPTPU{} 
runtime-system \etpu{} model-input instructions:
(1) Models embed a 120-byte general header that allows TPUs to recognize
the model-format version. The last 4 bytes of the header contain an
unsigned integer describing the size of the data section. (2) Following the header, the data section
contains binary-encoded 8-bit integers stored in row-major order. If the raw
data values exceed the scope of 8-bit integers or are non-integers, the
values must be scaled to fit in the 8-bit integer range.
(3) A metadata section following the data section
describes the data-section dimensions in terms of rows and
columns. The metadata section also contains the scaling factor ($f$), a floating-point number that the compiler uses to rescale raw data into 8-bit integers; that
is, an 8-bit integer value in the data section is calculated by multiplying its raw
value by $f$.
(4) The model encodes all values using little endian.

In addition to making the above observations, we determined that data
dimensions do not necessarily reflect the dimensions of raw data inputs. The \etpu{} compiler
adds zero padding to unused elements (depending on the instructions) to reflect the 
hardware microarchitecture. Taking the most optimized instruction in \etpu{} architecture as an 
example, the \etpu{} compiler reshapes all input data into 128\x{}128 sub-matrices. This implies that
the \etpu{}'s matrix unit is designed for computing on 128\x{}128\x{}8-bit matrices,
in contrast to the Cloud TPU matrix unit, which is designed for 256\x{}256\x{}8-bit
matrices.

\section{Overview of the \GPTPU{} System}
\label{sec:model}
\cfigure[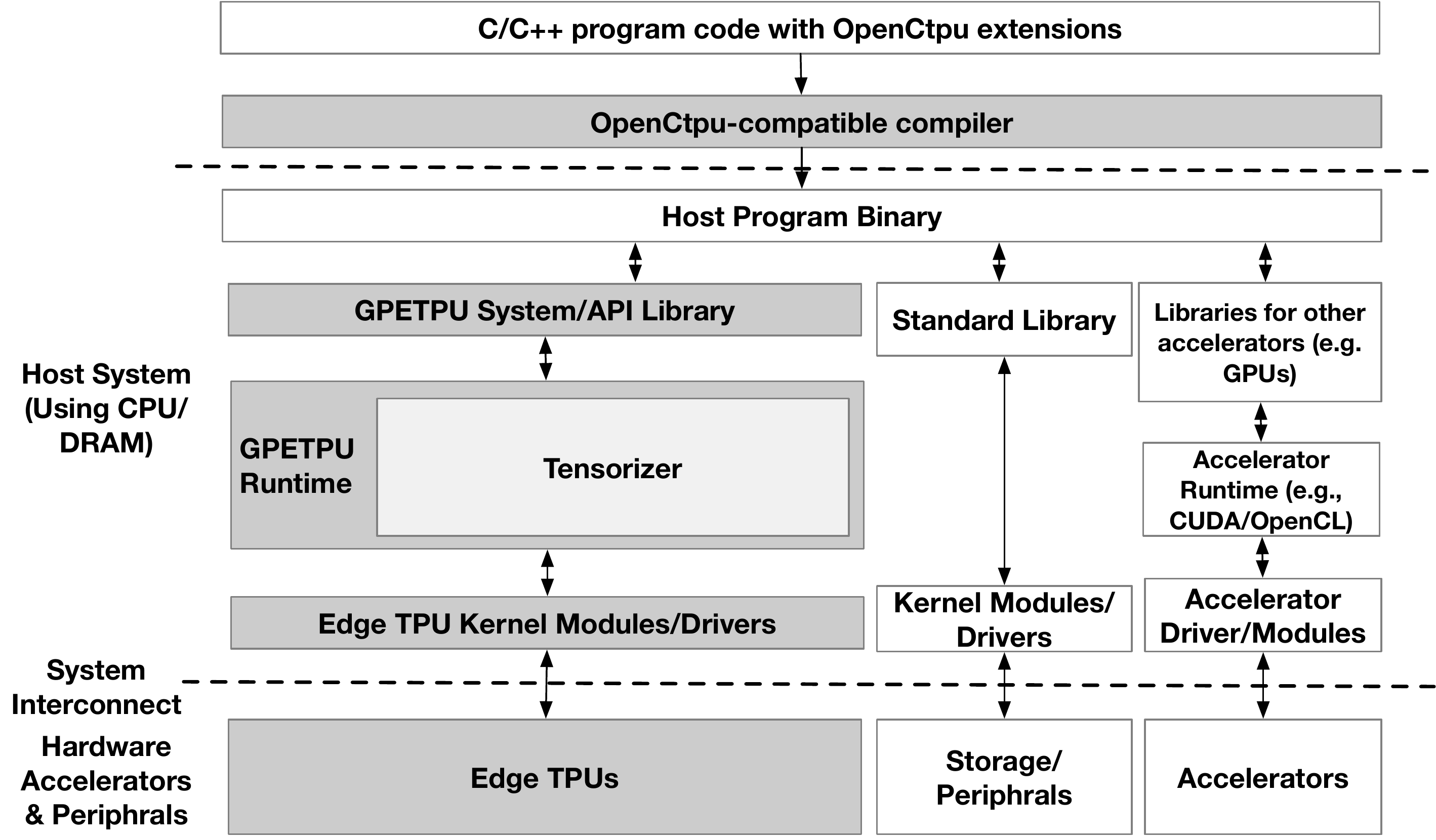, {The \GPTPU{} system overview
},fig:model]
Using insights learned from Section~\ref{sec:arch}, this paper presents the system stack of the \GPTPU{} framework
that Figure~\ref{fig:model} shows. 
\GPTPU{} maintains the original heterogeneous-computing system stack and
extends the programming-language front end. \GPTPU{} also provides a system library
that can trigger the runtime system to (1) transform data, (2) schedule
instructions for the underlying TPU hardware,
(3) communicate with the TPU hardware, and (4) use the TPU hardware to
accomplish computation tasks.

\CTPU{} serves as the programming-language front end for \GPTPU{}. A programmer can
use \CTPU{} to create a host program that describes TPU tasks and coordinates the
use of heterogeneous computing resources and data exchanges in the system.
A compiler supporting the \CTPU{} extensions will generate machine
binaries compatible with the host CPU architecture and will generate code that
transfers control to the \GPTPU{} runtime system.

The \GPTPU{} runtime system coordinates available TPU hardware.
The runtime system schedules TPU operations from programmer-defined TPU tasks and 
prepares the inputs/outputs for TPU operations. Task scheduling and data preparation
are left to the runtime system because doing so allows the \GPTPU{} system
to (1) adapt to changes in the underlying hardware without the need for
reprogramming, (2) flexibly utilize underlying hardware
resources, and (3) unburden the programmer of hardware-limitation details
(e.g., data precision).

The following sections describe 
the design of the \CTPU{} programming-language front end
(Section~\ref{sec:programming_model}), the \GPTPU{} runtime system
(Section~\ref{sec:runtime}), and optimized operators/library function/applications
(Section~\ref{sec:applications}).

\begin{table*}[t]
\centering
\scriptsize
\begin{tabular}{|p{2in}|p{4.6in}|}
\hline
Synopsis 
& Description \\
\hline
\inst{\ctpu{}\_dimension *\ctpu{}\_alloc\_dimension(int dimensions, ...)} & 
This function allocates an \inst{\ctpu{}\_dimension} data structure that
describes the dimensionality of data in an input/output buffer. Depending on the
input value of \inst{dimensions}, the function can accept additional
parameters that describe the dimensions.\\
\hline
\inst{\ctpu{}\_buffer\_t *\ctpu{}\_create\_buffer(
\ctpu{}\_dimension *dimension, void *data, unsigned flags)} & 
This function creates an input data buffer for TPU kernels. The pointer 
\inst{dimension} provides a data structure with information about the number of
data elements, the data type, and the dimensionality of the data. The pointer
\inst{data} provides the address for the raw data. The \inst{\ctpu{}\_buffer\_t} function returns a pointer to the created buffer. \\
\hline
\inst{int *\ctpu{}\_enqueue(void *(*func)(void *), ...)} & 
This function enqueues a TPU task described in \inst{func}. In addition to
\inst{func}, this function can accept an arbitrary number of arguments as
\inst{func} parameters. The function returns a task ID for the enqueued
task.\\
\hline
\inst{int *\ctpu{}\_invoke\_operator(enum tpu\_ops op, unsigned flags, ...)} & 
This function invokes a supported TPU operator (with operator arguments) and returns the operator output. The \inst{flags}
consist of parameters like the quantization method.
\\
\hline
\inst{int *\ctpu{}\_sync()} & 
This synchronization function requires all TPU tasks to complete before it returns.
\\
\hline
\inst{int *\ctpu{}\_wait(int task\_id)} & 
This function blocks the calling thread until the specified task returns. 
\\
\hline
\end{tabular}
\caption{Sample functions from the \CTPU{} API}
\vspace{-0.2in} 
\label{table:API}
\vspace{-0.2in} 
\end{table*}

\ignore{
\inst{int kfset(const char *restrict format, \ctpu{}\_operator*
op, FILE *restrict stream)} & 
The function generates the \inst{\ctpu{}\_start}
command by obtaining corresponding information from both the arguments and
the file system. \\
\hline
\inst{int kfread(void *restrict ptr, size\_t size, size\_t nitems, FILE *restrict stream,
\ctpu{}\_feedback *fb)} & 
The function reads data from the storage device using the previously set
operator for the file stream and provide the feedback.\\
\hline
\inst{int kfunset(FILE *restrict stream)} & 
The function generates the \inst{\ctpu{}\_stop}
command.\\
\hline
}

\section{\CTPU{}---The \GPTPU{} Programming Interface}
\label{sec:programming_model}

\cfigure[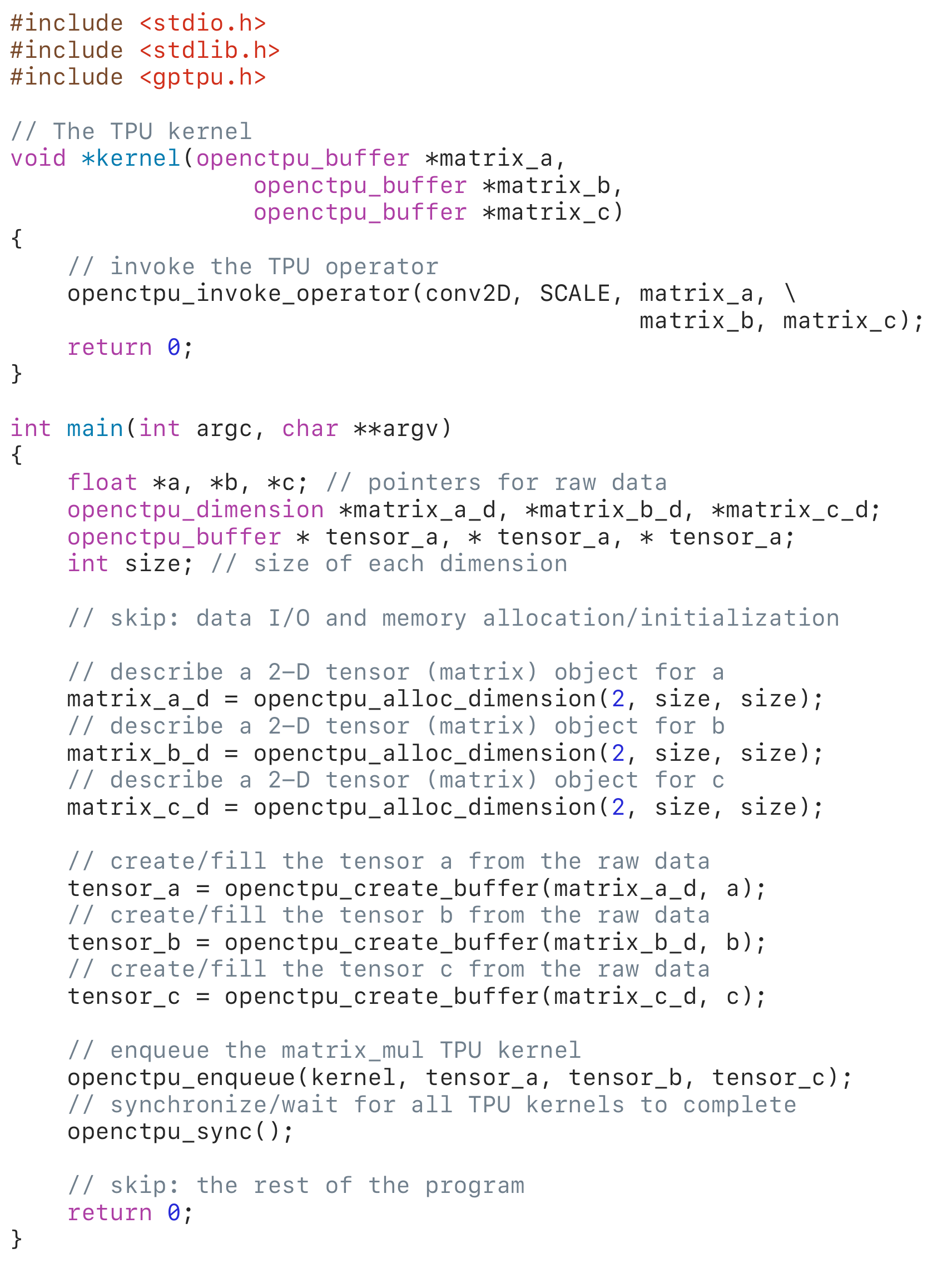, {An \CTPU{} code sample},fig:code_example]
\CTPU{} is a C/C++ extension for general-purpose programming with \GPTPU{}. \CTPU{} 
shares similarities with popular GPU programming models like CUDA~\cite{CUDA} and 
OpenCL~\cite{OpenCL} in that \CTPU{} (1) places the control of application flow and device
usage on the CPU-based host, (2) leverages virtual
memory abstraction so that applications can specify data locations, (3) requires
the programmer to explicitly manage data buffers for TPUs, and (4) provides functions that enable 
programmers to describe computation tasks for computation on TPUs.

A programmer can use \CTPU{} API functions and the C/C++ standard library to compose a TPU-accelerated program
(see Table~\ref{table:API} for a list of representative \CTPU{} API functions).
To create tasks for TPUs with the \CTPU{} API functions, a program 
needs to have the following: (1) kernel
functions that describe the desired computation for TPUs, (2) input/output data buffers/structures 
for TPU kernels, and (3) enqueuing kernel functions and their inputs/outputs as tasks
(\CTPU{} is similar to OpenCL in this respect). 
In the \CTPU{} programming model, all TPU operations within a task (i.e., an
instance of a TPU kernel function) will perform in serial, but tasks can
perform out of order in parallel. Therefore, the
programmer may need to invoke synchronized primitives to ensure
execution order and task completion. 

To use \etpu{} operators in the kernel function, \CTPU{} provides an API
function \\\inst{\ctpu{}\_invoke\_operator}. As the runtime system handles the
precision, the programmer simply needs to specify the desired
 quantization method. 
In addition to \inst{\ctpu{}\_invoke\_operator} that
directly invoke \etpu{} instructions, \CTPU{} also implemented optimized overloaded
operators on tensor data (e.g., matrix-add [+], matrix-sub [-], matrix-multiply
[*]) to perform pair-wise matrix addition, subtraction
and multiplication to further simplify programming.

The current \CTPU{} design brings several benefits to the \GPTPU{} system.
First, \CTPU{} gives the runtime system the flexibility to schedule and execute parallel tasks and to control 
the data movements associated with each task. Second,
\CTPU{} avoids hardware complexity related to managing data consistency/coherency; \CTPU{} does this by 
leaving data management to software, as with GPGPU programming models. 
Third, \CTPU{} is designed to be complementary to existing heterogeneous computing platforms (we have verified 
that CUDA/OpenCL are compatible with our \CTPU{} extensions when run in the same program).
\rv{We expect that CUDA/OpenCL can easily integrate our proposed extensions
into their programming interface. The purpose of \CTPU{} simply serves as a transition
for developers to easily exploit \etpu{} features and rethink/rewrite
algorithms for applications, rather than replacing any existing
heterogeneous programming standard.}

Figure~\ref{fig:code_example} shows an \CTPU{} code sample. \CMFdel{This function}The code contains 
a kernel function that uses the \inst{conv2D} operator.
Before creating a task instance from the kernel, the code must prepare two
tensors, \inst{a} and \inst{b}, as inputs and another tensor, \inst{c}, as the output. 
To describe the dimensionalities of these tensors, the program must call \inst{\ctpu{}\_alloc\_dimension} to create 
\inst{\ctpu{}\_dimension} data structures for each tensor. The program can then make calls to 
\inst{\ctpu{}\_create\_buffer}, which contains the \inst{\ctpu{}\_dimensions} values created for \inst{a} and \inst{b}, 
the pointers to the raw data for \inst{a} and \inst{b}, and the reserved data buffer for the product, \inst{c}. 
To perform the \inst{conv2D}
operation, the program calls the \inst{\ctpu{}\_invoke\_operator} function,
specifying \inst{SCALE} as the quantization method for input/output data,
\inst{a} and \inst{b} as the inputs, and \inst{c} as the output for the
\etpu{} operator
(currently a one-to-one mapping to a fixed set of \etpu{}/CPU instructions). The kernel function returns 
when the operator is complete.

\section{The \GPTPU{} Library and Runtime System}
\label{sec:runtime}
The \GPTPU{} runtime system receives tasks from the \CTPU{} front end, dynamically schedules tasks, and 
transforms input/output datasets for tasks. This section describes the design of the \GPTPU{} runtime system. 

\subsection{Task scheduling}
\label{sec:scheduling}
\cfigure[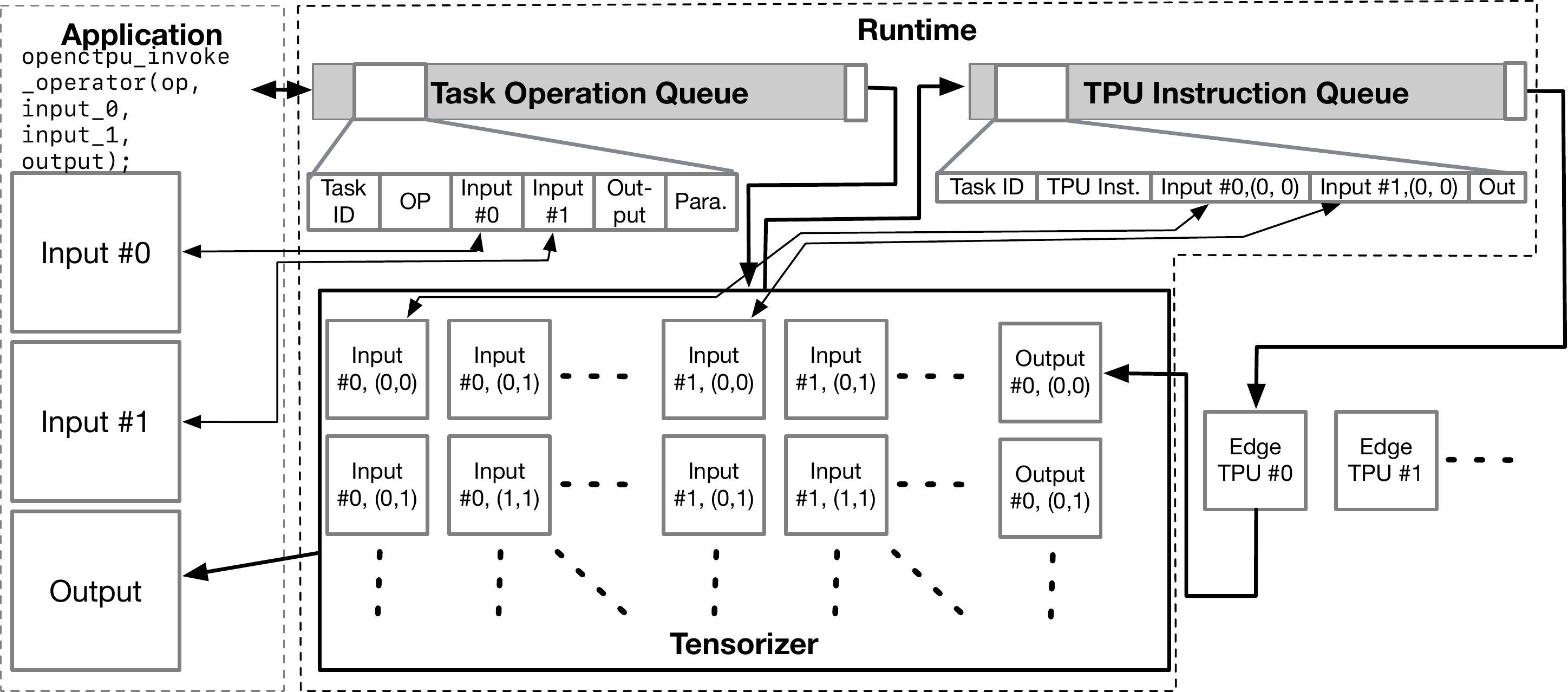,{\rv{An overview of \GPTPU{}'s runtime
system.}},fig:runtime]

The \GPTPU{} runtime task-scheduling policy is a 
dataflow-based algorithm
on a front-end task operation queue (OPQ) and a back-end instruction queue
(IQ) \rv{as Figure~\ref{fig:runtime} highlights}. 
An OPQ entry contains a task ID, the requested TPU operation, the input and output locations, 
and parameters like the quantization method. 

\GPTPU{} gradually fills the OPQ during the execution of the user
application.
When the program calls the \inst{\ctpu{}\_enqueue} function, the 
\GPTPU{} runtime system initiates a new task ID for the invoked kernel function. 
The runtime system then executes the code designated by the function pointer using the set of 
parameters from the \inst{\ctpu{}\_enqueue} call.
The above process ends when \inst{\ctpu{}\_invoke\_operator} is called to request the involvement of a TPU
operator/instruction. 

The \inst{\ctpu{}\_invoke\_operator} function triggers the runtime system to create an
OPQ entry with the task ID created from the current kernel function. The \GPTPU{} runtime system then fills the rest of the queue 
entry with information passed to the \inst{\ctpu{}\_invoke\_operator} function.  
As
the current \CTPU{} design serializes operators from a single kernel-function instance, kernel-function 
execution will be blocked until the operation finishes and each task has one operator
from the 
\inst{\ctpu{}\_invoke\_o
perator} function in the OPQ.
Since \CTPU{} allows all tasks to execute in parallel, the \GPTPU{} runtime system can
issue entries in the OPQ to \Tensorizer{} without considering their original order. 

After \Tensorizer{} optimizes, reshapes and transforms data and operations into
instructions, \Tensorizer{} divides a task into instructions in the IQ.
The runtime system then schedules 
to the same \etpu{} if they share 
the same input, quantization flags, and the same task ID, but have different output
locations---a scheduling approach that reduces movement overhead and the number of data transformations required. 
For other instructions, the \GPTPU{} runtime system will use a first-come-first-serve policy to assign them
to available \etpu{}s. 

\subsection{\Tensorizer{}}
\label{sec:chunking}
\Tensorizer{} is responsible for dynamic optimizations at the task level. 
\Tensorizer{} transforms and optimizes programmer-requested operations into 
instructions, input tensors and models that enable efficient use of \etpu{}s. 

Upon receiving a task from OPQ, \Tensorizer{} first partitions the programmer-requested 
operation into \etpu{} instructions into sub-problems where each instruction works on 
its optimal data/model shapes 
using insights from Section~\ref{sec:characteristics}.
\Tensorizer{} transforms the input data to minimize loss of accuracy due to the 8-bit precision of TPU matrix
units 
for each \etpu{} operator. 
\ignoreexact{ 
As \GPTPU{} supports programmer-accessible mechanisms 
to accommodate loss of accuracy (due to the 8-bit precision of TPU matrix units)
or exact computing and exposes these features
through \GPTPU{}'s library interface, \Tensorizer{} 
transforms the input data using the programmer-selected accuracy mode 
and generates a result using the same mode as the user specifies for each \etpu{} operator. 
We now describe how \Tensorizer{} partitions operators and generates inputs for each mode in turn.
}

\subsubsection{Mapping operators into instructions}
\label{sec:blocking}
As \CTPU{} hides the hardware details from the programmer, programmer's
tasks are agnostic to the granularity of inputs that optimize \etpu{}
instructions. \Tensorizer{} tackles this performance issue by dynamically
partitioning these tasks into \etpu{} instructions working on their optimal
data sizes/shapes (e.g., 128\x{}128 matrices in most arithmetic instructions). 
As \etpu{} supports limited numbers of instructions/operators, we creates a set 
of rules that guides \Tensorizer{} in rewriting tasks. 

For pair-wise operators that calculate on pairs of values from both input matrices, 
including \inst{add}, \inst{sub} and \inst{mul} or element-wise operators that
calculate on each value of an input matrix, including \inst{tanh} and
\inst{relu} the rule is straightforward. \Tensorizer{} simply needs to first divide
the input data into tensors and models that contain sub-matrices with the optimal 
shape. Then, \Tensorizer{} rewrites the operator into a set of \etpu{} instructions 
where each works on a sub-matrix or a pair of sub-matrices locating at the corresponding 
positions in the original inputs and collects the results in the corresponding memory
locations. 

For matrix-wise operators, including \inst{mean} and \inst{max}, \Tensorizer{}
still divides the input into sub-matrices with optimal shapes (i.e., both
instructions favor 64\x{}64 sub-matrices) and uses
instructions to work on each \rv{sub-matrix}. However, \Tensorizer{} will
additionally generate CPU code to aggregate the received values from
results of instructions to produce the final outcome. An alternative
approach is to create another sets of \etpu{} instructions and making the received
values an input tensor/model to iteratively use \etpu{} to produce the
result. \Tensorizer{} does not take this approach as (1) the first round of
executing \inst{mean} or \inst{max} instruction
already shrinks the values to aggregate by a factor of 4096, and (2) the latency of moving data
in the currently system architecture is significantly longer than
aggregating results
with CPU code.

For arithmetic operators, including \inst{FullyConnected} and \inst{conv2D}, \Tensorizer{}
applies mechanisms similar to the blocking algorithm for matrix
multiplications~\cite{dongarra1986linear} in rewriting tasks. If each input
matrix is partitioned into $P\times Q$ sub-matrices, The resulting
code will contain \etpu{} instructions that perform $P\times Q$ \inst{FullyConnected} or \inst{conv2D}
instructions and CPU code that aggregates results into the final
outcome. The CPU
code only needs to add received values that requires very short latency to
execute on modern processors. In addition, as CPU registers are wider than
\etpu{}'s data precision, aggregating results on CPU will allow the platform
to reduce precision loss in results.
\ignoreexact{ or more efficiently support exact computing. }

After rewriting operations into actual machine/accelerator code, \Tensorizer{}
will obtain the mapping between an input value and its location in the
transformed tensor/model. 
\ignoreexact{
The following sections will describe
\Tensorizer{}'s procedure in filling values in the input data
tensors/models for both approximate and exact modes. 
}

\ignore{
\subsubsection{Data transformation}
\label{sec:data_transformation}
When an operation enters the OPQ, the runtime system initiates data transformation to mitigate 
the latency of ongoing computation. 
Most \etpu{} arithmetic instructions accept two inputs: (1) the input vector/matrix that the runtime system can
easily transform and (2) the model that requires instruction-specific optimizations and metadata.

\subsubsection{Accuracy modes}
\label{sec:accuracy}
As noted above, the \GPTPU{} system programmer chooses
one of two accuracy modes: the approximate mode or the
exact mode. We now describe these modes in turn.
}

\subsubsection{Data transformation}
\label{sec:approximate}
To minimize the inaccuracy of computation, \Tensorizer{} carefully rescales values 
into fixed-point numbers and fill numbers into models or inference data
arrays that \etpu{}s can accept. 
\rv{\GPTPU{} determines the scaling factor for input datasets using (1) the
sequence of operators, (2) the number of operators, and (3) the range of
input data. }
As the data size of each \etpu{} instruction and \rv{the
sequence of operators} are known at runtime, 
the \GPTPU{} system can estimate the number of logical arithmetic operations
($num\_logical\_operations$) that the
instructions will generate. By discovering the maximum value \rv{($max$)} and the minimum
\rv{($min$)} value of the dataset, 
the runtime system can estimate the range of output values and derive the model/tensor scaling
factor. \rv{The general rule of the scaling factor $S$ of an operator is
\begin{equation}
\label{eqs}
\footnotesize
S=\frac{1}{max(|output_{max}|,|output_{min}|)} 
\end{equation}
where $output_{max}$ is the expected maximum output value and  $output_{min}$  is the expected minimum output
value.
For most datasets, sampling is efficient enough in large datasets as
previous work indicates that small subset of input data is representative
for the whole dataset~\cite{IRA}. As \GPTPU{} calculates $S$ using the maximum absolute
value of outputs, \GPTPU{} prevents the case of overflow. 
\ignore{
If the runtime system samples input data
and finds the data range has a positive normal distribution,
\GPTPU{} can further optimize the expected maximum value to be $\frac{n}{2}\times \frac{n}{2}\times N$
and calculate the scaling factor.}

\GPTPU{} applies different formulas for different types of
operators. If the input data is a pair of $N\times N$ matrix, \GPTPU{}
estimates the scaling factor for each \inst{conv2D} and \inst{FullyConnected}, as:
\begin{equation}
\label{eq3}
\footnotesize
S = \frac{1}{|max-min|^2 \times N}
\end{equation}
For pair wise \inst{add} and \inst{sub}, \GPTPU{} uses:
\begin{equation}
\label{eq4}
\footnotesize
S = \frac{1}{2\times |max-min|}
\end{equation}
as the scaling factor. For pair wise \inst{mul}, \GPTPU{} uses:
\begin{equation}
\label{eq5}
\footnotesize
S = \frac{1}{|max-min|^2}
\end{equation}
as the scaling factor, and for other operators, \GPTPU{} calculates the
scaling factor as: 
\begin{equation}
\label{eq6}
\footnotesize
S = \frac{1}{|max-min|}
\end{equation}
}

For example, consider a request that performs matrix multiplication \rv{and
then pairwisely add another matrix} on $N\times N$ matrices 
with data ranging from $0$ to $n-1$. The maximum output value in the resulting matrix will
be $2\times N \times (n-1)^2$. The runtime system can choose
$\frac{1}{2\times N (n-1)^2}$ as the scaling factor. 
\ignoreexact{
\subsubsection{Exact mode}
\label{sec:exact}
\cfigure[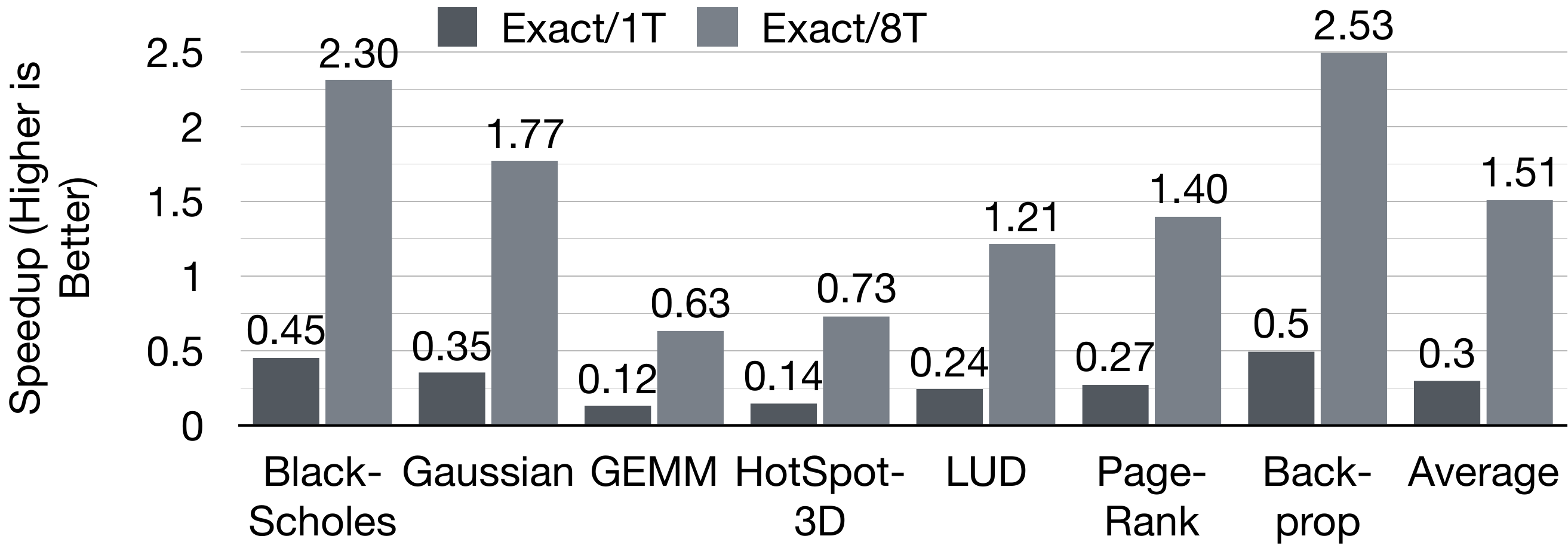, {An exemplary data layout for exact mode},fig:exact]
To accommodate applications that require high accuracy, \GPTPU{} uses \Tensorizer{}
to support the exact 
computation mode through rewriting operators into instructions with numerical methods (e.g., the Toom--Cook
algorithm, an algorithm frequently used in numerical methods with large-number 
multiplications~\cite{LargeNumberMM}) optimized for each \etpu{} operator. 
In the exact mode, \Tensorizer{} splits each number in the raw data into smaller numbers, with
each smaller number comprising an equal number of bits (from the original
number). \Tensorizer{} will implicitly spawn
more \etpu{} instructions from the original instructions that the steps in
Section~\ref{sec:blocking} created to accomplish exact computing. In
contrast, the approximate mode does not create additional \etpu{}
instructions, but precision will be lost if the range of input values or accumulated results is wide. 
Typically, an operator will spawn instructions in accordance with \GPTPU{} runtime-system scheduling, thereby allowing the runtime system to 
parallelize the operation 
in different locations and reuse the generated model.
Excepting this the last step, aggregation
of the final outcome from each partial result requires the instructions to
follow synchronization. 

To minimize the overhead of synchronizing each partial result, 
\Tensorizer{} proposes a synchronization free mechanism that places all split bits from the same number in the same sub-matrix
sequentially. Figure~\ref{fig:exact} illustrates an example data
layout for arithmetic operations like \inst{conv2D} and
\inst{FullyConnected}. In Figure~\ref{fig:exact}, we split a value
$A[0,0]$ into $A[0, 0, 0]$ -- $A[0, 0, 7]$ where each $A[0, 0, i]$
represents the $i$th portion of the raw data bits. \Tensorizer{} applies similar
partitioning to all elements in both input arrays and makes two rectangular
matrices, each rectangular matrix is $N$\x{} wide as the original input
matrix if the exact mode splits a number to $N$ different parts.
Figure~\ref{fig:exact} splits each number into 8 different parts. By
applying the requested instruction to rectangular matrices, each
$N\times N$ sub-matrix in the result matrix will contain all partial
products necessary in calculating the exact value in
the $\frac{i}{N}$th row and $\frac{j}{N}$th column of resulting matrix.
Therefore, \Tensorizer{} does not need any synchronization primitive when
calculating each result value in exact mode as all
values are presented at the same time in a sub-matrix.

To further optimize performance, \Tensorizer{} still divides the
operations on rectangular matrices into the requested instruction's optimal
shape. As the optimal shape of each instruction is larger than and the multiplies of
the size of an expanded number, such division incurs no additional
data synchronization but helps in computation throughput. 
}

\ignore{
\Tensorizer{} hides the complexities of error handling and precision adjustment from the 
programmer. 
In exact mode, \Tensorizer{} will implicitly spawn
more \etpu{} instructions in the runtime system from the original operator that the programmer created.
}

\ignore{
Typically, an operator will spawn instructions 
in accordance with \GPTPU{} runtime-system scheduling, thereby allowing the runtime system to 
parallelize the operation 
in different locations and reuse the generated model.
Excepting this 
the last step, aggregation
of the final outcome from each partial result requires the instructions to
follow synchronization. 
}

\subsubsection{The overhead of \Tensorizer{}}
\label{sec:tensorizer_performance}
Using the information we gained from reverse-engineering the \etpu{} model format as
described in Section~\ref{sec:decipher}, we implemented the proposed \Tensorizer{} 
to dynamically create models from arbitrary input data. The C-based  \Tensorizer{} can
bring the latency of generating a model from a 2K\x{}2K matrix down to
1.8~ms---a 1500\x{} speedup over the original Python-based \etpu{} TFLite
compiler and shorter than the latency of data transfer. 
The \GPTPU{} runtime system thus can overlap \etpu{} matrix-input data movements with 
\Tensorizer{} to reduce the total latency of executing \etpu{} instructions
from tasks. 

\section{Optimizing Applications for \GPTPU{}}
\label{sec:applications}
Mapping a problem into a \GPTPU{} application requires inputs/outputs to be transformed into tensors 
that \etpu{}s can operate on. Although many applications
use data in tensor form, the \etpu{} instructions are optimized for NN
workloads, meaning that naively applying the default tensor operators
may not improve performance. \Tensorizer{} helps to optimize performance in
the task level, but using the most efficient operator for a task still requires
programmer's optimization. 
This section describes \GPTPU{} application design and optimization 
using matrix multiplication as an example.


\subsection{General Matrix Multiply (GEMM)}
\label{sec:mm}
To demonstrate the importance of designing algorithms to wisely use \etpu{}
instructions, we explain the design of an efficient GEMM on \etpu{}s, a fundamental linear-algebra tool for
matrices. 
GEMM takes two 2-dimensional tensors (matrices) as inputs and produces a single
2-dimensional tensor as output. 
We can calculate each element in the result matrix, $C$, obtained from a set of pairwise 
multiplications and accumulations from an $M \times N$ matrix, $A$,
and an $N \times K$ matrix, $B$.
\ignore{
\begin{equation}
\label{eq4}
C_{i,k}=\sum_{j=0}^{N-1}A_{i,j}\cdot B_{j,k}, 
\forall\ 0\leq i< M, 0\leq k < K
\end{equation}
In short, GEMM yields a set of pairwise multiplications and accumulations. 
}
\subsubsection{GEMM and the \inst{FullyConnected} operator}
\label{app:cmf_gemm_fc}
\ignore{
A program can select either matrix $A$ or matrix $B$ 
and iterate through a column or row of the other
matrix to produce the result, and matrix multiplication will be performed via the $M$ or $K$ \inst{FullyConnected} operators.
}


The \etpu{} \inst{FullyConnected} instruction offers an intuitive way to implement \GPTPU{}
GEMM, as the operator essentially produces a matrix-vector product. 
\ignore{
For the typical
GEMM problem, a program can select either of the source matrices and iterate through a column or row of the other
matrix to produce the result columns or rows using \inst{FullyConnected}. 
}
A program can select either matrix $A$ or matrix $B$ 
and iterate through a column or row of the other
matrix to produce the result, and matrix multiplication will be performed via the $M$ or $K$ \inst{FullyConnected} operators. 


\subsubsection{The \inst{conv2D} operator/instruction}
\label{app:cmf_conv2d}

\etpu{}'s \inst{conv2D} instruction
can also perform multiplications and accumulations but in different orientations to derive the result.
In conventional architectures, programmers implement convolutions by
performing scalar-scalar or vector-vector multiplications and accumulations
for higher efficiency. However, Table~\ref{table:goodput} shows that the RPS of
convolution (i.e., \inst{conv2D}) is
25\x{} the RPS of matrix-vector multplications (i.e.
\inst{FullyConnected}) on \etpu{}s. Inspired by this observation, we therefore explore
the implementation by changing the layout of input data and using \inst{conv2D} to 
perform exactly the same number of multiplications and accumulations on the set of
input numbers to leverage the high RPS of \inst{conv2D}
for a more efficient GEMM implementation. 

The \inst{conv2D} instruction takes one of its inputs as
the kernel, multiplies each kernel element with an input 
element mapping to the corresponding location, and accumulates the result as an output element.
Each \inst{conv2D} instruction can produce a result matrix that has the same size as the non-kernel input. 

For an $M \times N$ input matrix, $A$, and an $L \times L$ 
kernel, $B'$, each element in the \inst{conv2D} $M \times N$ output matrix, $C$,
is:
\begin{equation}
\label{eq5}
C_{i,j}=\sum_{q={0}}^{L}\sum_{p={0}}^{L}A_{i+p,j+q}\cdot
B'_{p,q}
 (\forall\ 0\leq i< M, 0\leq j < N)
\end{equation}
\ignore{
The \inst{conv2D} instruction takes one of its inputs as
the kernel, rotates the kernel orientation by 180$^{\circ}$, multiplies each rotated kernel element with an input 
element mapping to the corresponding location, and accumulates the result as an output element.
Each \inst{conv2D} instruction can produce a result matrix that has the same size as the non-kernel input. 

For an $M \times N$ input matrix, $A$, and an $L \times L$ 
kernel, $B'$, each element in the \inst{conv2D} $M \times N$ output matrix, $C$,
is:
\begin{equation}
\label{eq5}
C_{i,j}=\sum_{q=-\lfloor\frac{L}{2}\rfloor}^{\lfloor\frac{L}{2}\rfloor}\sum_{p=-\lfloor\frac{L}{2}\rfloor}^{\lfloor\frac{L}{2}\rfloor}A_{i+p,j+q}\cdot
B'_{\lfloor\frac{L}{2}\rfloor-p,\lfloor\frac{L}{2}\rfloor-q}
\end{equation}
\begin{equation}
\forall\ 0\leq i< M, 0\leq j < N \nonumber
\end{equation}
}

Targeting AI/ML workloads that are error tolerant (and so permit
approximations), the \etpu{} \inst{conv2D} instruction allows a programmer to assign a \emph{stride} value ($s_x, s_y$) that treats
inputs as groups of $s_x$ \x{} $s_y$ sub-matrices and produces a corresponding result
value for them. 

\begin{figure}[t]
\begin{center}
\footnotesize
\begin{tabular}{cc}
\includegraphics[width=1.3in]{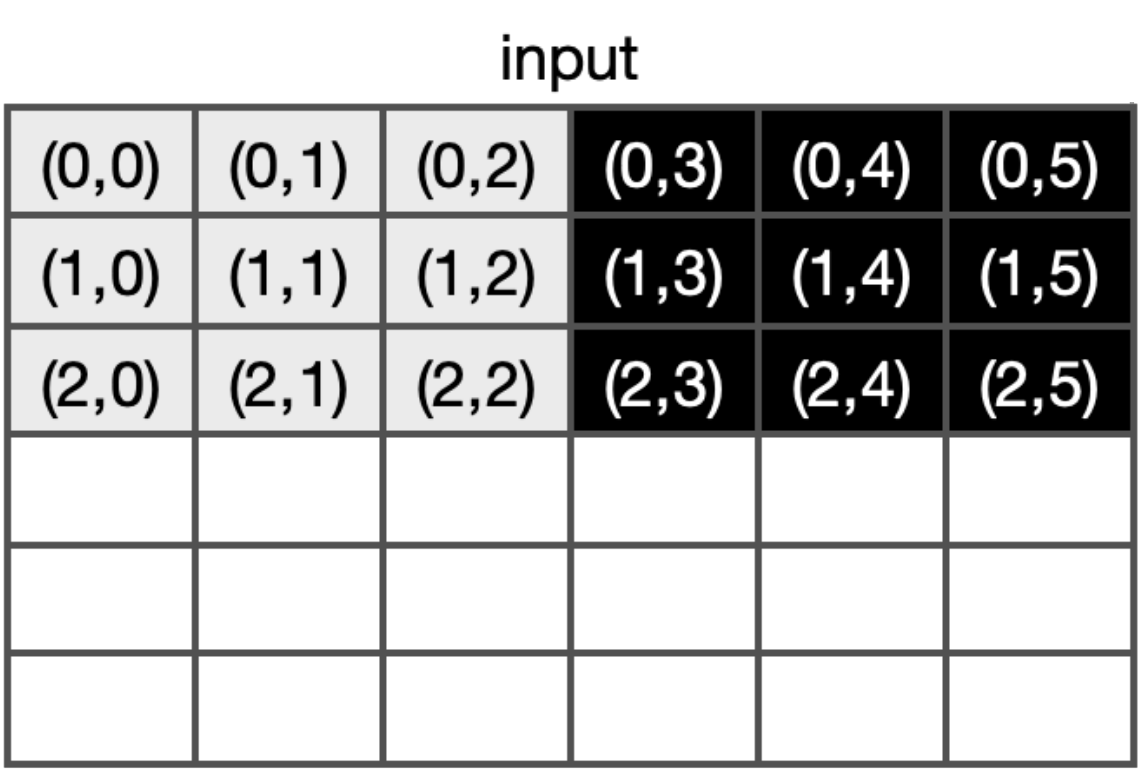} &
\includegraphics[width=1.3in]{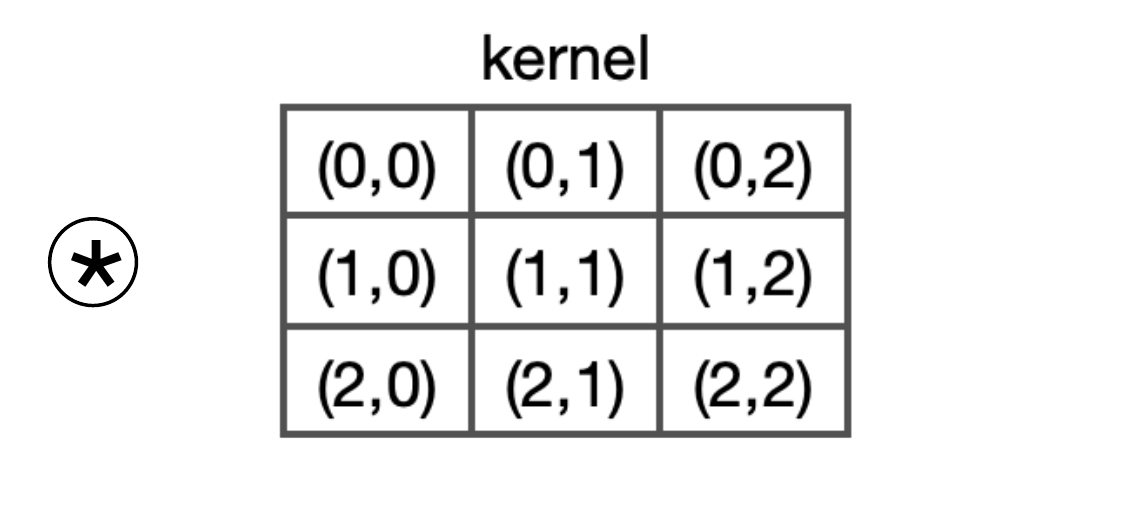} \\
(a) & (b)\\
\includegraphics[width=1.3in]{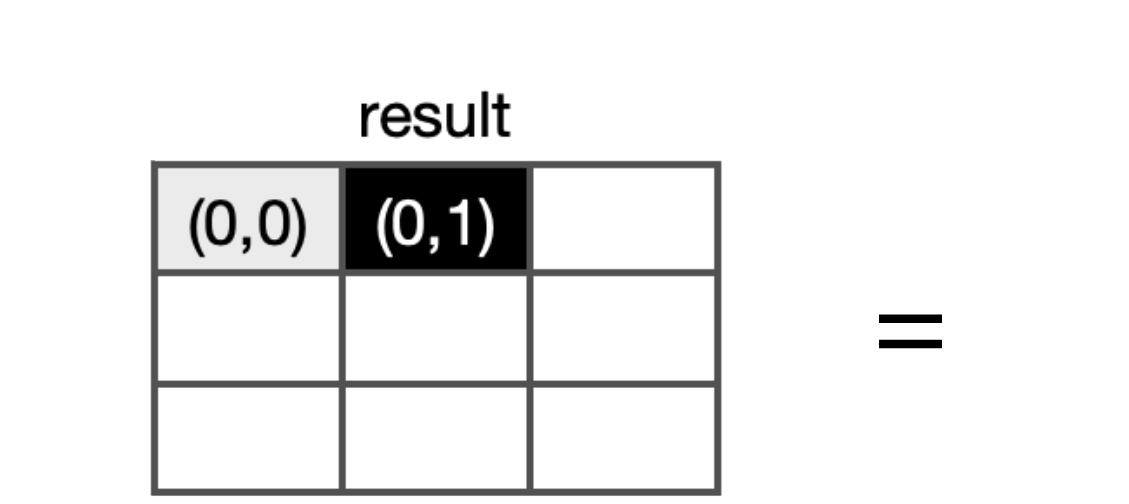} &
\includegraphics[width=1.3in]{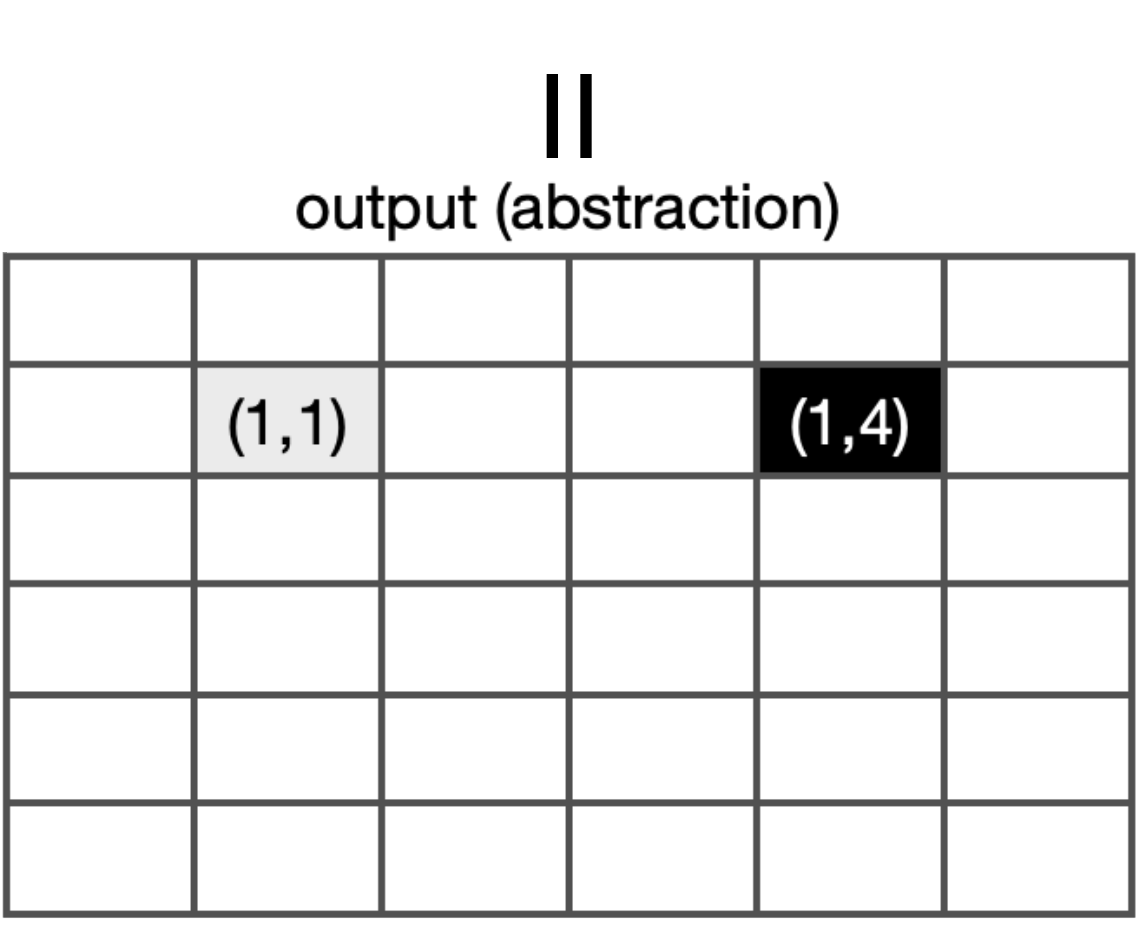} \\
(d) & (c)\\
\end{tabular}
\end{center}
\vspace{-0.1in}
\caption[]{The \inst{conv2D} as implemented with stride}
\label{fig:conv2D}
\vspace{-0.2in}
\end{figure}

\ignore{
Figure~\ref{fig:conv2D} illustrates the concept of
\inst{conv2D} with stride. We select (3, 3) as our
stride, restricting \inst{conv2D} to 9 numbers in a group; the \inst{conv2D} only produces a value for every 3 row/column elements
 in the abstracted outcome (Figure~\ref{fig:conv2D}(c)) from the source matrix (Figure~\ref{fig:conv2D}(a)) using the kernel
in Figure~\ref{fig:conv2D}(b). The final output of \inst{conv2D} is a
condensed matrix as in Figure~\ref{fig:conv2D}(d).
}

Figure~\ref{fig:conv2D} illustrates the concept of \inst{conv2D} with stride. We select (3, 3) as our
stride, restricting \inst{conv2D} to 9 numbers in a group; the \inst{conv2D} operator only produces a value for every 3 row/column 
elements in the abstracted outcome, as in Figure~\ref{fig:conv2D}(c), from the source matrix, as in Figure~\ref{fig:conv2D}(a), using the kernel
in Figure~\ref{fig:conv2D}(b). The final output of \inst{conv2D} is a condensed matrix, as in Figure~\ref{fig:conv2D}(d).

\GPTPU{} uses \inst{conv2D} and its striding feature to implement an efficient GEMM
algorithm. 
The algorithm starts by reshaping both inputs that transform each row in the
chosen source matrix into a sub-matrix whose size is determined by the selected stride $(s_x, s_y)$. Ordinarily, 
both $s_x$ and $s_y$ are \rv{the round-up of} the square root of the column dimension in the source
matrix. The other input matrix serves as a list of kernels, where
each kernel of size $s_x \times s_y$ contains a column from that matrix. When creating the kernels, the \GPTPU{} 
GEMM algorithm fills the kernel elements to match the desired element-wise multiplications for GEMM. 
\rv{In other words, for a matrix with $N$ columns and $K$ rows, the
resulting kernel matrix will contain $N$ kernels where each kernel contains
$\lceil \sqrt{K} \rceil \times \lceil \sqrt{K} \rceil$ elements. That being
said, the resulting kernel matrix still contains exactly the same or similar 
amount of elements (i.e., $N \times (\lceil \sqrt{K} \rceil)^2$ v.s. $N\times K$) as the original
input matrix. }
After transforming both inputs, \inst{conv2D} 
iterates through all sub-matrices over each kernel with the selected stride and generates output identical to that of 
conventional matrix multiplication. 

\ignore{
As Section~\ref{sec:decipher} shows, \inst{conv2D} performs
best when the inputs are in groups of 128\x{}128. The blocking algorithm and other conventional matrix-multiplication 
optimization techniques are still applicable to further improve
performance.}

\cfigure[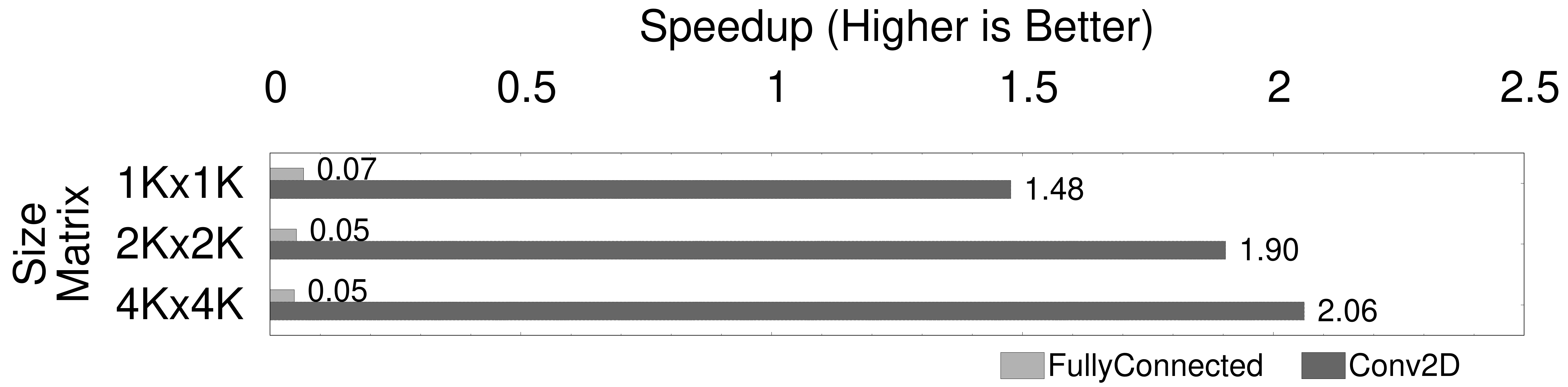, {Speedup of GEMM \GPTPU{} implementations
using \inst{FullyConnected} and \inst{conv2D}, relative to the baseline CPU
\rv{OpenBLAS} implementations.},fig:mvconv2D]
\subsubsection{\inst{FullyConnected} and \inst{conv2D} together}
\label{app:cmf_fc_conv2d}

Figure~\ref{fig:mvconv2D} shows the performance
of \GPTPU{} GEMM kernel implementations using \inst{FullyConnected} and
\inst{conv2D} compared to the CPU baseline using \rv{OpenBLAS}~\cite{OpenBLAS}. The \inst{conv2D}
implementation reveals a strong performance gain (a 2.06\x{} speedup in the
4K\x{}4K microbenchmark) over the CPU baseline. In
contrast, the \GPTPU{} GEMM implementation cannot beat the CPU baseline without \inst{conv2D} (i.e., when GEMM only uses \inst{FullyConnected}).

Though the \GPTPU{} GEMM algorithm incurs additional data-transformation
overhead, 
\GPTPU{}'s \inst{conv2D}-based GEMM significantly outperforms the conventional
vector-product-based algorithm by 43\x{} on our \GPTPU{} platform.
This is because the \etpu{} architecture highly optimizes \inst{conv2D} and
the favorable RPS of \inst{conv2D} compensates for the additional overhead.


Since GEMM is a widely used, fundamental linear-algebra tool for matrices,
\GPTPU{} makes the core GEMM algorithm available as an optimized library function,
\inst{tpuGemm}, that \GPTPU{} applications can invoke---just as CUDA invokes the cuBLAS GEMM implementation via the \inst{cublasGemm}
function~\cite{cuBLAS}.

\subsection{Other applications}
\label{sec:other_apps}
As with GEMM, our goal for all \GPTPU{} applications is to utilize
instructions with the
highest RPS. We now summarize how we extended the \GPTPU{} GEMM approach to other 
applications whose workloads we evaluate in the latter part of this paper. 
This section focuses on the \GPTPU{} instructions that the \GPTPU{} 
implementations use.


\subsubsection{PageRank}
\label{sec:pagerank}
The PageRank algorithm~\cite{PageRank} is a representative graph application.
\ignore{is a representative algorithm that measures the importance
of a node (e.g., a webpage) within a large graph (e.g., the links and nodes of all
webpages on the Internet). As with most graph applications,}PageRank takes an 
adjacency matrix representing a graph as input. Both the
baseline and the \GPTPU{}
implementations use the classic power method that iteratively performs matrix-vector
multiplications. In contrast to CPU/GPU PageRank implementations that perform pairwise or 
vector-wise multiplications, 
the \GPTPU{} PageRank implementation simply uses one \inst{FullyConnected}
instruction for each adjacency-matrix multiplication with a single vector.

\subsubsection{HotSpot3D}
\label{sec:hotspot}
HotSpot3D is a thermal-simulation tool for estimating the temperature of a chip made 
with 3D-stacking. The main algorithm gradually and iteratively updates 
each point on the chip, which is represented as a matrix with a weighted average of 
the point's closest neighbors in 8 different directions. The HotSpot3D algorithm can
naturally map to \inst{conv2d} with a 3\x{}3 kernel without striding. 

\subsubsection{LU Decomposition (LUD)}
\label{sec:lud}
LUD factors a matrix into a lower triangular matrix ($L$) and an upper triangular matrix ($U$) such that $L \times U$ yields the 
original matrix. 
Our \GPTPU{} LUD implementation uses the recursive
algorithm~\cite{I2A} via \inst{crop}, \inst{FullyConnected}, and
\inst{conv2D} to partition matrices and perform appropriate operations on
different combinations of the partitioned matrices.

\subsubsection{Gaussian elimination\CMFdel{s} (Gaussian)}
\label{sec:gaussian}
Like LUD, Gaussian is a method for solving a system of linear equations.
Gaussian combines row swaps, the scalar multiplication of rows, and row additions 
until the lower left-hand triangular matrix contains only zeroes. For Gaussian, \GPTPU{}  uses \inst{mul} to perform each row reduction. 

\subsubsection{Backpropagation (Backprop)}
\label{sec:backprop}
Backprop is foundational to NN supervised learning. We implemented a plain-vanilla version of Backprop to demonstrate the ML/AI-generalizable nature of 
\GPTPU{}.
For a feedforward NN, the \GPTPU{} Backprop uses (1) multiple layers of \inst{FullyConnected} 
and sigmoid activation functions in \inst{ReLu}, (2) \inst{add} for the actual backpropagation, and (3)
\inst{tpuGEMM} to derive weights for the delta matrix.
\subsubsection{Black--Scholes (BlackScholes)}
\label{sec:BlackScholes}
BlackScholes is a financial model for estimating the market price of stock
options. 
\GPTPU{} uses a ninth-degree polynomial function~\cite{CNDF} with the \inst{FullyConnected} instruction to
compute the cumulative normal distribution function.
\section{Experimental methodology}
\label{sec:methodology}

\subsection{The system platform}
\label{sec:hardware}
We use \rv{exactly the same} prototype machine described in Section~\ref{sec:arch} for all
experiments performed with \GPTPU{}. 
\ignore{
the prototype machine
runs Ubuntu Linux 16.04 with kernel version 4.15. }
When performing experiments for baseline applications, we removed the TPUs from the machine.

For each application, we measured the end-to-end latency. 
We also measure the total system power using a Watts
Up meter. 
When calculating energy consumption, we aggregate the \rv{total system
power} throughout the
application execution time.\ignore{
 and subtract the idle power (i.e., 40~W \x{}
execution time) from the result to reflect the pure power/energy consumption
due to executing the application. The measured active power includes the
computation power that all active CPU cores, GPUs, TPUs as well as the
storage system consume during
the program execution. 
The measured power includes 
computation power that all active CPU cores, GPUs, TPUs as well as the
storage system consume during
the program execution.}
On average, each \rv{active} \etpu{} adds only 0.9~W to 1.4~W of power consumption, while a loaded AMD Matisse
core in the \GPTPU{} hardware prototype consumes from 6.5~W to 12.5~W. 
\rv{As \GPTPU{} still relies on the CPU for the runtime system and data
transformation, both CPUs and \etpu{}s can be active when running
applications. The idle power of the experimental platform is 40~W, including the southbridge
chip on the motherboard, NVMe-based storage devices as well as other
peripherals connected to the system. }

\subsection{The baseline application implementations}
\label{sec:baseline}
\begin{table}
\centering
\scriptsize
\begin{tabular}{|l|c|c|c|c|}
\hline
  	&  Input                          & Data  & 		& Baseline\\ 
Benchmark  	&  Matrices                 & Size & Category	&
Implementation\\ 
\hline
Backprop	        & 1\x{}8K\x{}8K    &  512MB & Pattern Recognition &
~\cite{RodiniaBenchmark,8091072}\\ 
BlackScholes   & 1\x{}256M\x{}9            &  9GB & Finance & ~\cite{AxBench}\\
Gaussian	        & 1\x{}4K\x{}4K              &  64MB  & Linear Algebra &
~\cite{RodiniaBenchmark,8091072}\\ 
GEMM		& 2\x{}16K\x{}16K & 1GB & Linear Algebra & ~\cite{OpenBLAS,cuBLAS,FBGEMM}\\
HotSpot3D	& 8\x{}8K\x{}8K           &  2GB & Physics Simulation
&~\cite{RodiniaBenchmark,8091072} \\
LUD		        & 1\x{}4K\x{}4K   &  64MB &  Linear Algebra &
~\cite{RodiniaBenchmark,8091072}\\
PageRank	        &  1\x{}32K\x{}32K  &  4GB & Graph &
~\cite{PRGunRock}\\
\hline
\end{tabular}
\caption{The input dataset sizes for the \GPTPU{} benchmark applications}
\label{table:benchmark}
\vspace{-0.2in}
\end{table}

\ignore{
\begin{table}
\centering
\scriptsize
\begin{tabular}{|l|c|c|c|}
\hline
Benchmark  	&  Input Dataset                 & Input Range & Data Size\\ 
\hline
Gaussian	        & 1\x{}4K\x{}4K matrix              & real values           &  2GB \\ 
GEMM		& 2\x{}16K\x{}16K matrices & [0, 128) integers 		    &  1GB \\
Hotspot3D	& 8\x{}8K\x{}8K matrix              & real values                       &  2GB  \\
LUD		        & 1\x{}4K\x{}4K matrix              & [0, 1024) integers             &  64MB  \\
PageRank	        &  1\x{}32K\x{}32K matrix         & [0,1) real values                  &  4GB  \\
Backprop	        & 1\x{}4K\x{}4K matrix             & (-6, 6) real values              &  512MB   \\ 
BlackScholes   & 1\x{}256M\x{}9 matrix                  & real values                 &  9GB \\
\hline
\end{tabular}
\caption{The input dataset size\CMF{s for the \GPTPU{}}\CMFdel{ of} benchmark applications\CMFdel{.}}
\label{table:benchmark}
\end{table}

}
For each application described in
Sections~\ref{sec:applications}, we compared our \GPTPU{} implementations with 
(1) optimized CPU/GPU implementations from benchmark
suites~\cite{RodiniaBenchmark,AxBench} or (2) widely-used
distributions~\cite{OpenBLAS,cuBLAS,PRGunRock}. Table~\ref{table:benchmark}
lists the input datasets and the baseline implementations for each
application we used in our experiments. \rv{We only select a subset of
applictions from these benchmark suites because these are all applications
that (a) preserve the form of matrix inputs and (b) can map their core
algorithms to reasonable matrix operations. 
We do not expect \GPTPU{} and \etpu{}s to be effective for applications that
can only exploit vector arithmetics since \etpu{}'s architecture is
specialized for matrix operations.}
\ignore{
We selected the CPU/GPU implementations of
HotSpot3D, LUD, Gaussian and Backprop from the Rodinia benchmark
suite~\cite{RodiniaBenchmark}. Blackscoles from AxBench~\cite{AxBench}. 
We selected CPU/GPU versions of PageRank and GEMM implemented with
CBLAS~\cite{CBLAS} and cuBlas~\cite{cuBLAS}). }
We also use Facebook's GEMM (FBGEMM)~\cite{FBGEMM} for approximate computing on GEMM. 
\section{Results}
\label{sec:result}

This section describes the speedup, energy consumption, and accuracy observed for \GPTPU{} when running
different applications. Compared to modern CPU-based platforms running optimized code, \GPTPU{} exhibits 
improved performance and significantly reduced energy needs. 
In addition, the \GPTPU{} GEMM implementation yields more reliable results in approximation than a 
low-precision matrix-multiplication library run on a CPU.

\subsection{Single core performance: \GPTPU{} vs. CPU}
\label{sec:performance_tpu_cpu}
\ignoreexact{
\cfigure[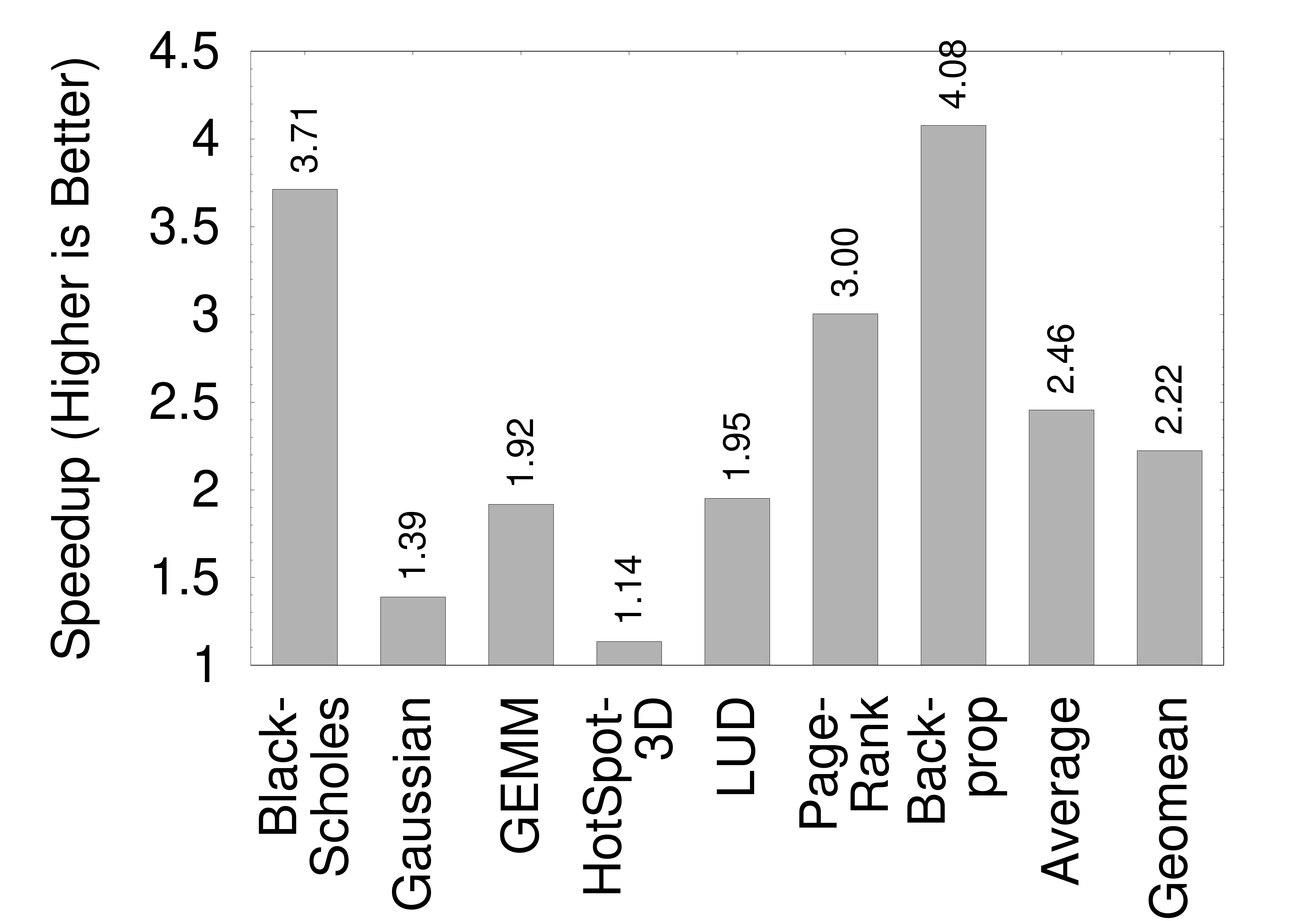, {The application speedup, energy consumption, and
energy-delay products for a single \etpu{}, relative to the baseline CPU implementations}
,fig:speedup]}

\begin{figure}[t]
\footnotesize
\begin{center}
\vspace*{-0.15in}
\begin{tabular}{cc}
\hspace*{-0.15in}
\includegraphics[width=1.8in]{Figures/speedup.pdf} &
\hspace*{-0.3in}
\includegraphics[width=1.8in]{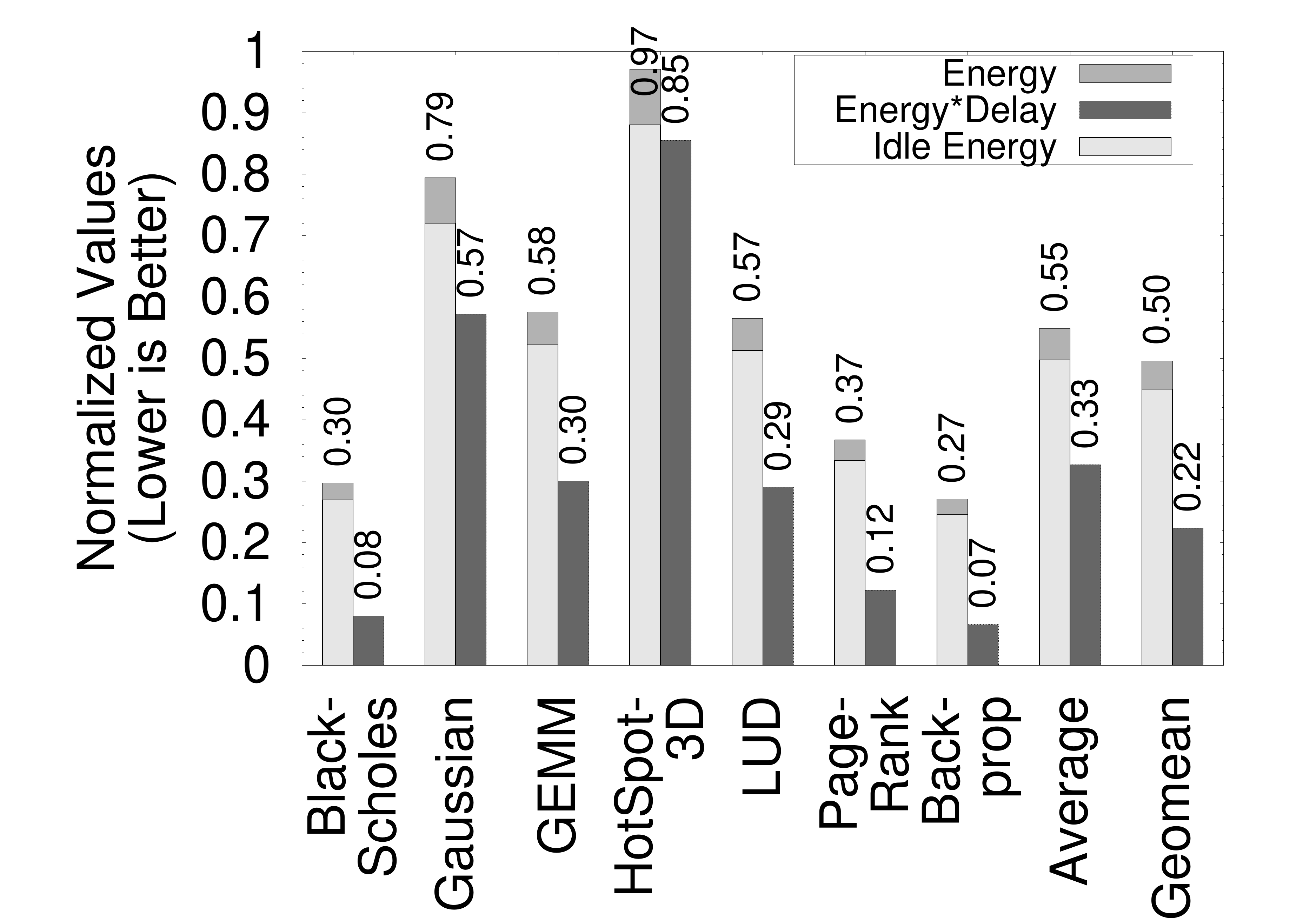} \\
(a) & (b)\\
\end{tabular}
\end{center}
\vspace{-0.1in}
\caption[]{The application (a) speedup, (b) energy consumption, and
energy-delay products for a single \etpu{}, relative to the baseline CPU
implementations}
\label{fig:speedup}
\vspace{-0.2in}
\end{figure}

Figure~\ref{fig:speedup} summarizes the speedup, energy consumption, and
energy-delay of \GPTPU{}-based applications. 
We used a single \etpu{} and a single CPU core to compare execution
of workloads in our baseline tests to compare the per-core raw hardware
capabilities.

Figure~\ref{fig:speedup}(a) compares end-to-end latency.
The \GPTPU{} system is, on average, \speedup{} faster than the CPU. For 
Backprop, the speedup is 4.08\x{} (not surprising given that the
\etpu{} was originally designed for applications like Backprop). Excluding Backprop, 
the average speedup is still 2.19\x{}. 
HotSpot3D 
actually experiences the least speedup with \GPTPU{}.
This is because \GPTPU{}'s HotSpot3D uses very small kernels and
large inputs accompany each iteration, the data-movement overhead dominates end-to-end
application latency. However, even under this scenario, \GPTPU{} can still
speed up the performance of HotSpot3D by 1.14\x{}. 

Figure~\ref{fig:speedup}(b) shows the relative energy consumption and energy-delay products
for \GPTPU{} applications vs. their CPU baseline implementations.
\GPTPU{} consumes only 5\% of the \rv{active} energy and \rv{only 51\% of the
idle energy }
that a CPU consumes (an energy savings of 45\%), and
even the worst-performing \GPTPU{} benchmark still saves \rv{3\% overall system
energy}.
For energy-delay products, which take both latency and energy
consumption into consideration, applications run with \GPTPU{} enjoy a
\rv{67\%} reduction over the baseline
CPU.
Excluding the top-performing
Backprop, \GPTPU{} still achieves an 40\% energy savings and a 62\% energy-delay improvement.

\begin{table}
\centering
\scriptsize
\begin{tabular}{|l|c|c|c|c|}
\hline
		& 	  &$-2^{7} \leq $  &  $-2^{15}\leq $ & $-2^{31}\leq $ \\
Benchmark	& default &$x< 2^{7} $     &  $x<2^{15}$     & $x<2^{31}$ \\
\hline
Backprop	&0.12\% &0.17\%	&0.10\%	&0.11\%	\\
Blackscholes	&0.18\% &0.18\%	&0.18\%	&0.18\%	\\
Gaussian	&0.00\% &0.00\%	&0.00\%	&0.00\%	\\
GEMM		&0.89\% &0.90\%	&0.90\%	&0.90\%	\\
HotSpot		&0.50\% &0.49\%	&0.46\%	&0.46\%	\\
LUD		&0.00\% &0.00\%	&0.00\%	&0.00\%	\\
PageRank	&0.61\%	&0.73\%	&0.73\%	&0.73\%	\\
\hline
Average		&0.33\%&0.35\%	&0.34\%	&0.34\%	\\
\hline
\end{tabular}
\\
(a)\\
\begin{tabular}{|l|c|c|c|c|}
\hline
		& &$-2^{7} \leq $  &  $-2^{15}\leq $ & $-2^{31}\leq $ \\
Benchmark	& default &$x< 2^{7} $  &  $x<2^{15}$ & $x<2^{31}$ \\
\hline
Backprop	&0.14\%&0.17\%	&0.12\%	&0.12\%\\
Blackscholes	&0.33\%&0.33\%	&0.33\%	&0.33\%\\
Gaussian	&0.00\% &0.00\%	&0.00\%	&0.00\%\\
GEMM		&0.98\%&0.91\%	&0.91\%	&0.91\%\\
HotSpot		&0.64\%&0.64\%	&0.59\%	&0.59\%\\
LUD		&0.00\%&0.00\%	&0.00\%	&0.00\%\\
PageRank	&0.41\% &0.91\%	&0.91\%	&0.91\%\\
\hline
Average		&0.41\%&0.42\%	&0.41\% &0.41\%\\
\hline
\end{tabular}
\\
(b)
\caption{\rv{The (a) MAPEs and (b) RMSEs for \GPTPU{} applications}}
\label{table:error}
\vspace{-0.2in}
\end{table}
\GPTPU{} sacrifices accuracy---but only to a limited
degree. Table~\ref{table:error} measured the mean absolute percentage error (MAPE) 
and the root mean square error (RMSE) between the \GPTPU{} and CPU application implementations
\rv{using the default dataset from the benchmark and our randomly generated
datasets with various ranges of values in their inputs. 
The MAPE is always less than 1\% across all applications, regardless their
ranges of input values. The average MAPE is 0.26\%--0.33\%. }
The largest RMSE we 
measured was an acceptable 0.98\%. \rv{In some cases, the
\GPTPU{} results in higher error rates in compute on default datasets than
on synthetic inputs with larger data ranges. This is because the input
values of synthetic datasets are typically normally distributed but the
real, default datasets are not always normally distributed.}

\subsection{\GPTPU{}-GEMM vs. \rv{8-bit CPU GEMM}}
\label{sec:fbgemm}
\ignoreexact{
\cfigure[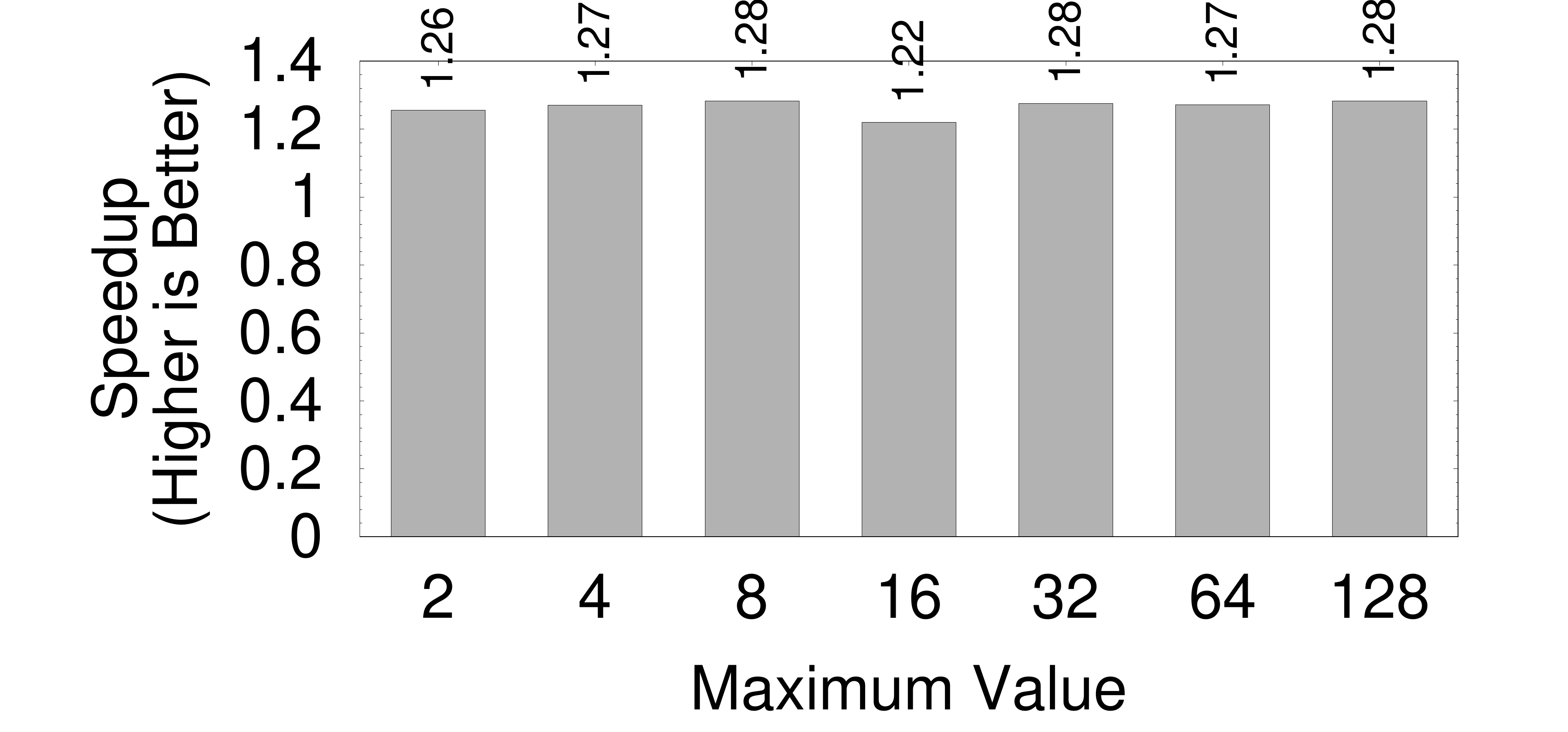, {Speedup and RMSE {for} \GPTPU{}'s
GEMM library function relative to FBGEMM},fig:fbgemm]
}
\begin{figure}[t]
\footnotesize
\begin{center}
\vspace*{-0.15in}
\begin{tabular}{cc}
\hspace*{-0.15in}
\includegraphics[width=1.8in]{Figures/FBGEMM.pdf} &
\hspace*{-0.3in}
\includegraphics[width=1.8in]{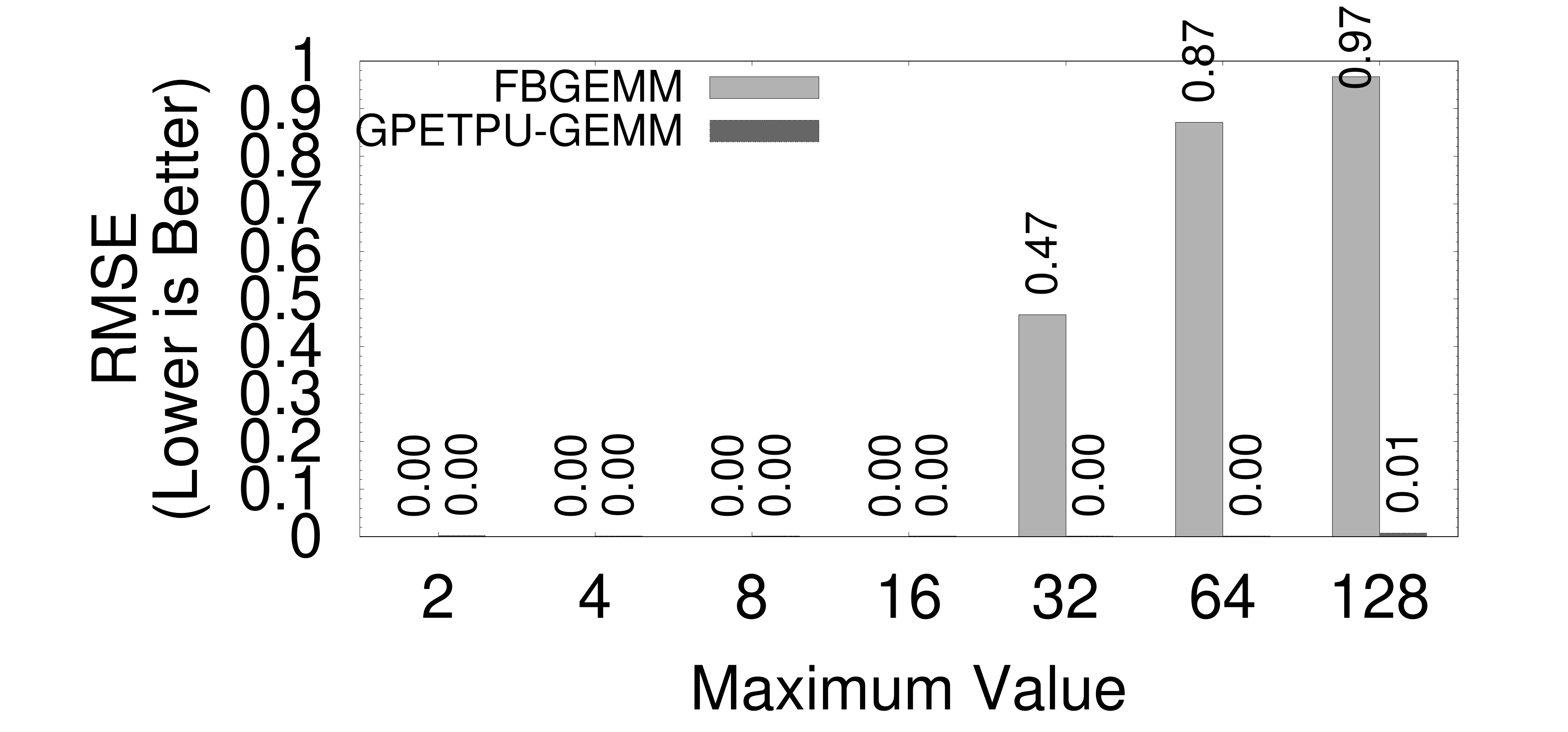} \\
(a) & (b)\\
\end{tabular}
\end{center}
\vspace{-0.1in}
\caption[]{(a) Speedup and (b) RMSE {for} \GPTPU{}'s
GEMM library function relative to FBGEMM}
\label{fig:fbgemm}
\vspace{-0.1in}
\end{figure}

\GPTPU{} allows single--\etpu{} performance to surpass single-CPU-core 
performance. That being said, the \etpu{} uses low-precision data types, whereas the baseline
CPU implementations do not. 
To account for this difference when using approximate computing 
with the CPU cores, we compared the \GPTPU{} implementation running with the state-of-the-art FBGEMM low-precision CPU matrix-multiplication library
\rv{that intensively uses the latest AVX instructions to support 8-bit
operations~\cite{FBGEMM}. We did not include other workloads in this part as
other workloads do not have implementations optimized for 8-bit CPU
operations.}

Figure~\ref{fig:fbgemm} 
shows the results for \GPTPU{}'s GEMM vs. FBGEMM using 1024\x{}1024 matrices with
positive integers and maximum input values
ranging from 2 to 128 (we chose this data size only to accommodate FBGEMM's limitations). 
As Figure~\ref{fig:fbgemm}(a) shows, \GPTPU{}-GEMM consistently outperforms FBGEMM on 
high-performance CPU cores with 1.22\x{} to 1.28\x{} across
all configurations. However, when the maximum matrix-entry value 
exceeds 16, FBGEMM's RMSE is poor as Figure~\ref{fig:fbgemm}(b)
presents, reaching 47\% when the largest value within the dataset is 32. Furthermore, the FBGEMM RMSE goes as high as 
97\%, meaning that most result values are not convincing when the largest value is 128. 
In contrast, \GPTPU{}-GEMM's RMSE is always less than 1\% (0.82\% when
maximum value is 128). \rv{This is because FB's GEMM targets at error-tolerant
ML applications but does not handle overflow cases. However, the performance
evaluation indicates that even if the CPU baseline uses 8-bit operations,
\GPTPU{}-GEMM is faster.}

\subsection{Parallel processing with multiple \etpu{}s}
\label{sec:multicore}
\begin{figure}[t]
\footnotesize
\begin{center}
\vspace*{-0.15in}
\begin{tabular}{cc}
\hspace*{-0.15in}
\includegraphics[width=1.8in]{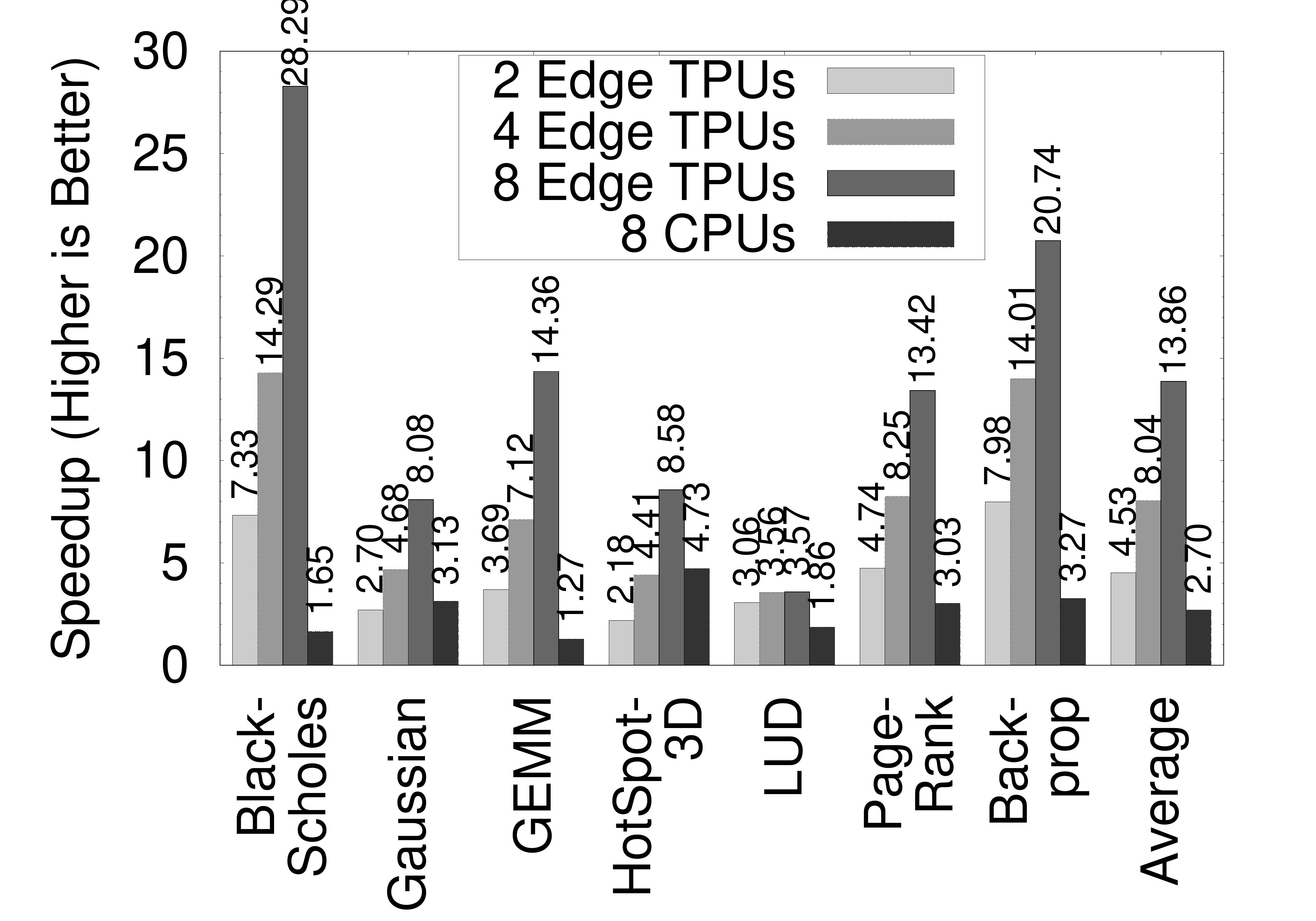} &
\hspace*{-0.3in}
\includegraphics[width=1.8in]{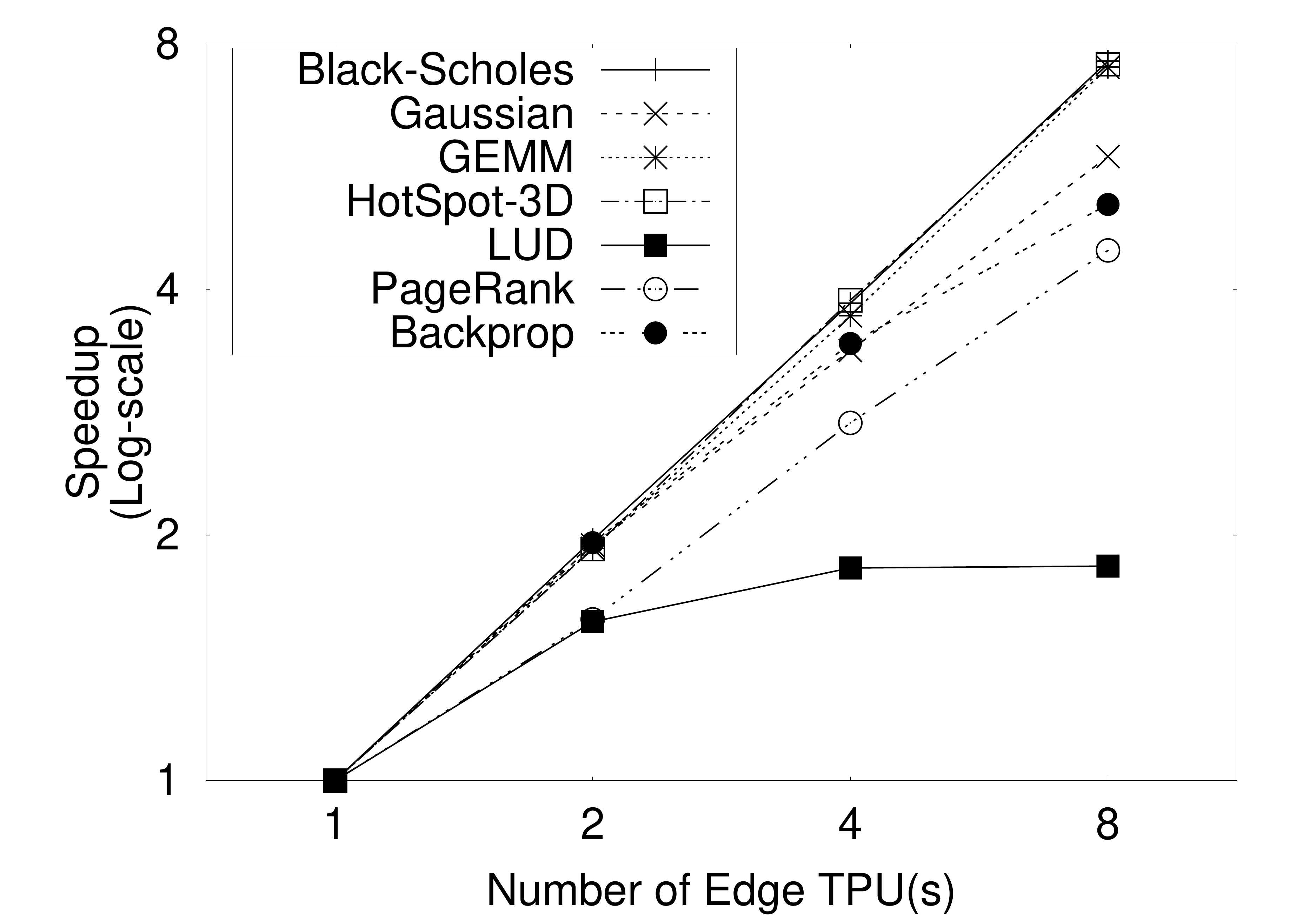} \\
(a) & (b)\\
\end{tabular}
\end{center}
\vspace{-0.1in}
\caption[]{\rv{Performance scaling for multiple \etpu{}s}}
\label{fig:gemm_mtpus}
\vspace{-0.1in}
\end{figure}
\ignore{
\begin{figure}[t]
\footnotesize
\begin{center}
\includegraphics[width=3.6in]{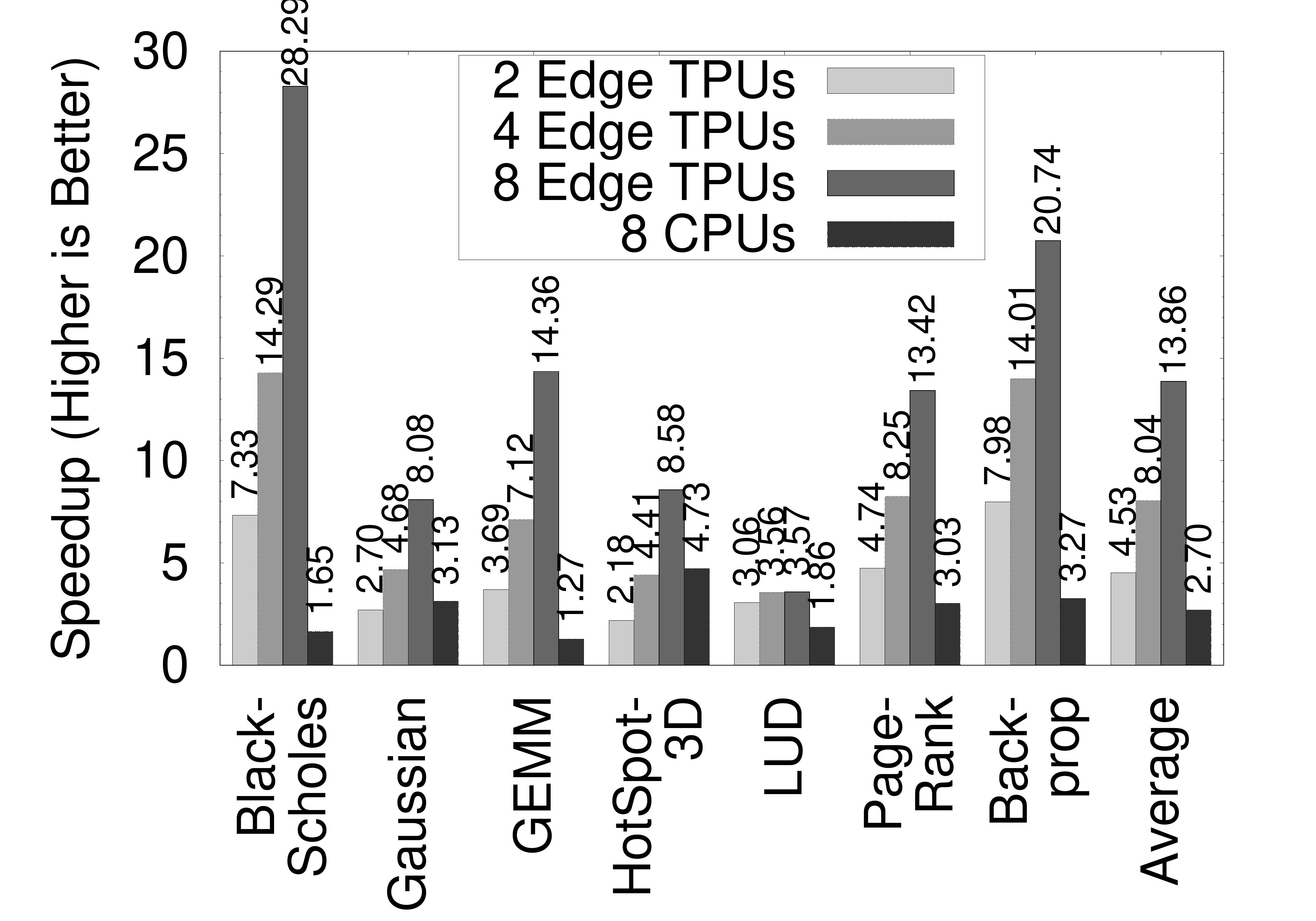}
\end{center}
\vspace{-0.1in}
\caption[]{\rv{Performance scaling for multiple \etpu{}s}}
\label{fig:gemm_mtpus}
\vspace{-0.1in}
\end{figure}

\begin{figure}[t]
\footnotesize
\begin{center}
\includegraphics[width=3.6in]{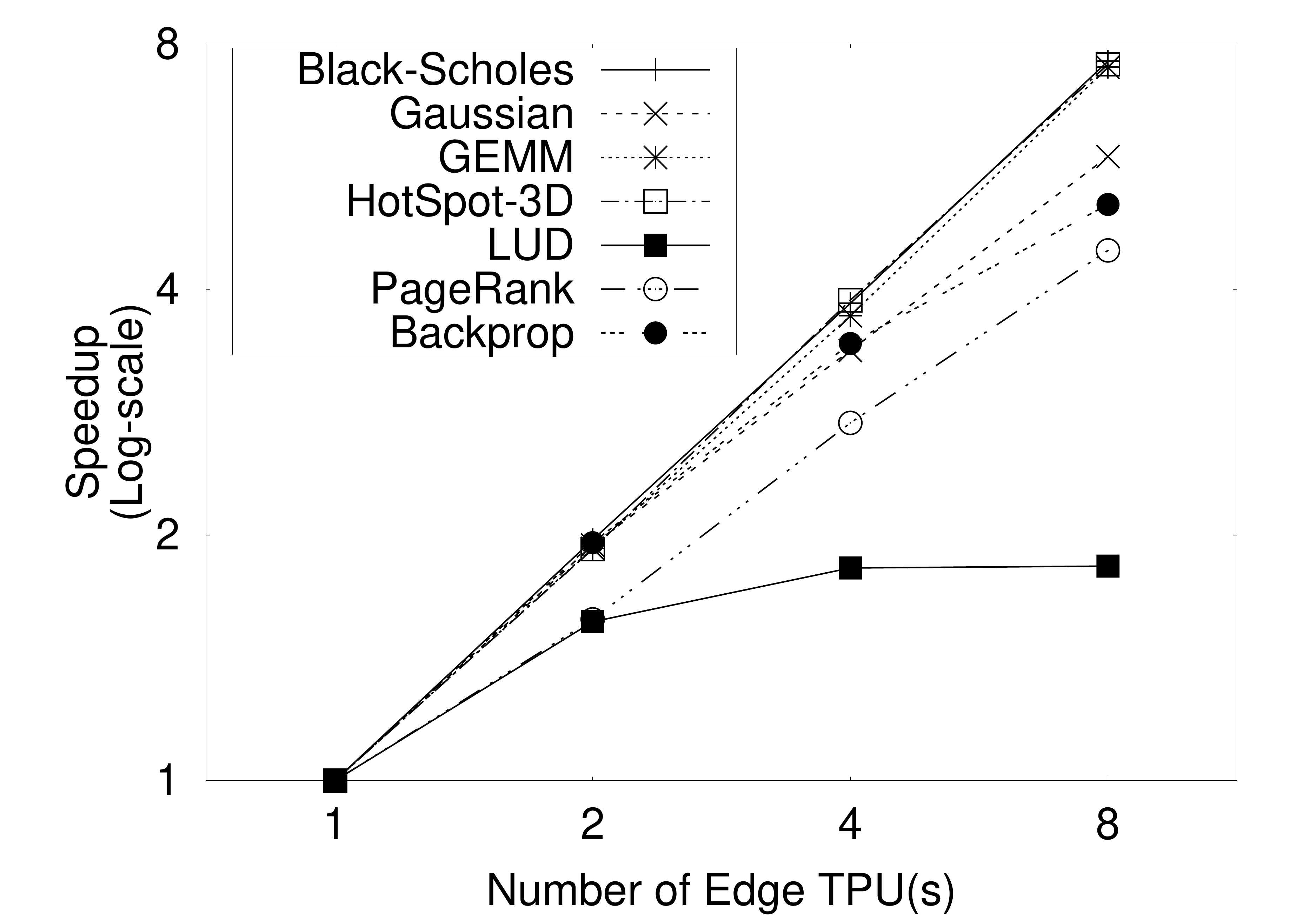}
\end{center}
\vspace{-0.1in}
\caption[]{\rv{Performance scaling for multiple \etpu{}s}}
\label{fig:gemm_mtpus_scaling}
\vspace{-0.1in}
\end{figure}
}

The \GPTPU{} runtime system uses a task queue that allows multiple \etpu{}s to process 
tasks in parallel. Even without programmer's explicit partitioning of tasks, \Tensorizer{}
also automatically generates parallel tasks from the user code. 
Figure~\ref{fig:gemm_mtpus}(a) shows the speedup of adding more \etpu{}s
into our system, without modifying the user code, compared with the
single-core CPU baseline. With 8 \etpu{}s that consume similar active
power as a single RyZen core, \GPTPU{} achieves an average 13.86\x{} speedup.
\rv{In constrast, the 8-core, OpenMP-based CPU implementations can only
achieve 2.70\x{} speedup over the baseline.}
Figure~\ref{fig:gemm_mtpus}(b) further shows log-scale performance 
with up to 8 \etpu{}s running \GPTPU{} tasks, compared with single \etpu{}. 
The linear plots reveal good performance scaling for 6 out of 7
applications when the \GPTPU{} runtime system
executes tasks in parallel. The only exception is LUD, which
already partitions matrices into four sub-matrices for computation using
matrix-wise operators, making
it difficult for \Tensorizer{} to scale the performance in only one of the
four partitions. 

\ignoreexact{
\subsection{\GPTPU{} performance in exact mode}
\label{sec:exact}

\cfigure[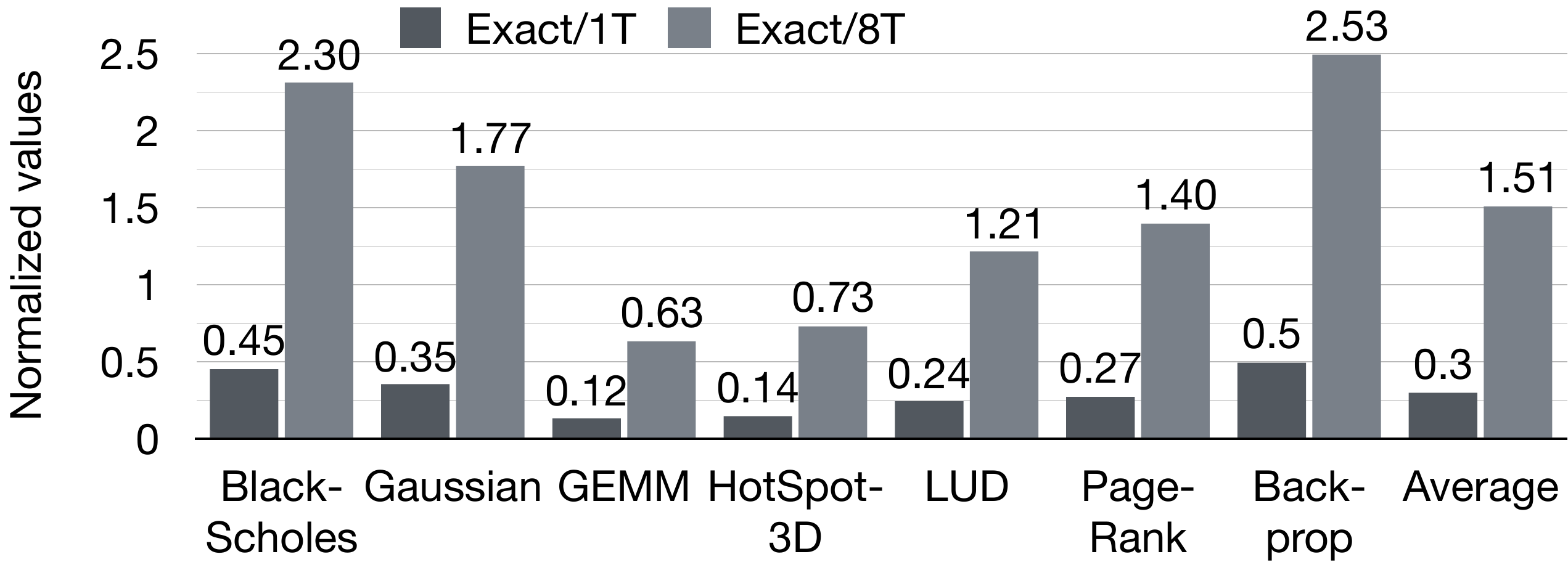, {The speedup of a single \etpu{} and 8 \etpu{}s running in
\GPTPU{}'s exact mode, relative to the baseline CPU implementations},fig:mtpus_chunking]
\cfigure[Figures/energy.pdf, {Application energy and energy-delay products for a single \etpu{}
and 8 \etpu{}s using \GPTPU{}'s exact mode, relative to the baseline CPU implementations},fig:energy]

The exact mode that the \GPTPU{} runtime provides is essential for accommodating general-purpose applications that
cannot work with approximate results. Due to the overhead from iterative numerical
computations,
a single \etpu{} (Exact/1T in Figure~\ref{fig:mtpus_chunking}) running in exact mode only achieves 0.28\x{}
the performance of a high-end CPU core due to the increased amount of
computation and aggregation. Because \etpu{} does not
provide the capability of aggregating results with more precised registers
(e.g., Cloud TPUs and NVIDIA's Tensor Cores supports computation in 16 bits
but accumulating results in 32 bits), \GPTPU{} has to split
a number into more finer-grained partitions to avoid overflow in each
partial result but also lead to more arithmetic operations. 
Nonetheless, 
the \GPTPU{} running in exact mode 
requires, on average, only 44\% of the energy that the CPU core requires---an impressive energy savings of
56\% (Exact/1T in Figure~\ref{fig:energy}). 

As the \GPTPU{} runtime can implicitly parallelize each operation (in exact computation
mode) with \Tensorizer{}'s synchronization free data placement, Figure~\ref{fig:mtpus_chunking} also shows the performance of 8 \etpu{}s running in exact computation 
mode compared with the baseline single-core CPU implementation. With 8
\etpu{}s, \GPTPU{} outperforms single-core CPU by 1.51\x{}. 
In spite of the additional power consumption from the 8 \etpu{}s, the \GPTPU{} runtime system still 
achieves a 31\% energy savings compared with the CPU\CMFdel{
implementations. T}s, and the energy-delay product is 27\% better than that of the CPUs.  

\subsection{The performance of \Tensorizer{}}
\label{sec:tensorizer_performance}
\Tensorizer{} serves as the core of performing task-level dynamic
optimizations. We compared the model/instruction/data that \Tensorizer{}
creates with the ones by not-open-sourced \etpu{} compiler in performing the
same \etpu{} task/operator of Section~\ref{sec:performance_tpu_cpu}. We exclude the compilation overhead but
only take into account for the computation time.
Note that \Tensorizer{} is 1500\x{} faster in compiling and optimizing data
layout and instructions. The result shows that \Tensorizer{} is
1.69\x{} faster on average for \etpu{} operators. 

We also evaluated the performance of \Tensorizer{}'s synchronization free
data placement for exact mode. Synchronization free data placement is
1.87\x{} faster than naive implementation that splits bits into different
sub-matrices. 
}

\subsection{Comparison with GPUs}
\label{sec:gpu}
\begin{table}
\centering
\scriptsize
\begin{tabular}{|l|c|c|l|}
\hline
  	&  Cost                 & Power Con. & Comment\\
\hline
Single \etpu{}	& USD 24.99	& 2~W& \\
RTX 2080& USD 699.66 & 215~W& Now USD 1399\\
Jetson Nano & USD 123.99 & 10~W& \\
8\x{} \etpu	& USD 159.96	& 16~W & Using 4\x{} dual \etpu{} modules \\
\hline
\end{tabular}
\caption{\rv{The cost and power consumption of hardware accelerators that we compared in this work}}
\label{table:cost}
\vspace{-0.2in}
\end{table}

\begin{figure}[t]
\footnotesize
\begin{center}
\begin{tabular}{cc}
\hspace*{-0.15in}
\includegraphics[width=1.8in]{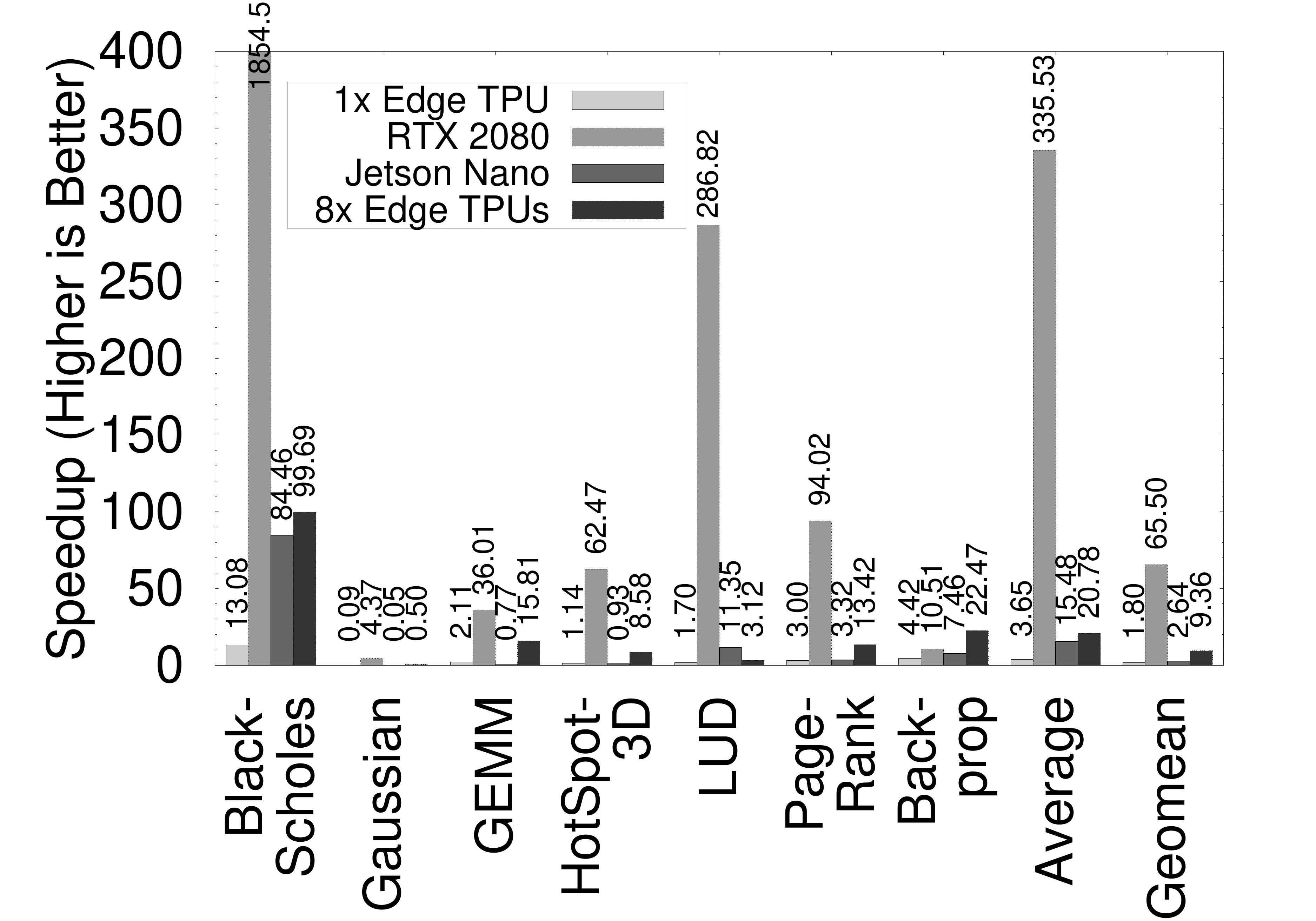} &
\hspace*{-0.3in}
\includegraphics[width=1.8in]{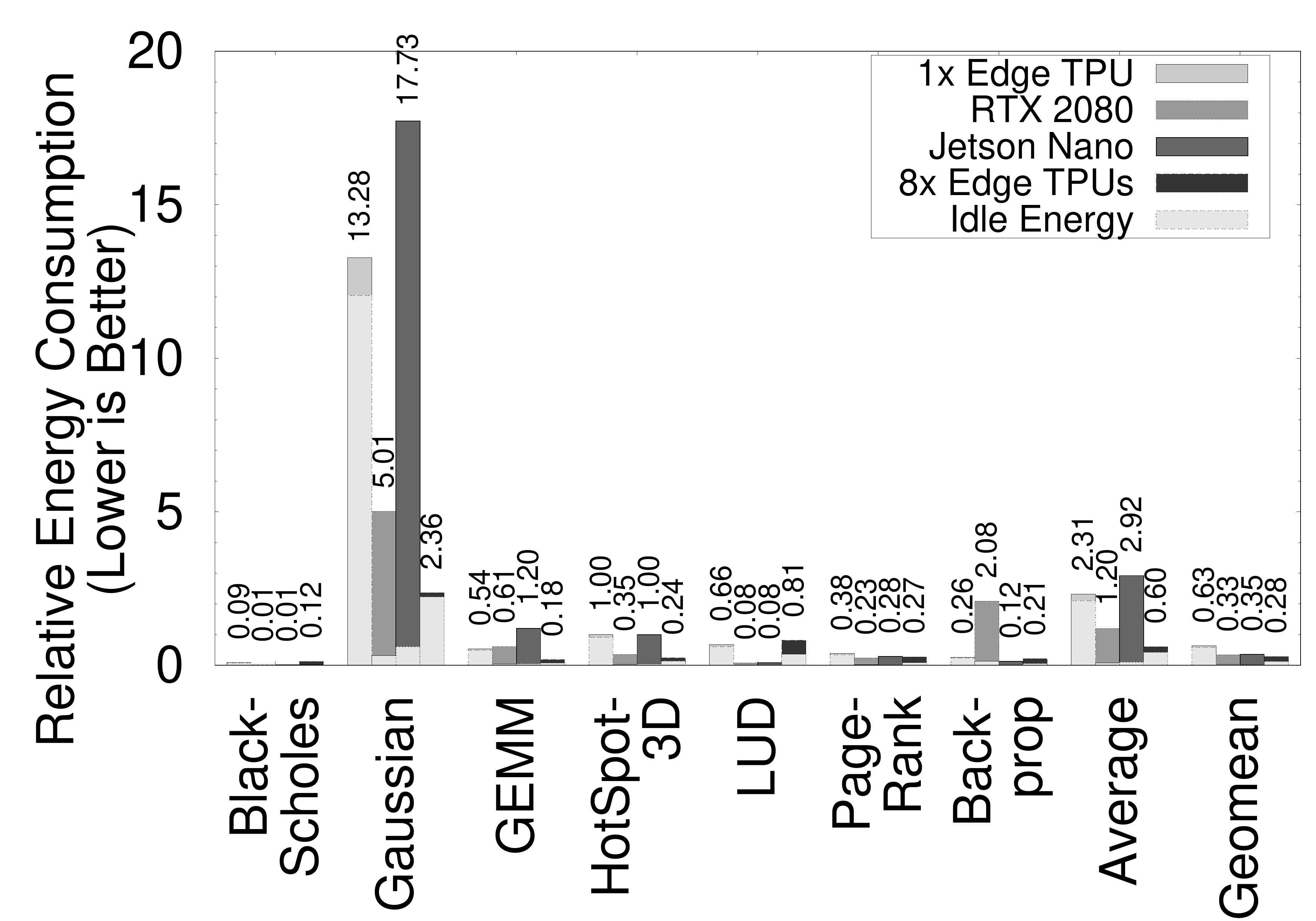} \\
(a) & (b)\\
\end{tabular}
\end{center}
\vspace{-0.1in}
\caption[]{\rv{The relative (a) performance and (b) energy for \GPTPU{} with
1\x{} and 8\x{} \etpu{}s vs. the GTX 2080
GPU and Jetson Nano}}
\label{fig:GPU}
\vspace*{-0.2in}
\end{figure}


Because an increasing number of workloads leverage GPU parallelism, we
compared the \GPTPU{} to NVIDIA's high-end Turing-architecture-based GTX 2080
and NVIDIA's embedded Jetson Nano platform. \rv{Table~\ref{table:cost} lists the
cost and power consumption of evaluated GPUs along with \etpu{}s. Due to the
limitation of Jetson Nano's available memory capacity, we have to scale down
the input datasets of Blackscholes, Gaussian, GEMM, LUD and PageRank by 25\%
to 50\% to not crash the GPU kernel.} 
Figure~\ref{fig:GPU}(a) compares the performance for the RTX 2080
and Jetson Nano, using a 
\rv{single Ryzen 3700X CPU core as the baseline}, for Rodinia benchmark 
applications and GEMM using cuBLAS. \rv{We enabled RTX-2080's 16-bit ALUs
for Gaussian, HotSpot3D, Backprop and Tensor Cores in 8-bit mode for GEMM.}
The GTX 2080 GPU is \rv{364\x{} faster than a CPU core and 69\x{}
faster than the \etpu{}}. 
The embedded GPU on Jetson Nano is still \rv{15\x{} faster than a CPU core} and 2.30\x{} faster than
an \etpu{} on average. However, with 8\x{} \etpu{}s, the \GPTPU{} can
outperform the CPU core by 3.65\x{} and Jetson Nano by 2.48\x{}. 

Figure~\ref{fig:GPU}(b) compares the energy consumption of evaluated
platforms. \rv{Including idle energy, the 8\x{}-\etpu{} system is the most
energy-efficient as the platform can save energy by 40\% from the CPU
baseline but achieve reasonable speedup. In constrast, the GTX 2080 platform consumes
9\% more energy than
the CPU baseline. Even though the idle power of the Jetson nano development
kit is simply 0.5~W, 
Jetson nano is
still more energy-consuming than GTX 2080 due to the limited speedup. }
\ignore{
However, if we take the geometric mean that
discounts the outliers, all evaluated platforms are more
energy-efficient than a CPU system. }

\rv{
If we only consider the active power consumption to exclude the factor of
various idle power in different system settings, the GTX 2080 
consumes 14\x{} the energy of 1\x{} \etpu{} on average, due to the GPU's 
195\x{} average active power consumption compared with the \etpu{}, translating 
to 4.96\x{} worse energy-delay than the baseline. Jetson Nano consumes 23.55\x{} more energy than
1\x{} \etpu{}, making the energy-delay of nano 15.54\x{} worse than
1\x{} \etpu{}. 
8\x{}-\etpu{} system consumes just 75\% more active energy than
1\x{} \etpu{}, even though the active power consumption is almost 8\x{} of a single \etpu{}.
With 8\x{} \etpu{}s, \GPTPU{} offers even better energy-delay (i.e., 46\%
lower) than the baseline. }
This result shows
that \GPTPU{} offers better energy-efficiency than
the current GPU-based solution on embedded/edge platforms. 


\ignore{
using a single \etpu{} based \GPTPU{} as the baseline. 
The GTX 2080 
consumes 15.63\x{} the energy of the
\etpu{} on average, due to the GPU's 195\x{} average active power consumption compared with
the \etpu{}, translating to 5\x{} worse energy-delay than the baseline. 
Jetson Nano consumes 23.55\x{} more energy than
\GPTPU{}, making the energy-delay of nano 15.54\x{} worse than \GPTPU{}. With
8\x{} \etpu{}s, \GPTPU{} consumes the 75\% more energy, even
though the active power consumption is almost 8\x{} of a single \etpu{}.
With 8\x{} \etpu{}s, \GPTPU{} offers even better energy-delay (i.e., 46\%
lower) than the baseline. 
}
\ignoreexact{
In addition, \GPTPU{} still excels Nano in three out of 7 evaluated applications,
and PageRank is a tie on both platforms. With geometric mean discounting the
outliers, the performance of these two platforms is very close. }

\section{Related work}
\label{sec:related_works}

\rv{Neural processing units (NPUs)~\cite{HadiNeuralApproximation,YazdanbakhshNGPU}
work by using pre-trained models that predicts the outcome of code blocks and
map the user program to these models. }
The \GPTPU{}-based approach is fundamentally different from approaches that rely on the acceleration of 
approximate programs via NPU in three important ways: (1) \GPTPU{} can accelerate any user-defined algorithm by mapping tensor/matrix operations
to supported operators, whereas NPUs can only accelerate a limited set
of algorithms that match previously trained NN models.
(2) \GPTPU{} can leverage the \etpu{} microarchitecture and NN hardware to implement exact tensor/matrix operations for applications, whereas NPUs
use NNs to produce approximate results for applications. 
(3) \GPTPU{} can achieve the desired level of precision by iteratively computing on different 
portions of raw input numbers, whereas NPUs are always limited by the approximate outcomes of NN models. 

ASICs can be used like TPUs to accelerate NN applications, as can existing fine-tuned architecture components.
Industry data centers~\cite{caulfield2016cloud,FBML,Catapult} take advantage of heterogeneous hardware
components by using different processors and reconfigurable hardware for different ML tasks.
EFLOPS~\cite{EFLOPS}, Richins et. al.~\cite{EdgeDataCentersAI}, and FlexTensor~\cite{FlexTensor} optimize algorithms
and task allocations for network traffic in data-center-scale edge
computing or single-server computing to reduce infrastructure costs.
Language frameworks like ApproxHPVM~\cite{10.1145/3360612} and
ApproxTuner~\cite{ApproxTuner} further helps programmer to estimate and
optimize the loss of accruacy in ML workloads.
The \GPTPU{} framework is orthogonal to the aforementioned research because \GPTPU{}
 is compatible with existing heterogeneous computing platforms; \etpu{}s can function as complementary hardware 
accelerators within the system. Ultimately, emerging tensor-processing hardware will inspire the development of
related algorithms and associated software~\cite{Chou:2018:FAS:3288538.3276493,
10.1145/3329785.3329932, 10.1145/3330345.3331057}. 
We have seen work extending the application of TPUs to medical image
processing~\cite{49748}. We expect \GPTPU{} can further facilitate this
trend. 
\GPTPU{} can exist in parallel to
such future research and potentially extend newly developed algorithms to work in additional application
domains.

This paper does not focus on sparse matrices, as many NN accelerators
 implicitly optimize for sparse matrices. Examples include SCNN~\cite{SCNN}, SparTen~\cite{SparTen}, Sparch~\cite{zhang2020sparch}, Scalpel~\cite{Scalpel}, SIGMA~\cite{SIGMA}, Cambricon-X~\cite{Cambricon}, Bit-Tactical~\cite{Bit-Tactical}, Bit-Pragmatic~\cite{Bit-Pragmatic}, OuterSPACE~\cite{OuterSPACE}, Laconic ~\cite{Laconic}, Bit Fusion~\cite{BitFusion},
Sparse Tensor Core~\cite{SparseTensorCore}, PermDNN~\cite{PermDNN}, Park et al.~\cite{Outlier}, Song et al.~\cite{PBEDNN}, and Rhu et al.~\cite{rhu2018compressing}.
\section{Conclusion}
\label{sec:conclude}
This paper presents \GPTPU{} to bridge the gap between NN
accelerators and general-purpose programming. 
By reverse engineering the commercially available, low-profile NN accelerator, the Google \etpu{}, 
to uncover important architectural characteristics and the data-exchange
protocol, we implement an efficient runtime system, including \Tensorizer{}
that dynamically optimizes data layout and instructions, as the \GPTPU{} platform's
backend. Using the \GPTPU{} platform and the derived performance numbers, we
re-designed the algorithms for a set of important, non-AI/ML related applications. 
The prototype \GPTPU{} system exhibits a \speedup{} speedup over modern high-end CPUs
with \energysaving{} energy reduction. Though single \etpu{} performance is not yet competitive
with high-end GPUs, but the strong scalability of multiple \etpu{}s reveals
the potential of future extensions of this line of accelerators. \rv{As the
demand of ML applications keep growing, we expect manufacturers to keep
advancing the microarchitecture of ML accelerators for higher performance and
energy-efficiency. }
\GPTPU{} thus represents an important exploration of general-purpose computing on NN
accelerators and is complementary to existing work. 
The insights presented in this paper will also help  extend the range of NN accelerator applications \rv{as well as guiding
the algorithm design and code optimization for future NN accelerators.}
\section*{Acknowledgments}
The authors would like to thank the anonymous reviewers
for their helpful comments.  
This work was sponsored by an National Science Foundation (NSF) award, 2007124.
This work was also supported by new faculty start-up funds from University of California, Riverside. 
We also owe a debt of gratitude to 
Christopher Fraser for his
excellent copyediting skills.



%

\bibliographystyle{ieeetr}
\bibliography{paper}

\begin{thebibliography}{100}

\bibitem{EdgeTPUM2}
{Google LLC}, ``Coral {M}.2 accelerator datasheet.''
  \url{https://coral.withgoogle.com/static/files/Coral-M2-datasheet.pdf}, 2019.

\bibitem{AppleM1}
{Apple}, ``{Small chip. Giant leap.}.'' \url{https://www.apple.com/mac/m1/}.

\bibitem{ryoo2008optimization}
S.~Ryoo, C.~I. Rodrigues, S.~S. Baghsorkhi, S.~S. Stone, D.~B. Kirk, and
  W.-m.~W. Hwu, ``Optimization principles and application performance
  evaluation of a multithreaded {GPU} using {CUDA},'' in {\em Proceedings of
  the 13th ACM SIGPLAN Symposium on Principles and practice of parallel
  programming}, pp.~73--82, 2008.

\bibitem{volkov2008benchmarking}
V.~Volkov and J.~W. Demmel, ``Benchmarking gpus to tune dense linear algebra,''
  in {\em SC'08: Proceedings of the 2008 ACM/IEEE conference on
  Supercomputing}, pp.~1--11, IEEE, 2008.

\bibitem{garland2008parallel}
M.~Garland, S.~Le~Grand, J.~Nickolls, J.~Anderson, J.~Hardwick, S.~Morton,
  E.~Phillips, Y.~Zhang, and V.~Volkov, ``Parallel computing experiences with
  cuda,'' {\em IEEE micro}, vol.~28, no.~4, pp.~13--27, 2008.

\bibitem{lee2009openmp}
S.~Lee, S.-J. Min, and R.~Eigenmann, ``Openmp to gpgpu: a compiler framework
  for automatic translation and optimization,'' {\em ACM Sigplan Notices},
  vol.~44, no.~4, pp.~101--110, 2009.

\bibitem{narasiman2011improving}
V.~Narasiman, M.~Shebanow, C.~J. Lee, R.~Miftakhutdinov, O.~Mutlu, and Y.~N.
  Patt, ``Improving gpu performance via large warps and two-level warp
  scheduling,'' in {\em Proceedings of the 44th Annual IEEE/ACM International
  Symposium on Microarchitecture}, pp.~308--317, 2011.

\bibitem{yang2010gpgpu}
Y.~Yang, P.~Xiang, J.~Kong, and H.~Zhou, ``A gpgpu compiler for memory
  optimization and parallelism management,'' {\em ACM Sigplan Notices},
  vol.~45, no.~6, pp.~86--97, 2010.

\bibitem{ryoo2008program}
S.~Ryoo, C.~I. Rodrigues, S.~S. Stone, S.~S. Baghsorkhi, S.-Z. Ueng, J.~A.
  Stratton, and W.-m.~W. Hwu, ``Program optimization space pruning for a
  multithreaded gpu,'' in {\em Proceedings of the 6th annual IEEE/ACM
  international symposium on Code generation and optimization}, pp.~195--204,
  2008.

\bibitem{baskaran2008compiler}
M.~M. Baskaran, U.~Bondhugula, S.~Krishnamoorthy, J.~Ramanujam, A.~Rountev, and
  P.~Sadayappan, ``A compiler framework for optimization of affine loop nests
  for gpgpus,'' in {\em Proceedings of the 22nd annual international conference
  on Supercomputing}, pp.~225--234, 2008.

\bibitem{zhang2011fly}
E.~Z. Zhang, Y.~Jiang, Z.~Guo, K.~Tian, and X.~Shen, ``On-the-fly elimination
  of dynamic irregularities for gpu computing,'' {\em ACM SIGPLAN Notices},
  vol.~46, no.~3, pp.~369--380, 2011.

\bibitem{jablin2011automatic}
T.~B. Jablin, P.~Prabhu, J.~A. Jablin, N.~P. Johnson, S.~R. Beard, and D.~I.
  August, ``Automatic cpu-gpu communication management and optimization,'' in
  {\em Proceedings of the 32nd ACM SIGPLAN conference on Programming language
  design and implementation}, pp.~142--151, 2011.

\bibitem{CUDA}
{NVIDIA Corporation}, ``{CUDA} {C} programming guide v6.0.''
  \url{http://docs.nvidia.com/cuda/pdf/CUDA_C_Programming_Guide.pdf}, 2014.

\bibitem{OpenCL}
{Khronos Group}, ``{OpenCL}.'' \url{http://www.khronos.org/opencl/}.

\bibitem{HyGCN}
M.~{Yan}, L.~{Deng}, X.~{Hu}, L.~{Liang}, Y.~{Feng}, X.~{Ye}, Z.~{Zhang},
  D.~{Fan}, and Y.~{Xie}, ``{HyGCN}: A {GCN} accelerator with hybrid
  architecture,'' in {\em 2020 IEEE International Symposium on High Performance
  Computer Architecture (HPCA)}, pp.~15--29, 2020.

\bibitem{Caffeine}
C.~Zhang, Z.~Fang, P.~Zhou, P.~Pan, and J.~Cong, ``Caffeine: Towards uniformed
  representation and acceleration for deep convolutional neural networks,'' in
  {\em 2016 IEEE/ACM International Conference on Computer-Aided Design
  (ICCAD)}, IEEE Press.

\bibitem{Chain-NN}
S.~Wang, D.~Zhou, X.~Han, and T.~Yoshimura, ``Chain-{NN}: An energy-efficient
  1{D} chain architecture for accelerating deep convolutional neural
  networks,'' in {\em Design, Automation Test in Europe Conference Exhibition
  (DATE), 2017}, DATE, '17, pp.~1032--1037, IEEE Press, 2017.

\bibitem{Tensaurus}
N.~{Srivastava}, H.~{Jin}, S.~{Smith}, H.~{Rong}, D.~{Albonesi}, and
  Z.~{Zhang}, ``Tensaurus: A versatile accelerator for mixed sparse-dense
  tensor computations,'' in {\em IEEE International Symposium on High
  Performance Computer Architecture (HPCA)}, HPCA '20, pp.~689--702, IEEE
  Press, 2020.

\bibitem{Eyeriss}
Y.~{Chen}, J.~{Emer}, and V.~{Sze}, ``Eyeriss: A spatial architecture for
  energy-efficient dataflow for convolutional neural networks,'' in {\em 2016
  ACM/IEEE 43rd Annual International Symposium on Computer Architecture
  (ISCA)}, pp.~367--379, 2016.

\bibitem{Tangram}
M.~Gao, X.~Yang, J.~Pu, M.~Horowitz, and C.~Kozyrakis, ``Tangram: Optimized
  coarse-grained dataflow for scalable nn accelerators,'' in {\em Proceedings
  of the Twenty-Fourth International Conference on Architectural Support for
  Programming Languages and Operating Systems}, ASPLOS ’19, (New York, NY,
  USA), p.~807–820, Association for Computing Machinery, 2019.

\bibitem{SNNAP}
T.~{Moreau}, M.~{Wyse}, J.~{Nelson}, A.~{Sampson}, H.~{Esmaeilzadeh},
  L.~{Ceze}, and M.~{Oskin}, ``{SNNAP}: Approximate computing on programmable
  {SoC}s via neural acceleration,'' in {\em 2015 IEEE 21st International
  Symposium on High Performance Computer Architecture (HPCA)}, pp.~603--614,
  2015.

\bibitem{AccPar}
L.~{Song}, F.~{Chen}, Y.~{Zhuo}, X.~{Qian}, H.~{Li}, and Y.~{Chen}, ``{AccPar}:
  Tensor partitioning for heterogeneous deep learning accelerators,'' in {\em
  2020 IEEE International Symposium on High Performance Computer Architecture
  (HPCA)}, pp.~342--355, 2020.

\bibitem{auto}
X.~Wei, C.~H. Yu, P.~Zhang, Y.~Chen, Y.~Wang, H.~Hu, Y.~Liang, and J.~Cong,
  ``Automated systolic array architecture synthesis for high throughput {CNN}
  inference on {FPGA}s,'' in {\em 2017 54th ACM/EDAC/IEEE Design Automation
  Conference (DAC)}, DAC, '17, (New York, NY, USA), IEEE Press, 2017.

\bibitem{PSCNN}
H.~Kung, B.~McDanel, and S.~Q. Zhang, ``Packing sparse convolutional neural
  networks for efficient systolic array implementations: Column combining under
  joint optimization,'' in {\em Proceedings of the Twenty-Fourth International
  Conference on Architectural Support for Programming Languages and Operating
  Systems}, ASPLOS ’19, (New York, NY, USA), p.~821–834, Association for
  Computing Machinery, 2019.

\bibitem{DianNao}
T.~Chen, Z.~Du, N.~Sun, J.~Wang, C.~Wu, Y.~Chen, and O.~Temam, ``{DianNao}: A
  small-footprint high-throughput accelerator for ubiquitous
  machine-learning,'' {\em SIGARCH Comput. Archit. News}, vol.~42,
  pp.~269--284, Feb. 2014.

\bibitem{DaDianNao}
Y.~{Chen}, T.~{Luo}, S.~{Liu}, S.~{Zhang}, L.~{He}, J.~{Wang}, L.~{Li},
  T.~{Chen}, Z.~{Xu}, N.~{Sun}, and O.~{Temam}, ``{DaDianNao}: A
  machine-learning supercomputer,'' in {\em 2014 47th Annual IEEE/ACM
  International Symposium on Microarchitecture}, pp.~609--622, 2014.

\bibitem{MAERI}
H.~Kwon, A.~Samajdar, and T.~Krishna, ``{MAERI}: Enabling flexible dataflow
  mapping over dnn accelerators via reconfigurable interconnects,'' in {\em
  Proceedings of the Twenty-Third International Conference on Architectural
  Support for Programming Languages and Operating Systems}, ASPLOS ’18, (New
  York, NY, USA), p.~461–475, Association for Computing Machinery, 2018.

\bibitem{CambriconISA}
S.~{Liu}, Z.~{Du}, J.~{Tao}, D.~{Han}, T.~{Luo}, Y.~{Xie}, Y.~{Chen}, and
  T.~{Chen}, ``Cambricon: An instruction set architecture for neural
  networks,'' in {\em 2016 ACM/IEEE 43rd Annual International Symposium on
  Computer Architecture (ISCA)}, pp.~393--405, 2016.

\bibitem{FlexFlow}
W.~{Lu}, G.~{Yan}, J.~{Li}, S.~{Gong}, Y.~{Han}, and X.~{Li}, ``Flex{F}low: A
  flexible dataflow accelerator architecture for convolutional neural
  networks,'' in {\em 2017 IEEE International Symposium on High Performance
  Computer Architecture (HPCA)}, pp.~553--564, 2017.

\bibitem{ScaleDeep}
S.~{Venkataramani}, A.~{Ranjan}, S.~{Banerjee}, D.~{Das}, S.~{Avancha},
  A.~{Jagannathan}, A.~{Durg}, D.~{Nagaraj}, B.~{Kaul}, P.~{Dubey}, and
  A.~{Raghunathan}, ``Scale{D}eep: A scalable compute architecture for learning
  and evaluating deep networks,'' in {\em 2017 ACM/IEEE 44th Annual
  International Symposium on Computer Architecture (ISCA)}, pp.~13--26, 2017.

\bibitem{MnnFast}
H.~{Jang}, J.~{Kim}, J.~{Jo}, J.~{Lee}, and J.~{Kim}, ``Mnn{F}ast: A fast and
  scalable system architecture for memory-augmented neural networks,'' in {\em
  2019 ACM/IEEE 46th Annual International Symposium on Computer Architecture
  (ISCA)}, pp.~250--263, 2019.

\bibitem{TIE}
C.~{Deng}, F.~{Sun}, X.~{Qian}, J.~{Lin}, Z.~{Wang}, and B.~{Yuan}, ``{TIE}:
  Energy-efficient tensor train-based inference engine for deep neural
  network,'' in {\em 2019 ACM/IEEE 46th Annual International Symposium on
  Computer Architecture (ISCA)}, pp.~264--277, 2019.

\bibitem{UCNN}
K.~{Hegde}, J.~{Yu}, R.~{Agrawal}, M.~{Yan}, M.~{Pellauer}, and C.~{Fletcher},
  ``{UCNN}: Exploiting computational reuse in deep neural networks via weight
  repetition,'' in {\em 2018 ACM/IEEE 45th Annual International Symposium on
  Computer Architecture (ISCA)}, pp.~674--687, 2018.

\bibitem{CirCNN}
C.~Ding, S.~Liao, Y.~Wang, Z.~Li, N.~Liu, Y.~Zhuo, C.~Wang, X.~Qian, Y.~Bai,
  G.~Yuan, X.~Ma, Y.~Zhang, J.~Tang, Q.~Qiu, X.~Lin, and B.~Yuan, ``Cir{CNN}:
  Accelerating and compressing deep neural networks using block-circulant
  weight matrices,'' in {\em Proceedings of the 50th Annual IEEE/ACM
  International Symposium on Microarchitecture}, MICRO-50 '17, (New York, NY,
  USA), p.~395–408, Association for Computing Machinery, 2017.

\bibitem{HyPar}
L.~{Song}, J.~{Mao}, Y.~{Zhuo}, X.~{Qian}, H.~{Li}, and Y.~{Chen}, ``Hy{P}ar:
  Towards hybrid parallelism for deep learning accelerator array,'' in {\em
  2019 IEEE International Symposium on High Performance Computer Architecture
  (HPCA)}, pp.~56--68, 2019.

\bibitem{Scale-Out}
J.~{Park}, H.~{Sharma}, D.~{Mahajan}, J.~K. {Kim}, P.~{Olds}, and
  H.~{Esmaeilzadeh}, ``Scale-out acceleration for machine learning,'' in {\em
  2017 50th Annual IEEE/ACM International Symposium on Microarchitecture
  (MICRO)}, pp.~367--381, 2017.

\bibitem{DNNFPGA}
H.~{Sharma}, J.~{Park}, D.~{Mahajan}, E.~{Amaro}, J.~K. {Kim}, C.~{Shao},
  A.~{Mishra}, and H.~{Esmaeilzadeh}, ``From high-level deep neural models to
  {FPGA}s,'' in {\em 2016 49th Annual IEEE/ACM International Symposium on
  Microarchitecture (MICRO)}, pp.~1--12, 2016.

\bibitem{FLCNN}
M.~{Alwani}, H.~{Chen}, M.~{Ferdman}, and P.~{Milder}, ``Fused-layer {CNN}
  accelerators,'' in {\em 2016 49th Annual IEEE/ACM International Symposium on
  Microarchitecture (MICRO)}, pp.~1--12, 2016.

\bibitem{song2018towards}
M.~Song, J.~Zhang, H.~Chen, and T.~Li, ``Towards efficient microarchitectural
  design for accelerating unsupervised {GAN}-based deep learning,'' in {\em
  2018 IEEE International Symposium on High Performance Computer Architecture
  (HPCA)}, pp.~66--77, IEEE, 2018.

\bibitem{ShortcutMining}
A.~{Azizimazreah} and L.~{Chen}, ``Shortcut mining: Exploiting cross-layer
  shortcut reuse in {DCNN} accelerators,'' in {\em 2019 IEEE International
  Symposium on High Performance Computer Architecture (HPCA)}, pp.~94--105,
  2019.

\bibitem{VIP}
S.~{Hurkat} and J.~F. {Martínez}, ``{VIP}: A versatile inference processor,''
  in {\em 2019 IEEE International Symposium on High Performance Computer
  Architecture (HPCA)}, pp.~345--358, 2019.

\bibitem{Simba}
Y.~S. Shao, J.~Clemons, R.~Venkatesan, B.~Zimmer, M.~Fojtik, N.~Jiang,
  B.~Keller, A.~Klinefelter, N.~Pinckney, P.~Raina, S.~G. Tell, Y.~Zhang, W.~J.
  Dally, J.~Emer, C.~T. Gray, B.~Khailany, and S.~W. Keckler, ``Simba: Scaling
  deep-learning inference with multi-chip-module-based architecture,'' in {\em
  Proceedings of the 52nd Annual IEEE/ACM International Symposium on
  Microarchitecture}, MICRO ’52, (New York, NY, USA), p.~14–27, Association
  for Computing Machinery, 2019.

\bibitem{NeuralCache}
C.~Eckert, X.~Wang, J.~Wang, A.~Subramaniyan, R.~Iyer, D.~Sylvester, D.~Blaauw,
  and R.~Das, ``Neural cache: Bit-serial in-cache acceleration of deep neural
  networks,'' in {\em Proceedings of the 45th Annual International Symposium on
  Computer Architecture}, ISCA'18, pp.~383--396, IEEE Press, 2018.

\bibitem{TensorDIMM}
Y.~Kwon, Y.~Lee, and M.~Rhu, ``Tensor{DIMM}: A practical near-memory processing
  architecture for embeddings and tensor operations in deep learning,'' in {\em
  Proceedings of the 52nd Annual IEEE/ACM International Symposium on
  Microarchitecture}, MICRO ’52, (New York, NY, USA), p.~740–753,
  Association for Computing Machinery, 2019.

\bibitem{Manna}
J.~R. Stevens, A.~Ranjan, D.~Das, B.~Kaul, and A.~Raghunathan, ``Manna: An
  accelerator for memory-augmented neural networks,'' in {\em Proceedings of
  the 52nd Annual IEEE/ACM International Symposium on Microarchitecture}, MICRO
  ’52, (New York, NY, USA), p.~794–806, Association for Computing
  Machinery, 2019.

\bibitem{DRISA}
S.~{Li}, D.~{Niu}, K.~T. {Malladi}, H.~{Zheng}, B.~{Brennan}, and Y.~{Xie},
  ``{DRISA}: A dram-based reconfigurable in-situ accelerator,'' in {\em 2017
  50th Annual IEEE/ACM International Symposium on Microarchitecture (MICRO)},
  pp.~288--301, 2017.

\bibitem{TETRIS}
M.~Gao, J.~Pu, X.~Yang, M.~Horowitz, and C.~Kozyrakis, ``{TETRIS}: Scalable and
  efficient neural network acceleration with {3D} memory,'' {\em SIGPLAN Not.},
  vol.~52, p.~751–764, Apr. 2017.

\bibitem{NAND-Net}
H.~{Kim}, J.~{Sim}, Y.~{Choi}, and L.~{Kim}, ``{NAND-Net}: Minimizing
  computational complexity of in-memory processing for binary neural
  networks,'' in {\em 2019 IEEE International Symposium on High Performance
  Computer Architecture (HPCA)}, pp.~661--673, 2019.

\bibitem{SCOPE}
S.~{Li}, A.~O. {Glova}, X.~{Hu}, P.~{Gu}, D.~{Niu}, K.~{T. Malladi},
  H.~{Zheng}, B.~{Brennan}, and Y.~{Xie}, ``{SCOPE}: A stochastic computing
  engine for {DRAM}-based in-situ accelerator,'' in {\em 2018 51st Annual
  IEEE/ACM International Symposium on Microarchitecture (MICRO)}, pp.~696--709,
  2018.

\bibitem{BPICA}
X.~{Wang}, J.~{Yu}, C.~{Augustine}, R.~{Iyer}, and R.~{Das}, ``Bit prudent
  in-cache acceleration of deep convolutional neural networks,'' in {\em 2019
  IEEE International Symposium on High Performance Computer Architecture
  (HPCA)}, pp.~81--93, 2019.

\bibitem{PIM}
J.~{Liu}, H.~{Zhao}, M.~A. {Ogleari}, D.~{Li}, and J.~{Zhao},
  ``Processing-in-memory for energy-efficient neural network training: A
  heterogeneous approach,'' in {\em 2018 51st Annual IEEE/ACM International
  Symposium on Microarchitecture (MICRO)}, pp.~655--668, 2018.

\bibitem{DLANYT}
M.~{Imani}, M.~{Samragh Razlighi}, Y.~{Kim}, S.~{Gupta}, F.~{Koushanfar}, and
  T.~{Rosing}, ``Deep learning acceleration with neuron-to-memory
  transformation,'' in {\em 2020 IEEE International Symposium on High
  Performance Computer Architecture (HPCA)}, pp.~1--14, 2020.

\bibitem{FPSA}
Y.~Ji, Y.~Zhang, X.~Xie, S.~Li, P.~Wang, X.~Hu, Y.~Zhang, and Y.~Xie, ``{FPSA}:
  A full system stack solution for reconfigurable {ReRAM}-based {NN}
  accelerator architecture,'' in {\em Proceedings of the Twenty-Fourth
  International Conference on Architectural Support for Programming Languages
  and Operating Systems}, ASPLOS'19, (New York, NY, USA), pp.~733--747,
  Association for Computing Machinery, 2019.

\bibitem{LerGAN}
H.~{Mao}, M.~{Song}, T.~{Li}, Y.~{Dai}, and J.~{Shu}, ``{LerGAN}: a zero-free,
  low data movement and {PIM}-based {GAN} architecture,'' in {\em 2018 51st
  Annual IEEE/ACM International Symposium on Microarchitecture (MICRO)},
  pp.~669--681, 2018.

\bibitem{SReRAME}
T.~{Yang}, H.~{Cheng}, C.~{Yang}, I.~{Tseng}, H.~{Hu}, H.~{Chang}, and H.~{Li},
  ``Sparse {ReRAM} engine: joint exploration of activation and weight sparsity
  in compressed neural networks,'' in {\em 2019 ACM/IEEE 46th Annual
  International Symposium on Computer Architecture (ISCA)}, pp.~236--249, 2019.

\bibitem{PRIME}
P.~{Chi}, S.~{Li}, C.~{Xu}, T.~{Zhang}, J.~{Zhao}, Y.~{Liu}, Y.~{Wang}, and
  Y.~{Xie}, ``{PRIME}: a novel processing-in-memory architecture for neural
  network computation in {ReRAM}-based main memory,'' in {\em 2016 ACM/IEEE
  43rd Annual International Symposium on Computer Architecture (ISCA)},
  pp.~27--39, 2016.

\bibitem{PipeLayer}
L.~{Song}, X.~{Qian}, H.~{Li}, and Y.~{Chen}, ``{PipeLayer}: A pipelined
  {ReRAM}-based accelerator for deep learning,'' in {\em 2017 IEEE
  International Symposium on High Performance Computer Architecture (HPCA)},
  pp.~541--552, 2017.

\bibitem{PUMA}
A.~Ankit, I.~E. Hajj, S.~R. Chalamalasetti, G.~Ndu, M.~Foltin, R.~S. Williams,
  P.~Faraboschi, W.-m.~W. Hwu, J.~P. Strachan, K.~Roy, and D.~S. Milojicic,
  ``{PUMA}: A programmable ultra-efficient memristor-based accelerator for
  machine learning inference,'' in {\em Proceedings of the Twenty-Fourth
  International Conference on Architectural Support for Programming Languages
  and Operating Systems}, ASPLOS '19, (New York, NY, USA), pp.~715---731,
  Association for Computing Machinery, 2019.

\bibitem{MBM}
M.~N. {Bojnordi} and E.~{Ipek}, ``Memristive boltzmann machine: A hardware
  accelerator for combinatorial optimization and deep learning,'' in {\em 2016
  IEEE International Symposium on High Performance Computer Architecture
  (HPCA)}, pp.~1--13, 2016.

\bibitem{EHECNP}
X.~{Zhang}, S.~L. {Song}, C.~{Xie}, J.~{Wang}, W.~{Zhang}, and X.~{Fu},
  ``Enabling highly efficient capsule networks processing through a {PIM}-based
  architecture design,'' in {\em 2020 IEEE International Symposium on High
  Performance Computer Architecture (HPCA)}, pp.~542--555, 2020.

\bibitem{FloatPIM}
M.~{Imani}, S.~{Gupta}, Y.~{Kim}, and T.~{Rosing}, ``{FloatPIM}: In-memory
  acceleration of deep neural network training with high precision,'' in {\em
  2019 ACM/IEEE 46th Annual International Symposium on Computer Architecture
  (ISCA)}, pp.~802--815, 2019.

\bibitem{Khadas}
{Khadas, Shenzhen Wesion Technology Co., Ltd.}, ``{VIM3}.''
  \url{https://www.khadas.com/vim3l}, 2019.

\bibitem{RockchipRK1808}
{Fuzhou Rockchip Electronics Co., Ltd.}, ``{Rockchip RK1808}.''
  \url{https://www.rock-chips.com/a/en/products/RK18_Series/2019/0529/989.html},
  2019.

\bibitem{BM1880}
{Sophon Technology (Beijing) Co., Ltd. }, ``{Tensor Computing Processor
  BM1880}.'' \url{https://www.sophon.ai/product/introduce/bm1880.html}, 2018.

\bibitem{hikey970}
{Shenzhen LeMaker Technology Co., Ltd}, ``{HiKey 970}.''
  \url{http://www.lemaker.org/product-hikey970-specification.html}, 2018.

\bibitem{jetsonnano}
{NVIDIA Corporation}, ``{Jetson Nano Developer Kit}.''
  \url{https://developer.nvidia.com/embedded/jetson-nano-developer-kit}, 2019.

\bibitem{QNAP}
{QNAP}, ``{QM2 Expansion Card (Add M.2 SSD Slots)}.''
  \url{https://www.qnap.com/en/product/qm2-m.2ssd}, 2020.

\bibitem{9177369}
V.~{Sze}, Y.~H. {Chen}, T.~J. {Yang}, and J.~S. {Emer}, ``How to evaluate deep
  neural network processors: Tops/w (alone) considered harmful,'' {\em IEEE
  Solid-State Circuits Magazine}, vol.~12, no.~3, pp.~28--41, 2020.

\bibitem{dongarra1986linear}
J.~J. Dongarra and D.~C. Sorensen, ``Linear algebra on high performance
  computers,'' {\em Applied mathematics and computation}, vol.~20, no.~1-2,
  pp.~57--88, 1986.

\bibitem{IRA}
M.~A. Laurenzano, P.~Hill, M.~Samadi, S.~Mahlke, J.~Mars, and L.~Tang, ``Input
  responsiveness: Using canary inputs to dynamically steer approximation,'' in
  {\em Proceedings of the 37th ACM SIGPLAN Conference on Programming Language
  Design and Implementation}, PLDI '16, (New York, NY, USA), pp.~161--176, ACM,
  2016.

\bibitem{OpenBLAS}
{Zhang Xianyi and Martin Kroeker}, ``{OpenBLAS: An optimized BLAS library}.''
  \url{https://www.openblas.net/}, 2021.

\bibitem{cuBLAS}
{NVIDIA}, ``{cuBLAS}.'' \url{https://docs.nvidia.com/cuda/cublas/index.html},
  2019.

\bibitem{PageRank}
L.~Page, S.~Brin, R.~Motwani, and T.~Winograd, ``The {PageRank} citation
  ranking: Bringing order to the web.,'' Technical Report 1999-66, Stanford
  InfoLab, November 1999.
\newblock Previous number = SIDL-WP-1999-0120.

\bibitem{I2A}
T.~H. Cormen, C.~Stein, R.~L. Rivest, and C.~E. Leiserson, {\em Introduction to
  Algorithms}.
\newblock McGraw-Hill Higher Education, 2nd~ed., 2001.

\bibitem{CNDF}
K.~Aludaat and M.~Alodat, ``A note on approximating the normal distribution
  function,'' {\em Applied Mathematical Sciences (Ruse)}, 01 2008.

\bibitem{RodiniaBenchmark}
M.~B. S.~Che, J.~Meng, D.~Tarjan, J.~W. Sheaffer, S.-H. Lee, and K.~Skadron,
  ``Rodinia: A benchmark suite for heterogeneous computing,'' in {\em
  Proceedings of the IEEE International Symposium on Workload
  Characterization}, IISWC '09, pp.~44--54, Oct 2009.

\bibitem{8091072}
N.-M. Ho and W.-F. Wong, ``Exploiting half precision arithmetic in nvidia
  gpus,'' in {\em 2017 IEEE High Performance Extreme Computing Conference
  (HPEC)}, pp.~1--7, 2017.

\bibitem{AxBench}
A.~Yazdanbakhsh, D.~Mahajan, H.~Esmaeilzadeh, and P.~Lotfi-Kamran, ``{AxBench:
  A Multiplatform Benchmark Suite for Approximate Computing},'' {\em IEEE
  Design Test}, vol.~34, pp.~60--68, April 2017.

\bibitem{FBGEMM}
{Daya S Khudia and Protonu Basu and Summer Deng}, ``{Open-sourcing FBGEMM for
  state-of-the-art server-side inference}.''
  \url{https://engineering.fb.com/ml-applications/fbgemm/}, 2018.

\bibitem{PRGunRock}
{Carl Yang and Aydin Buluc and Yangzihao Wang and John D. Owens},
  ``{GraphBLAST}.'' \url{https://github.com/gunrock/graphblast}, 2019.

\bibitem{HadiNeuralApproximation}
H.~{Esmaeilzadeh}, A.~{Sampson}, L.~{Ceze}, and D.~{Burger}, ``Neural
  acceleration for general-purpose approximate programs,'' in {\em 2012 45th
  Annual IEEE/ACM International Symposium on Microarchitecture}, pp.~449--460,
  2012.

\bibitem{YazdanbakhshNGPU}
A.~Yazdanbakhsh, J.~Park, H.~Sharma, P.~Lotfi-Kamran, and H.~Esmaeilzadeh,
  ``Neural acceleration for {GPU} throughput processors,'' in {\em Proceedings
  of the 48th International Symposium on Microarchitecture}, MICRO-48, (New
  York, NY, USA), pp.~482--493, ACM, 2015.

\bibitem{caulfield2016cloud}
A.~M. {Caulfield}, E.~S. {Chung}, A.~{Putnam}, H.~{Angepat}, J.~{Fowers},
  M.~{Haselman}, S.~{Heil}, M.~{Humphrey}, P.~{Kaur}, J.~{Kim}, D.~{Lo},
  T.~{Massengill}, K.~{Ovtcharov}, M.~{Papamichael}, L.~{Woods}, S.~{Lanka},
  D.~{Chiou}, and D.~{Burger}, ``A cloud-scale acceleration architecture,'' in
  {\em 2016 49th Annual IEEE/ACM International Symposium on Microarchitecture
  (MICRO)}, pp.~1--13, IEEE, 2016.

\bibitem{FBML}
K.~{Hazelwood}, S.~{Bird}, D.~{Brooks}, S.~{Chintala}, U.~{Diril},
  D.~{Dzhulgakov}, M.~{Fawzy}, B.~{Jia}, Y.~{Jia}, A.~{Kalro}, J.~{Law},
  K.~{Lee}, J.~{Lu}, P.~{Noordhuis}, M.~{Smelyanskiy}, L.~{Xiong}, and
  X.~{Wang}, ``Applied machine learning at {F}acebook: A datacenter
  infrastructure perspective,'' in {\em 2018 IEEE International Symposium on
  High Performance Computer Architecture (HPCA)}, pp.~620--629, 2018.

\bibitem{Catapult}
A.~Putnam, A.~Caulfield, E.~Chung, D.~Chiou, K.~Constantinides, J.~Demme,
  H.~Esmaeilzadeh, J.~Fowers, G.~Gopal, J.~Gray, M.~Haselman, S.~Hauck,
  S.~Heil, A.~Hormati, J.-Y. Kim, S.~Lanka, J.~Larus, E.~Peterson, S.~Pope,
  A.~Smith, J.~Thong, P.~Xiao, and D.~Burger, ``A reconfigurable fabric for
  accelerating large-scale datacenter services,'' in {\em Computer Architecture
  (ISCA), 2014 ACM/IEEE 41st International Symposium on}, pp.~13--24, June
  2014.

\bibitem{EFLOPS}
J.~{Dong}, Z.~{Cao}, T.~{Zhang}, J.~{Ye}, S.~{Wang}, F.~{Feng}, L.~{Zhao},
  X.~{Liu}, L.~{Song}, L.~{Peng}, Y.~{Guo}, X.~{Jiang}, L.~{Tang}, Y.~{Du},
  Y.~{Zhang}, P.~{Pan}, and Y.~{Xie}, ``{EFLOPS}: Algorithm and system
  co-design for a high performance distributed training platform,'' in {\em
  2020 IEEE International Symposium on High Performance Computer Architecture
  (HPCA)}, pp.~610--622, 2020.

\bibitem{EdgeDataCentersAI}
D.~{Richins}, D.~{Doshi}, M.~{Blackmore}, A.~{Thulaseedharan Nair},
  N.~{Pathapati}, A.~{Patel}, B.~{Daguman}, D.~{Dobrijalowski}, R.~{Illikkal},
  K.~{Long}, D.~{Zimmerman}, and V.~{Janapa Reddi}, ``Missing the forest for
  the trees: End-to-end {AI} application performance in edge data centers,'' in
  {\em 2020 IEEE International Symposium on High Performance Computer
  Architecture (HPCA)}, pp.~515--528, 2020.

\bibitem{FlexTensor}
S.~Zheng, Y.~Liang, S.~Wang, R.~Chen, and K.~Sheng, ``Flextensor: An automatic
  schedule exploration and optimization framework for tensor computation on
  heterogeneous system,'' in {\em Proceedings of the Twenty-Fifth International
  Conference on Architectural Support for Programming Languages and Operating
  Systems}, ASPLOS '20, (New York, NY, USA), pp.~859--873, Association for
  Computing Machinery, 2020.

\bibitem{10.1145/3360612}
H.~Sharif, P.~Srivastava, M.~Huzaifa, M.~Kotsifakou, K.~Joshi, Y.~Sarita,
  N.~Zhao, V.~S. Adve, S.~Misailovic, and S.~Adve, ``Approxhpvm: A portable
  compiler ir for accuracy-aware optimizations,'' vol.~3, no.~OOPSLA, 2019.

\bibitem{ApproxTuner}
H.~Sharif, Y.~Zhao, M.~Kotsifakou, A.~Kothari, B.~Schreiber, E.~Wang,
  Y.~Sarita, N.~Zhao, K.~Joshi, V.~S. Adve, S.~Misailovic, and S.~Adve,
  ``{Approximating APSP without Scaling: Equivalence of Approximate Min-plus
  and Exact Min-Max},'' in {\em Symposium on Principles and Practice of
  Parallel Programming}, PPoPP 2021, 2021.

\bibitem{Chou:2018:FAS:3288538.3276493}
S.~Chou, F.~Kjolstad, and S.~Amarasinghe, ``Format abstraction for sparse
  tensor algebra compilers,'' {\em Proc. ACM Program. Lang.}, vol.~2,
  pp.~123:1--123:30, Oct. 2018.

\bibitem{10.1145/3329785.3329932}
P.~Holanda and H.~M\"{u}hleisen, ``Relational queries with a tensor processing
  unit,'' in {\em Proceedings of the 15th International Workshop on Data
  Management on New Hardware}, DaMoN'19, (New York, NY, USA), Association for
  Computing Machinery, 2019.

\bibitem{10.1145/3330345.3331057}
A.~Dakkak, C.~Li, J.~Xiong, I.~Gelado, and W.-m. Hwu, ``Accelerating reduction
  and scan using tensor core units,'' in {\em Proceedings of the ACM
  International Conference on Supercomputing}, ICS '19, (New York, NY, USA),
  p.~46?V57, Association for Computing Machinery, 2019.

\bibitem{49748}
C.~Ma, T.~Marin, T.~Lu, Y.~fan Chen, and Y.~Zhuo, ``Accelerating mri
  reconstruction on tpus,'' 2020.

\bibitem{SCNN}
A.~Parashar, M.~Rhu, A.~Mukkara, A.~Puglielli, R.~Venkatesan, B.~Khailany,
  J.~Emer, S.~W. Keckler, and W.~J. Dally, ``{SCNN}: An accelerator for
  compressed-sparse convolutional neural networks,'' in {\em Proceedings of the
  44th Annual International Symposium on Computer Architecture}, ISCA ’17,
  (New York, NY, USA), p.~27–40, Association for Computing Machinery, 2017.

\bibitem{SparTen}
A.~Gondimalla, N.~Chesnut, M.~Thottethodi, and T.~N. Vijaykumar, ``{SparTen}: A
  sparse tensor accelerator for convolutional neural networks,'' in {\em
  Proceedings of the 52nd Annual IEEE/ACM International Symposium on
  Microarchitecture}, MICRO '19, Association for Computing Machinery, 2019.

\bibitem{zhang2020sparch}
Z.~Zhang, H.~Wang, S.~Han, and W.~J. Dally, ``{SpArch}: Efficient architecture
  for sparse matrix multiplication,'' in {\em 2020 IEEE International Symposium
  on High Performance Computer Architecture (HPCA)}, pp.~261--274, IEEE, 2020.

\bibitem{Scalpel}
J.~{Yu}, A.~{Lukefahr}, D.~{Palframan}, G.~{Dasika}, R.~{Das}, and S.~{Mahlke},
  ``Scalpel: Customizing {DNN} pruning to the underlying hardware
  parallelism,'' in {\em 2017 ACM/IEEE 44th Annual International Symposium on
  Computer Architecture (ISCA)}, pp.~548--560, 2017.

\bibitem{SIGMA}
E.~{Qin}, A.~{Samajdar}, H.~{Kwon}, V.~{Nadella}, S.~{Srinivasan}, D.~{Das},
  B.~{Kaul}, and T.~{Krishna}, ``{SIGMA}: A sparse and irregular {GEMM}
  accelerator with flexible interconnects for {DNN} training,'' in {\em 2020
  IEEE International Symposium on High Performance Computer Architecture
  (HPCA)}, pp.~58--70, 2020.

\bibitem{Cambricon}
S.~{Zhang}, Z.~{Du}, L.~{Zhang}, H.~{Lan}, S.~{Liu}, L.~{Li}, Q.~{Guo},
  T.~{Chen}, and Y.~{Chen}, ``Cambricon-{X}: an accelerator for sparse neural
  networks,'' in {\em 2016 49th Annual IEEE/ACM International Symposium on
  Microarchitecture (MICRO)}, pp.~1--12, 2016.

\bibitem{Bit-Tactical}
A.~Delmas~Lascorz, P.~Judd, D.~M. Stuart, Z.~Poulos, M.~Mahmoud, S.~Sharify,
  M.~Nikolic, K.~Siu, and A.~Moshovos, ``{Bit-Tactical}: A software/hardware
  approach to exploiting value and bit sparsity in neural networks,'' in {\em
  Proceedings of the Twenty-Fourth International Conference on Architectural
  Support for Programming Languages and Operating Systems}, ASPLOS ’19, (New
  York, NY, USA), p.~749–763, Association for Computing Machinery, 2019.

\bibitem{Bit-Pragmatic}
J.~Albericio, A.~Delm\'{a}s, P.~Judd, S.~Sharify, G.~O'Leary, R.~Genov, and
  A.~Moshovos, ``Bit-pragmatic deep neural network computing,'' in {\em
  Proceedings of the 50th Annual IEEE/ACM International Symposium on
  Microarchitecture}, MICRO-50 '17, (New York, NY, USA), p.~382–394,
  Association for Computing Machinery, 2017.

\bibitem{OuterSPACE}
S.~{Pal}, J.~{Beaumont}, D.~{Park}, A.~{Amarnath}, S.~{Feng}, C.~{Chakrabarti},
  H.~{Kim}, D.~{Blaauw}, T.~{Mudge}, and R.~{Dreslinski}, ``{OuterSPACE}: An
  outer product based sparse matrix multiplication accelerator,'' in {\em 2018
  IEEE International Symposium on High Performance Computer Architecture
  (HPCA)}, pp.~724--736, 2018.

\bibitem{Laconic}
S.~{Sharify}, A.~D. {Lascorz}, M.~{Mahmoud}, M.~{Nikolic}, K.~{Siu}, D.~M.
  {Stuart}, Z.~{Poulos}, and A.~{Moshovos}, ``Laconic deep learning inference
  acceleration,'' in {\em 2019 ACM/IEEE 46th Annual International Symposium on
  Computer Architecture (ISCA)}, pp.~304--317, 2019.

\bibitem{BitFusion}
H.~{Sharma}, J.~{Park}, N.~{Suda}, L.~{Lai}, B.~{Chau}, V.~{Chandra}, and
  H.~{Esmaeilzadeh}, ``{Bit Fusion}: Bit-level dynamically composable
  architecture for accelerating deep neural networks,'' in {\em 2018 ACM/IEEE
  45th Annual International Symposium on Computer Architecture (ISCA)},
  pp.~764--775, 2018.

\bibitem{SparseTensorCore}
M.~Zhu, T.~Zhang, Z.~Gu, and Y.~Xie, ``Sparse tensor core: Algorithm and
  hardware co-design for vector-wise sparse neural networks on modern {GPU}s,''
  in {\em Proceedings of the 52nd Annual IEEE/ACM International Symposium on
  Microarchitecture}, MICRO '52, (New York, NY, USA), pp.~359--371, Association
  for Computing Machinery, 2019.

\bibitem{PermDNN}
C.~{Deng}, S.~{Liao}, Y.~{Xie}, K.~K. {Parhi}, X.~{Qian}, and B.~{Yuan},
  ``Perm{DNN}: Efficient compressed {DNN} architecture with permuted diagonal
  matrices,'' in {\em 2018 51st Annual IEEE/ACM International Symposium on
  Microarchitecture (MICRO)}, pp.~189--202, 2018.

\bibitem{Outlier}
E.~{Park}, D.~{Kim}, and S.~{Yoo}, ``Energy-efficient neural network
  accelerator based on outlier-aware low-precision computation,'' in {\em 2018
  ACM/IEEE 45th Annual International Symposium on Computer Architecture
  (ISCA)}, pp.~688--698, 2018.

\bibitem{PBEDNN}
M.~{Song}, J.~{Zhao}, Y.~{Hu}, J.~{Zhang}, and T.~{Li}, ``Prediction based
  execution on deep neural networks,'' in {\em 2018 ACM/IEEE 45th Annual
  International Symposium on Computer Architecture (ISCA)}, pp.~752--763, 2018.

\bibitem{rhu2018compressing}
M.~Rhu, M.~O'Connor, N.~Chatterjee, J.~Pool, Y.~Kwon, and S.~W. Keckler,
  ``Compressing {DMA} engine: Leveraging activation sparsity for training deep
  neural networks,'' in {\em 2018 IEEE International Symposium on High
  Performance Computer Architecture (HPCA)}, pp.~78--91, IEEE, 2018.

\end{thebibliography}

\ifdefined\showappendices
\clearpage
\pagebreak
\newpage
\begin{appendices}
\vspace{0.1in}
\setstretch{0.96}
\section{General Matrix Multiply (GEMM) and \etpu{}'s conv2D instruction}
\label{app:gemm}
\subsection{GEMM and the \inst{FullyConnected} operator}
\label{app:cmf_gemm_fc}
GEMM takes two 2-dimensional tensors (matrices) as inputs and produces a single
2-dimensional tensor as output. Assume $X_{i,j}$ represents an element in the $i$th row
and the $j$th column. We can calculate each element in the result matrix, $C$, obtained from multiplying an $M \times N$ matrix, $A$,
and an $N \times K$ matrix, $B$, as follows:
\vspace{.05in}
\begin{equation}
\label{eq4}
C_{i,k}=\sum_{j=0}^{N-1}A_{i,j}\cdot B_{j,k}, 
\forall\ 0\leq i< M, 0\leq k < K
\end{equation}
In short, GEMM yields a set of pairwise multiplications and accumulations. A program can select either matrix $A$ or matrix $B$ 
and iterate through a column or row of the other
matrix to produce the result, and matrix multiplication will be performed via the $M$ or $K$ \inst{FullyConnected} operators. 

\CMFdel{
The results in Section~\ref{sec:characteristics} \CMF{of this paper }show that the goodput of \inst{conv2D} is
27\x{} the goodput of \inst{FullyConnected}. 
The \inst{conv2D}  instruction also performs multiplications and accumulations,
but in different orientations to derive the result. Therefore, 
by changing the shape of input data and using \inst{conv2D} to perform exactly the 
same number of multiplications and accumulations on the set of
input numbers, we can potentially leverage the high goodput of \inst{conv2D}
to implement a more efficient GEMM function. 
}
\subsection{The \inst{conv2D} operator/instruction}
\label{app:cmf_conv2d}
\CMFdel{The \inst{conv2D} instruction takes one of its inputs as
the kernel, rotates the kernel orientation by 180$^{\circ}$, multiplies each rotated 
kernel element with an input element mapping to the same location, and accumulates 
the result as an output element. }For an $M \times N$ input matrix, $A$, and an $L \times L$ 
kernel, $B'$, each element in the \inst{conv2D} $M \times N$ output matrix, $C$,
is:
\vspace{.05in}
\begin{equation}
\label{eq5}
C_{i,j}=\sum_{q=-\lfloor\frac{L}{2}\rfloor}^{\lfloor\frac{L}{2}\rfloor}\sum_{p=-\lfloor\frac{L}{2}\rfloor}^{\lfloor\frac{L}{2}\rfloor}A_{i+p,j+q}\cdot
B'_{\lfloor\frac{L}{2}\rfloor-p,\lfloor\frac{L}{2}\rfloor-q}
\end{equation}
\begin{equation}
\forall\ 0\leq i< M, 0\leq j < N \nonumber
\end{equation}
\vspace{.05in}

\begin{figure}[t]
\begin{center}
\includegraphics[width=3.4in]{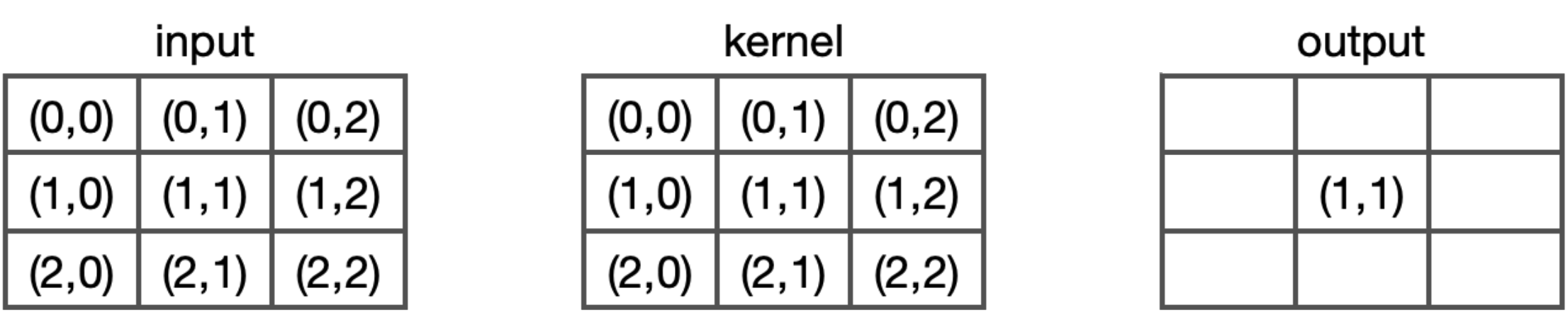}\\(a)\\
\includegraphics[width=3.4in]{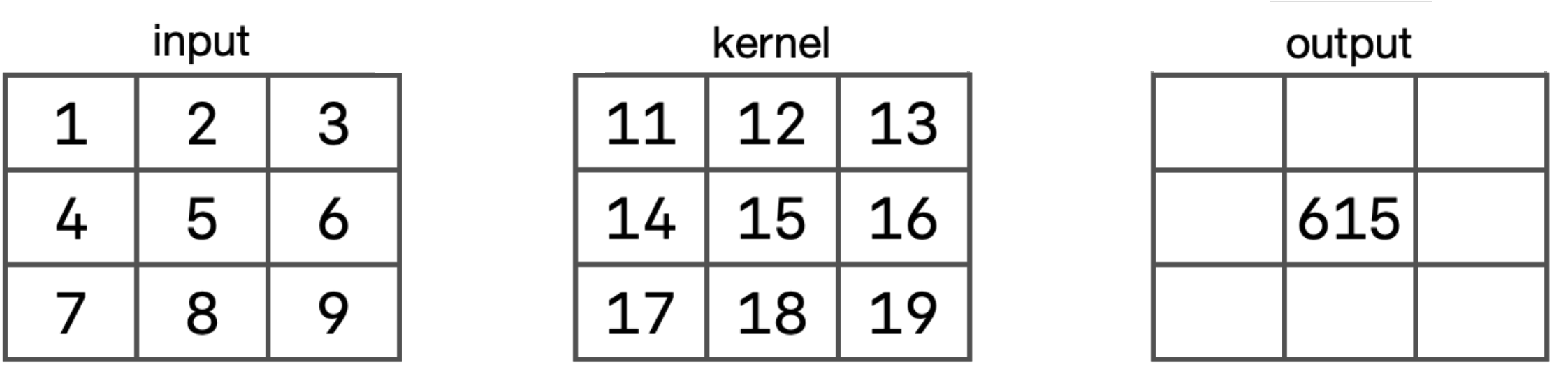}\\(b)\\
\end{center}
\vspace{-0.2in} 
\caption[]{(a) Conceptualization of the \inst{conv2D} operator with 3 $\times$ 3 input and a 3 $\times$ 3 kernel; (b) 
implementation of the \inst{conv2D} instruction with 3 $\times$ 3 input and a 3 $\times$ 3
kernel}
\vspace{.4in}
\label{fig:conv2D_appendix}
\end{figure}

Figure~\ref{fig:conv2D_appendix}(a) illustrates the basic concept of \inst{conv2D} using a 3\x{}3 input matrix and a 3\x{}3
kernel,
with the value of $output[1,1]$
derived as follows:
\vspace{0.1in}

\vspace{.05in}
\noindent$input[0,0]\times kernel[2,2] + input[0,1]\times kernel[2,1] + input[0,2]\times
kernel[2,0] + 
input[1,0]\times kernel[1,2] + input[1,1]\times kernel[1,1] + input[1,2]\times
kernel[0,0] + 
input[2,0]\times kernel[0,2] + input[2,1]\times 
kernel[0,1] + 
input[2,2] \;\; \times \;\; kernel[0,0]$. 
\vspace{0.15in}

\noindent For Figure~\ref{fig:conv2D_appendix}(b), the value of
$output[1,1]$ is 615. 

\begin{figure}[t]
\begin{center}
\begin{tabular}{cc}
\includegraphics[width=1.5in]{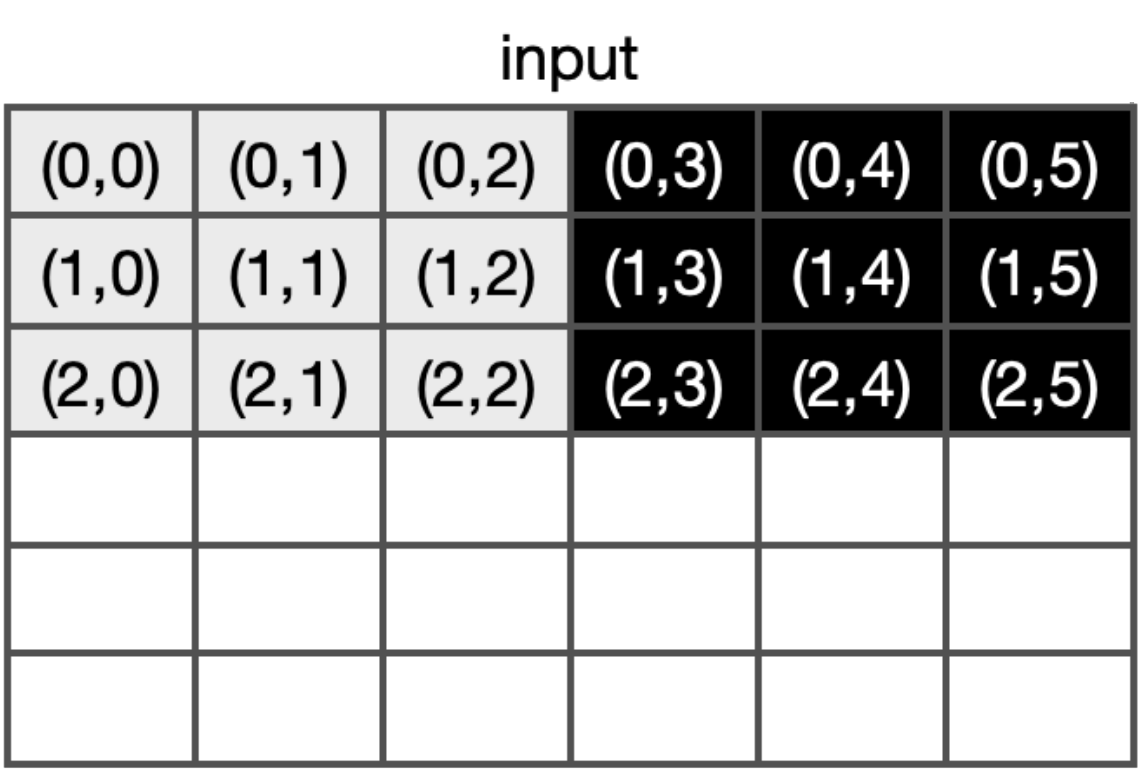} &
\includegraphics[width=1.5in]{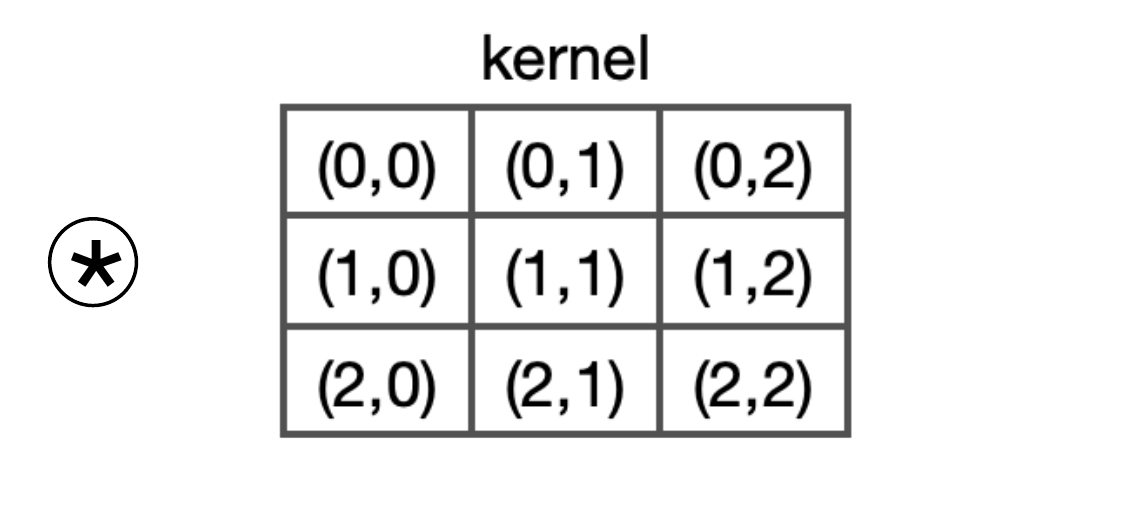} \\
(a) & (b)\\
\includegraphics[width=1.5in]{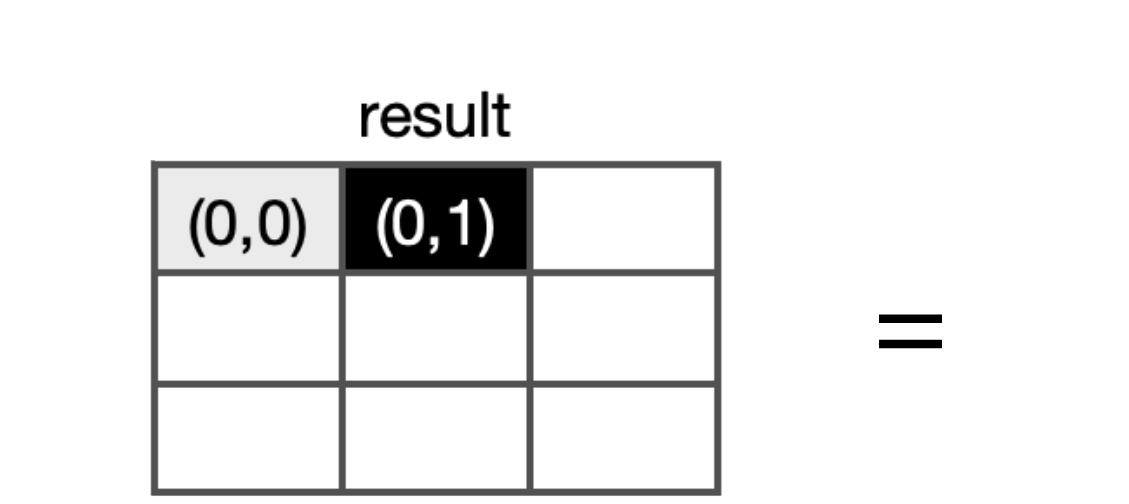} &
\includegraphics[width=1.5in]{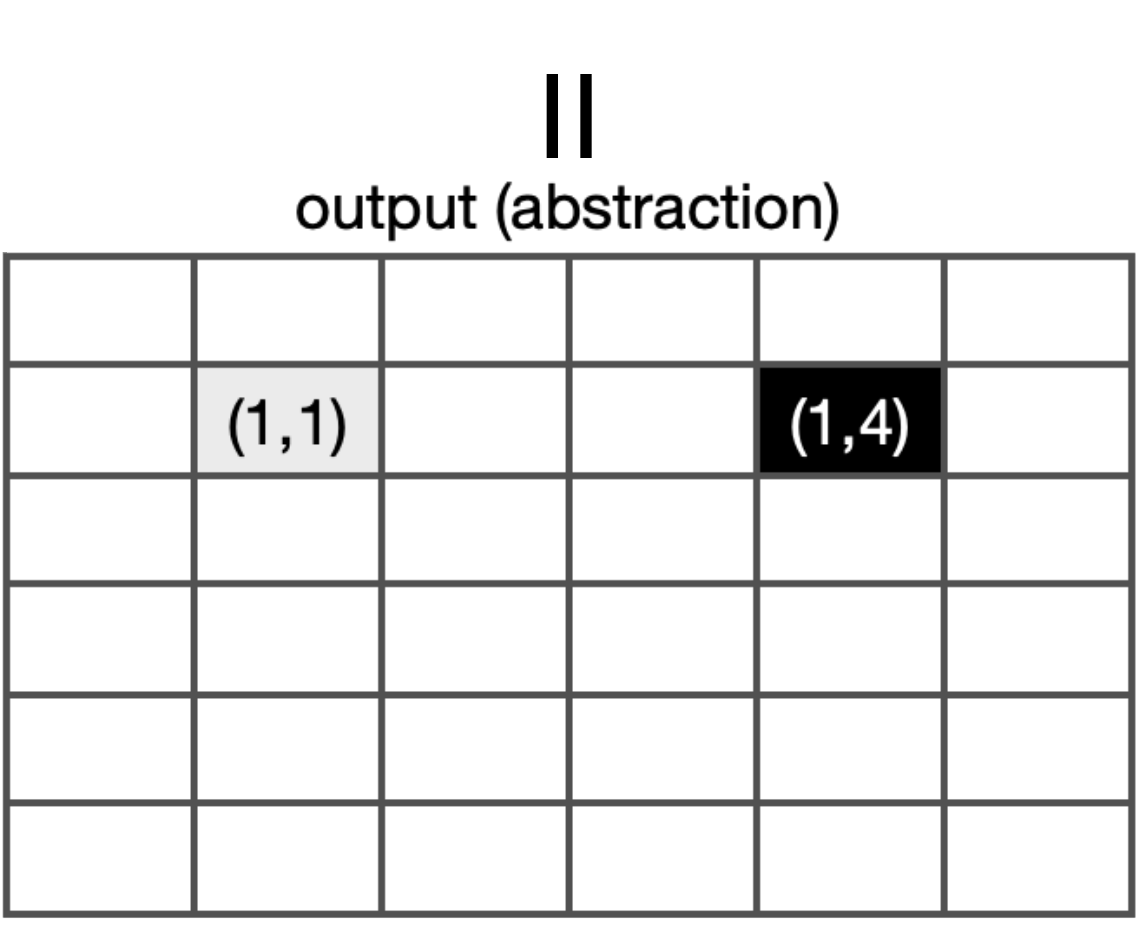} \\
(d) & (c)\\
\end{tabular}
\end{center}
\vspace{-0.2in}
\caption[]{The \inst{conv2D} as implemented with stride}
\label{fig:conv2D}
\vspace{.2in}
\end{figure}

Figure~\ref{fig:conv2D} illustrates the concept of \inst{conv2D} with stride. We select (3, 3) as our
stride, restricting \inst{conv2D} to 9 numbers in a group; the \inst{conv2D} operator only produces a value for every 3 row/column 
elements in the abstracted outcome, as in Figure~\ref{fig:conv2D}(c), from the source matrix, as in Figure~\ref{fig:conv2D}(a), using the kernel
in Figure~\ref{fig:conv2D}(b). The final output of \inst{conv2D} is a condensed matrix, as in Figure~\ref{fig:conv2D}(d).

\ignore{
\CMFdel{. Figure~\ref{fig:conv2D_appendix}(c) extends the example with strides} 
In Figure~\ref{fig:conv2D_appendix}(c), we select (3, 3) as our
stride, \CMFdel{making \inst{conv2D} only perform}\CMF{restricting} \inst{conv2D} \CMFdel{on every}\CMF{to} 9 numbers in a group\CMF{; the \inst{conv2D} operator}\CMFdel{ and} only produce\CMF{s} a value for every 3 \CMF{row/column }elements \CMFdel{for each row and column }in the \CMFdel{``abstracted''}\CMF{abstracted} outcome. \CMF{The }\inst{conv2D} \CMF{operator}\CMFdel{will} condense\CMF{s} the
abstracted outcome into a dense matrix \CMFdel{as the result matrix in the}\CMF{as in}
Figure~\ref{fig:conv2D_appendix}(c).
}
\CMFdel{
\GPTPU{} uses the \inst{conv2D} instruction and the striding feature as an efficient 
GEMM algorithm that initially reshapes both inputs. \GPTPU{} transforms each row in the 
$M \times N$ source matrix into a sub-matrix whose size is determined by the selected 
stride $(s_x, s_y)$. Ordinarily, both $s_x$ and $s_y$ are the square root of $N$. The other
input matrix, sized $N \times K$, serves as a list of kernels, where
each kernel of size $s_x \times s_y$ contains a column from that matrix. 
When creating the kernels, the \GPTPU{} GEMM algorithm flips the original column elements
by filling the kernel elements from the end of the
column. After transforming both inputs, the \inst{conv2D} operator iterates through all 
sub-matrices over each kernel with the selected stride and generates 
output comparable to that of conventional matrix multiplication.
}

\CMFdel{
Though the \GPTPU{} GEMM algorithm incurs additional data-transformation overhead and
is inefficient when implemented in conventional architectures, \CMFdel{however, the highly
optimized}the highly-optimized \etpu{} \inst{conv2D} allows the proposed
algorithm  to significantly outperforms the conventional matrix multiplication and compensates for
the the additional overhead. }
\subsection{\inst{FullyConnected} and \inst{conv2D} together}
\label{app:cmf_fc_conv2d}

Figure~\ref{fig:mvconv2D} shows the performance
of \GPTPU{} GEMM kernel implementations using \inst{FullyConnected} and
\inst{conv2D} compared to the CPU baseline. The \inst{conv2D}
implementation reveals a strong performance gain (a 2.06\x{} speedup in the
4K\x{}4K microbenchmark) over the CPU baseline. In
contrast, the \GPTPU{} GEMM implementation cannot beat the CPU baseline without \inst{conv2D} (i.e., when GEMM only uses \inst{FullyConnected}).


\CMFdel{
As Section~\ref{sec:decipher} shows, \inst{conv2D} performs
best when the inputs are in groups of 128\x{}128. The \CMFdel{programmer may use
similar techniques as} blocking algorithm and other conventional matrix-multiplication optimization techniques may be used to further improve performance
or create parallelism among multiple \etpu{}s. }
\vspace{.3in}

\cfigure[Figures/mvconv2D.pdf, {Speedup of GEMM \GPTPU{} implementations
using \inst{FullyConnected} and \inst{conv2D}, relative to the baseline CPU
CBLAS implementations.}
\vspace{.2in}
,fig:mvconv2D]

\section{Other \GPTPU{} Applications}
\label{app:other_apps}

We provide implementation details complementary to Section~\ref{sec:other_apps}
for PageRank, HotSpot3D, LUD, Backprop, and BlackScholes. 

\subsection{PageRank}
\label{sec:pagerank}
As with most graph applications, PageRank takes as input an adjacency matrix where the element in the $i$th row and $j$th column represents the in-degrees from
node $j$ to node $i$. During the first iteration of the power method, \GPTPU{} initializes each element in
a vector as $1/$($total\ number\ of\ nodes$). Subsequent
iterations multiply the adjacency matrix by the vector from the previous iteration. \GPTPU{}'s PageRank implementation simply uses the \inst{FullyConnected} operator for adjacency-matrix multiplication.

\subsection{HotSpot3D}
\label{sec:hotspot}
A key point regarding \GPTPU{}'s HotSpot3D implementation: \GPTPU{} will split the input into sub-matrices that multiple \etpu{}s can process in parallel when the HotSpot3D problem size scales over the granularity the \etpu{} supports.

\subsection{LUD}
\label{sec:lud}
LUD is essential to
solutions of square systems of linear equations, matrix inversions, and
determinant calculations. When \GPTPU{} executes each iteration of its recursive algorithm~\cite{I2A} for LUD, the algorithm partitions the input into four sub-matrices and
calculates the bottom right-hand corner sub-matrix ($A_{BR}$) using the remaining three sub-matrices.
The resulting combined matrix serves as the input for the next iteration. 
The input for the first iteration is the initial user input with a 1\x{}1 top left-hand corner sub-matrix ($A_{TL}$), a 1\x{}$N$ top right-hand sub-matrix
($A_{TR}$), and an $M$\x{}1 bottom left-hand sub-matrix ($A_{BL}$). The new $A_{BR}$ is calculated as $A_{BR}
- A_{BL}A_{TR}$. 
After each iteration, the \GPTPU{} LUD algorithm gradually shrinks the size of the bottom right-hand sub-matrix while expanding the other three (the \inst{crop}
operation generates each sub-matrix in each iteration). 
Our implementation uses \inst{FullyConnected} for sub-matrices with one dimension less than 8 and \inst{conv2D} otherwise.
\ignore{
\subsection{\CMFdel{Gaussian elimination (}Gaussian\CMFdel{)}}
\label{sec:gaussian}
Like LUD, Gaussian is a method for solving a system of linear equations.
Gaussian combines row swaps, the scalar multiplication of rows, and row additions until the lower left-hand triangular
matrix contains only zeroes. For Gaussian, \GPTPU{}  uses \inst{MUL} to perform each row reduction. 
}
\subsection{Backprop}
\label{sec:backprop}
A key point regarding the derivation of weights for Backprop's delta matrix: \GPTPU{} uses the the matrix multiplication library from Section~\ref{sec:mm} to calculate the weights from the weighted average of the old weight matrix and the outer product of hidden units and output units. 
\subsection{BlackScholes}
\label{sec:BlackScholes}
BlackSholes involves a series of basic arithmetic operations and the execution of \inst{exp} and \inst{log} on each data point. BlackSchole's reliance on the cumulative normal distribution function is critical to performance. \GPTPU{} speeds up processing substantially by using a ninth-degree polynomial function~\cite{CNDF} with \inst{FullyConnected} to approximate the cumulative normal distribution function. 

\end{appendices}

\fi
\end{document}